\documentclass[12pt]{article} 
\usepackage[sectionbib]{natbib}
\usepackage{array,epsfig,rotating}
\usepackage[]{hyperref}  
\usepackage{sectsty, secdot}
\sectionfont{\fontsize{12}{14pt plus.8pt minus .6pt}\selectfont}
\renewcommand{\theequation}{\thesection\arabic{equation}}
\subsectionfont{\fontsize{12}{14pt plus.8pt minus .6pt}\selectfont}

\usepackage{amsmath}
\usepackage{amssymb}
\usepackage{amsfonts}
\usepackage{multirow}
\usepackage{amsthm}
\usepackage{threeparttable}
\usepackage{rotating}
\usepackage{bm}
\usepackage{apalike}
\usepackage{caption}
\usepackage{enumerate}
\usepackage{mathrsfs}
\usepackage{algorithm}
\usepackage{graphicx}
\usepackage{setspace}

\usepackage[usenames,dvipsnames,svgnames,table]{xcolor}

\usepackage{soul,color}

\usepackage{booktabs}
\usepackage{epstopdf}
\usepackage[left=1in,right=1in,top=1in,bottom=1in]{geometry}

\numberwithin{equation}{section}

\newtheorem{thm}{Theorem}

\newtheorem{assumption}{Assumption}
\theoremstyle{definition}

\newtheorem{remark}{Remark}

\begin{document}


\renewcommand{\baselinestretch}{2}

%
%


\fontsize{12}{14pt plus.8pt minus .6pt}\selectfont \vspace{0.8pc}

\centerline{\large\bf  WEAK SIGNAL IDENTIFICATION AND INFERENCE}
\vspace{2pt}
\centerline{\large\bf IN PENALIZED LIKELIHOOD MODELS }
\vspace{2pt}
\centerline{\large\bf  FOR CATEGORICAL RESPONSES}
\vspace{.4cm} 
\centerline{Yuexia Zhang, Peibei Shi, Zhongyi Zhu, Linbo Wang and Annie Qu} 
\vspace{.4cm}
 \centerline{\it
The University of Texas at San Antonio, Meta, Fudan University} 
\centerline{\it
University of Toronto and University of California, Irvine} 
\vspace{.55cm} 
\fontsize{9}{11.5pt plus.8pt minus
	.6pt}\selectfont


\begin{quotation}
\noindent {\it Abstract:}\\
Penalized likelihood models are widely used to simultaneously select variables and estimate model parameters.  However, the existence of weak signals can lead to inaccurate variable selection, biased parameter estimation, and invalid inference. Thus, identifying weak signals accurately and making valid inferences are crucial in penalized likelihood models.  We develop a unified approach to identify weak signals and make inferences in penalized likelihood models, including the special case when the responses are categorical. To identify weak signals, we use the estimated selection probability of each covariate as a measure of the signal strength and formulate a signal identification criterion. To construct confidence intervals, we propose a two-step inference procedure. Extensive simulation studies show that the proposed  procedure outperforms several existing methods.  We illustrate the proposed method by applying it to the Practice Fusion diabetes data set.

\vspace{9pt}
\noindent {\it Key words and phrases:}
adaptive lasso, de-biased method, model selection, post-selection inference
\par
\end{quotation}\par

\def\thefigure{\arabic{figure}}
\def\thetable{\arabic{table}}

\renewcommand{\theequation}{\thesection.\arabic{equation}}

\fontsize{12}{14pt plus.8pt minus .6pt}\selectfont

\setcounter{section}{1} 
\setcounter{equation}{0} 

\noindent {\bf 1. Introduction}

In the big data era, massive data are collected with large-dimensional covariates. However,  only some of the covariates might be important.  To select the important variables and estimate their effects on the response variable, various penalized likelihood models have been proposed, such as the penalized least squares regression model \citep{tibshirani1996regression,zou2005regularization,tibshirani2005sparsity,yuan2006model,zou2006adaptive,zhang2010nearly}, penalized logistic regression model \citep{park2008penalized,zhu2004classification,wu2009genome}, and penalized Poisson regression model \citep{lambert2005bayesian,jia2019sparse}.

To achieve model selection consistency or the variable screening property for a high-dimensional problem,  a  common  condition  is the ``beta-min'' condition, which requires the nonzero regression coefficients to be sufficiently large \citep{zhao2006model,huang2007asymptotic,van2011adaptive,tibshirani2011regression,zhang2022elastic}. Therefore, classical methods for variable selection often  focus on strong signals that satisfy such a condition. However, if the ``beta-min'' condition is violated, the important variables and unimportant variables may be inseparable, and the true  important variables might not  be selected, even if the sample size goes to infinity \citep{zhang2013multi}.  In finite samples, the estimators shrink the true regression coefficients, owing to the penalty function. When the  signal strength is weak, its coefficient  is more likely to shrink to zero \citep{shi2017weak,liu2020bootstrap}.   Inaccurate variable selection and biased parameter estimation could lead to a poor post-selection inference, for example, the  estimation of  the confidence intervals could be inaccurate. Thus, both strong and weak signals need to be considered. Identification and inference for weak signals can also help  discover potentially important variables in practice. For example, in genome-wide association studies (GWAS),  overlooked risk factors  for  a disease may be recovered by incorporating weak signals \citep{liu2020bootstrap}.

For  linear regression models, studies have been done on weak signals.  In more extreme cases, \cite{jin2014optimality} assumed all signals were individually weak and proposed graphlet screening for variable selection. \cite{zhang2017recovery} proposed the perturbed lasso, where signals were strengthened by adding random perturbations to the design matrix. However, these methods  focused only on  variable selection consistency, and did not aim to identify weak signals or provide statistical inference. For weak signal identification and inference, \cite{shi2017weak}  proposed a weak signal identification procedure in finite samples, and introduced a two-step inference method for constructing confidence intervals after signal identification. However, their derivation relies on a crucial assumption that  the design matrix is orthogonal, which may not hold in practice. On the other hand, \cite{li2019weak} took advantage of the correlations between covariates,  detecting  weak signals  through the partial correlations between strong and weak signals.
However, they did not study weak signal inference. 
Recently, \cite{liu2020bootstrap} proposed a  method that combines  the bootstrap lasso and a partial ridge regression  for  constructing confidence intervals when there are weak signals in the covariates. 
However, as stated in their paper, the confidence intervals of the coefficients, with magnitudes of order $1/\sqrt{n}$, may be invalid.

To the best of our knowledge, there has been little work on weak signals in likelihood-based models for categorical responses.  One exception is \cite{reangsephet2020weak}, who proposed  variable selection methods for logistic regression models with weak signals. However, they did not conduct weak signal identification or inference. 

We address these gaps by developing a new unified approach to weak signal identification and inference in penalized likelihood models, including the special case when the responses are categorical. Specifically, the estimated probability of each covariate being selected by the one-step adaptive lasso estimator is used to measure the signal strength. After signal identification, a two-step inference procedure is proposed to construct the confidence intervals for the regression coefficients. 
The proposed method has several advantages.
First, we extend the method of  \cite{shi2017weak} from  linear regression models to  likelihood-based models, including generalized linear models. However, our extension is not trivial.  For example, in \cite{shi2017weak}, the selection probability has an explicit expression. For the proposed likelihood-based method, such an explicit expression  does not exist for categorical responses. Thus, we propose a new method to estimate the selection probability. 
Second, in \cite{shi2017weak}, the selection probability for the covariate $\bm{X}_j$ is an increasing function of $|\beta_{j0}|$, where $\beta_{j0}$ is the corresponding coefficient of $\bm{X}_j$. Under our current general framework, such a conclusion is not necessarily true. Thus, our signal identification criterion is based directly on the estimated selection probability, in contrast to \cite{shi2017weak}. We also discuss how each signal's selection probability is influenced by other covariates, owing to  nonlinear modeling or collinearity among the covariates;  in \cite{shi2017weak}, the selection probability of one covariate is independent of those of other covariates. 
Third, \cite{shi2017weak} assumed that the design matrix  in a linear regression model is orthogonal, whereas the proposed method relaxes this constraint.
Fourth, the proposed inference method differs  from that of  \cite{shi2017weak}. Specifically, we construct  confidence intervals for the noise variables as well, whereas their method does not. Simulation results show  that our proposed two-step inference method outperforms the  two-step inference method based on \cite{shi2017weak}. In particular, the proposed confidence intervals  achieve accurate  coverage probabilities for all signal strength levels.

The remainder of this paper is organized as follows. In Section \ref{sec:estimator}, we introduce the one-step adaptive lasso estimator and derive the variable selection condition. In Section  \ref{sec:weakdefiden}, we propose the weak signal identification criterion. In Section \ref{sec:weakinfer}, we develop a two-step inference procedure for constructing confidence intervals. In Section \ref{sec:simu}, we conduct simulation studies  to assess the finite-sample performance of the proposed method. In Section \ref{sec:realdata}, we apply the proposed method to an analysis  of diabetes data.  In Section \ref{sec:discussion}, we provide  brief concluding remarks. We provide the  technical  proofs,  implementation details of several methods, and some additional results  in  the  Supplementary Material.

\section{One-Step Adaptive Lasso Estimator and Variable Selection Condition}
\label{sec:estimator}
In this section, we introduce  the one-step penalized likelihood estimator and derive the condition  for variable selection, which we use later for  weak signal identification and inference. 

Let $(\mathbf{x}_{1}^\top,y_{1})^\top,\ldots,(\mathbf{x}_{n}^\top,y_{n})^\top$ be $n$ independent and identically distributed (i.i.d.) random vectors, where $\mathbf{x}_{i}=(x_{i1},\ldots,x_{ip})^\top$ is a $p\times1$ vector of predictors and $y_{i}$ is a response variable. Assume that $y_{i}$ depends on $\mathbf{x}_{i}$ through a linear combination $\mathbf{x}_{i}^\top\bm{\beta}_{0}$, and the conditional log-likelihood of $y_{i}$ given $\mathbf{x}_{i}$ is $\ell_{i}(\bm{\gamma}_{0})=\ell_{i}(\alpha_{0}+\mathbf{x}_{i}^\top\bm{\beta}_{0},y_{i})$, where $\bm{\gamma}_{0}=(\alpha_{0},\bm{\beta}_{0}^\top)^\top$, $\alpha_{0}$ is an unknown true location parameter, and $\bm{\beta}_{0}=(\beta_{10},\cdots,\beta_{p0})^\top$ is an unknown $p\times1$ vector of covariate effects. Note that for a likelihood-based model, it is not always possible to eliminate the location parameter  by centering the covariates and the response variable.  For simplicity, assume $p<n$ and $p$ is fixed. Let  $\ell(\bm{\gamma})=\sum_{i=1}^{n}\ell_{i}(\bm{\gamma})$ denote the log-likelihood. Assume $\bm{\gamma}^{(0)}$ is the maximum likelihood estimator of $\bm{\gamma}_{0}$; then, $\bm{\gamma}^{(0)}=(\alpha^{(0)},\bm{\beta}^{(0)\top})^\top={\rm argmax}_{\bm{\gamma}}\ell(\bm{\gamma})$.  In matrix notation, we set $\mathbf{X}=(\mathbf{x}_{1},\ldots,\mathbf{x}_{n})^\top=(\bm{X}_{1},\ldots,\bm{X}_{p})$, with $\bm{X}_{j}=(x_{1j},\ldots,x_{nj})^\top$ and $\bm{Y}=(y_{1},\ldots,y_{n})^\top$. Furthermore, denote $\widetilde{\mathbf{x}}_{i}=(1,\mathbf{x}_{i}^\top)^\top$ and $\widetilde{\mathbf{X}}=(\bm{1},\mathbf{X})$, where  $\bm{1}$ is an $n\times 1$ vector with all  elements equal to one. Throughout this paper, we assume that ${\rm E}(x_{ij})=0$ and ${\rm Var}(x_{ij})=1$, for all  $i\in\{1,\ldots,n\}$ and $j\in\{1,\ldots,p\}$, which can be realized by standardizing the covariate matrix $\mathbf{X}$, in practice. 

Assume that some components of  $\bm{\beta}_{0}$ are zero. In order to estimate the model parameters and select important variables simultaneously,  we consider the penalized likelihood function  $\ell(\bm{\gamma})/n-\sum_{j=1}^{p}p_{\lambda_{j}}(|\beta_{j}|)$,  where $p_{\lambda_{j}}(\cdot)$ is a penalty function controlled by the tuning parameter $\lambda_{j}$. One popular penalty function is derived from the adaptive lasso estimator \citep{zou2006adaptive}, where $p_{\lambda_{j}}(|\beta_{j}|)=\lambda |\beta_{j}|/|\beta_{j}^{(0)}|$.
Maximizing the penalized likelihood function is equivalent to minimizing 
\begin{equation}
	-\frac{1}{n}\ell(\bm{\gamma})+\sum_{j=1}^{p}p_{\lambda_{j}}(|\beta_{j}|)
	\label{eq:originalobj}
\end{equation}
with respect to $\bm{\gamma}$. According to \cite{wang2007unified} and \cite{zou2008one},  if the log-likelihood function has first and second derivatives, then it can be approximated by a  Taylor expansion. Furthermore,  the objective function (\ref{eq:originalobj}) can be approximated by
\begin{equation}
	Q_{1}(\bm{\gamma})=-\frac{1}{2n}(\bm{\gamma}-\bm{\gamma}^{(0)})^\top\ddot{\ell}(\bm{\gamma}^{(0)})(\bm{\gamma}-\bm{\gamma}^{(0)})+\sum_{j=1}^{p}p_{\lambda_{j}}(|\beta_{j}|),
	\label{eq:objective1}
\end{equation}
where $\ddot{\ell}(\cdot)$ is the second  derivative of function $\ell(\cdot)$. The one-step penalized likelihood estimator is $\bm{\gamma}^{(1)}=(\alpha^{(1)},\bm{\beta}^{(1)\top})^\top={\rm argmin}_{\bm{\gamma}}Q_{1}(\bm{\gamma})$.

Denote $\mu_{i}(\bm{\gamma})=\mu_{i}=\widetilde{\mathbf{x}}_{i}^\top \bm{\gamma}$ and $\ell_{i}\{\mu_{i}(\bm{\gamma})\}=\ell_{i}(\widetilde{\mathbf{x}}_{i}^\top \bm{\gamma},y_{i})$. Let $\mathbf{D}(\bm{\gamma})$ be an $n\times n$ diagonal matrix with  the $(i,i)$th element  $D_{ii}(\bm{\gamma})=-\partial^{2}\ell_{i}\{\mu_{i}(\bm{\gamma})\}/\partial \mu_{i}^{2}$, for $i=1,\ldots,n$. Then, $\ddot{\ell}(\bm{\gamma})=-\widetilde{\mathbf{X}}^\top\mathbf{D}(\bm{\gamma})\widetilde{\mathbf{X}}$. Furthermore, we assume $D_{ii}(\bm{\gamma})$ is a continuous function of $\bm{\gamma}$. For simplicity, denote $\mathbf{D}(\bm{\gamma}^{(0)})$,  $\mathbf{D}(\bm{\gamma}_{0})$, $D_{ii}(\bm{\gamma}^{(0)})$,  and $D_{ii}(\bm{\gamma}_{0})$ as $\mathbf{D}^{(0)}$,  $\mathbf{D}_{0}$,  $D_{ii}^{(0)}$, and  $D_{0,ii}$, respectively.  By solving the equation $\partial Q_{1}(\bm{\gamma})/\partial \alpha=0$, we  obtain that 
\begin{equation}
	\alpha-\alpha^{(0)}=(\bm{1}^\top\mathbf{D}^{(0)}\bm{1})^{-1}\bm{1}^\top\mathbf{D}^{(0)}\mathbf{X}(\bm{\beta}^{(0)}-\bm{\beta}).
	\label{eq:alpha}
\end{equation}
Replacing $\alpha-\alpha^{(0)}$ by  \eqref{eq:alpha}  in  (\ref{eq:objective1}), we obtain the  following objective function $Q_{2}(\bm{\beta})$: 
\begin{equation}
	\begin{aligned}
		Q_{2}(\bm{\beta})
		&=\frac{1}{2n}(\bm{\beta}-\bm{\beta}^{(0)})^\top\mathbf{X}^\top\mathbf{D}^{\dagger(0)}\mathbf{X}(\bm{\beta}-\bm{\beta}^{(0)})+\sum_{j=1}^{p}p_{\lambda_{j}}(|\beta_{j}|)\\
		&=\frac{1}{2n}(\bm{\beta}-\bm{\beta}^{(0)})^\top\mathbf{X}^\top\mathbf{D}^{\star(0)\top}\mathbf{D}^{\star(0)}\mathbf{X}(\bm{\beta}-\bm{\beta}^{(0)})+\sum_{j=1}^{p}p_{\lambda_{j}}(|\beta_{j}|),
		\label{eq:beta}
	\end{aligned}
\end{equation}
where $\mathbf{D}^{\dagger(0)}=\mathbf{D}^{(0)}-\mathbf{D}^{(0)}\bm{1}(\bm{1}^\top\mathbf{D}^{(0)}\bm{1})^{-1} \bm{1}^{\top}\mathbf{D}^{(0)}$ and
$\mathbf{D}^{\star(0)}=(\mathbf{D}^{(0)})^{1/2}-(\mathbf{D}^{(0)})^{1/2}\bm{1}(\bm{1}^\top\mathbf{D}^{(0)}\bm{1})^{-1}\\\times \bm{1}^{\top}\mathbf{D}^{(0)}$. 
Denote $\mathbf{D}_{0}^{\dagger}=\mathbf{D}_{0}-\mathbf{D}_{0}\bm{1}(\bm{1}^\top\mathbf{D}_{0}\bm{1})^{-1}\bm{1}^{\top}\mathbf{D}_{0}$ and
$\mathbf{D}_{0}^{\star}=\mathbf{D}_{0}^{1/2}-\mathbf{D}_{0}^{1/2}\bm{1}(\bm{1}^\top\mathbf{D}_{0}\bm{1})^{-1}\bm{1}^{\top}\mathbf{D}_{0}$, correspondingly.

We focus mainly on weak signal identification using the one-step adaptive lasso estimator. However, our method can  be extended to other  penalized likelihood estimators.  Following  the idea of \cite{zou2008one}, the algorithm for computing the one-step adaptive lasso estimator $\bm{\gamma}^{(1)}$ is as follows:
\begin{enumerate}[Step 1.]
	\item  Create the working data by $	\mathbf{X}^{\star}=\mathbf{D}^{\star(0)}\mathbf{X}\mathbf{W}$ and  $\bm{Y}^{\star}=\mathbf{D}^{\star(0)}\mathbf{X}\bm{\beta}^{(0)}$,
	where $\mathbf{W}={\rm diag}\{|\beta_{1}^{(0)}|,\ldots,|\beta_{p}^{(0)}|\}$.
	\item Apply the coordinate descent algorithm to solve
	\begin{equation}
		\hat{\bm{\beta}}^{\star}=\mathop{\rm argmin}_{\bm{\beta}}\left\{\frac{1}{2n}\sum_{i=1}^{n}\left(y_{i}^{\star}-\sum_{j=1}^{p}{x}_{ij}^{\star}\beta_{j}\right)^{2}+\lambda\sum_{j=1}^{p}|\beta_{j}|\right\},
		\label{eq:cdobject}
	\end{equation}
	where $\hat{\bm{\beta}}^{\star}=(\hat{\beta}_{1}^{\star},\ldots,\hat{\beta}_{p}^{\star})^{\top}$, $y_{i}^{\star}$ is the $i$th element of $\bm{Y}^{\star}$ and $x_{ij}^{\star}$ is the $(i,j)$th element of $\mathbf{X}^{\star}$.
	\item Obtain the value of $\bm{\beta}^{(1)}=(\beta_{1}^{(1)},\ldots,\beta_{p}^{(1)})^{\top}$ using $\beta_{j}^{(1)}=\hat{\beta}_{j}^{\star}|\beta_{j}^{(0)}|,$ for $j=1\ldots,p$.
	\item Obtain the value of $\alpha^{(1)}$ as $\alpha^{(1)}=(\bm{1}^\top\mathbf{D}^{(0)}\bm{1})^{-1}\bm{1}^\top\mathbf{D}^{(0)}\mathbf{X}(\bm{\beta}^{(0)}-\bm{\beta}^{(1)})+\alpha^{(0)}$.
\end{enumerate}

From the above algorithm, if $\hat{\beta}_{j}^{\star}\neq 0$, then the covariate $\bm{X}_{j}$ will be  selected. According to (\ref{eq:cdobject}), by using the coordinate descent algorithm, we  obtain that 
\[
\hat{\beta}_{j}^{\star}=s\left\{ \frac{\sum\limits_{i=1}^{n}\Big(y_{i}^{\star}-\sum\limits_{k\neq j}x_{ik}^{\star}\hat{\beta}_{k}^{\star}\Big)x_{ij}^{\star}}{\sum\limits_{i=1}^{n}(x_{ij}^{\star})^{2}   },\frac{n\lambda}{\sum\limits_{i=1}^{n}(x_{ij}^{\star})^{2} } \right\},
\]
where $s(z,r)={\rm sgn}(z)(|z|-r)_{+}$. Then, the condition for $\hat{\beta}_{j}^{\star}\neq 0$  ($\beta_{j}^{(1)}\neq 0$) is 
\begin{equation}
	\left| \frac{\sum\limits_{i=1}^{n}\Big(y_{i}^{\star}-\sum\limits_{k\neq j}x_{ik}^{\star}\hat{\beta}_{k}^{\star}\Big)x_{ij}^{\star}}{\sum\limits_{i=1}^{n}(x_{ij}^{\star})^{2}}\right|>\frac{n\lambda}{\sum\limits_{i=1}^{n}(x_{ij}^{\star})^{2} }.
	\label{condition}
\end{equation}
For each $i\in\{1,\ldots,n\}$ and $s\in\{1,\ldots,n\}$, let $d_{is}^{(0)}$ be the $(i,s)$th element of $\mathbf{D}^{\star(0)}$.  Then the variable selection condition (\ref{condition}) is equivalent to 
\begin{multline}
	\left|\sum_{i=1}^{n}\Big(\sum\limits_{s=1}^{n}d_{is}^{(0)}x_{sj}\Big)^2(\beta_{j}^{(0)})^2+\sum\limits_{k\neq j}\sum\limits_{i=1}^{n}\Big(\sum\limits_{s=1}^{n}d_{is}^{(0)}x_{sk}\Big)\Big(\sum\limits_{s=1}^{n}d_{is}^{(0)}x_{sj}\Big)\beta_{j}^{(0)}(\beta_{k}^{(0)}-\beta_{k}^{(1)})\right|\\
	>n\lambda.
	\label{condition2}
\end{multline}

Similarly to the proof in  \cite{zou2008one}, we obtain that if the tuning parameter $\lambda$ satisfies the conditions of $\sqrt{n}\lambda\rightarrow 0$ and $n\lambda\rightarrow \infty$, then the one-step adaptive lasso estimator enjoys  model selection consistency, and  the nonzero one-step adaptive lasso estimators have the property of asymptotic normality.

\section{Weak Signal Definition and Identification}
\label{sec:weakdefiden}
\subsection{Weak signal definition}
\label{ss:weakdef}
Suppose a model contains both strong  and weak signals. Without loss of generality, assume the  covariate matrix $\mathbf{X}$ consists of three components, that is, $\mathbf{X}=\{\mathbf{X}^{(S)},\mathbf{X}^{(W)},\mathbf{X}^{(N)}\}$, where $\mathbf{X}^{(S)}$, $\mathbf{X}^{(W)}$, and $\mathbf{X}^{(N)}$ represent the subsets of strong signals, weak signals, and noise variables, respectively. Following  \cite{shi2017weak}, we use  the selection probability of each covariate to measure the signal strength. 
\
Specifically, for any penalized model selection estimator $\hat{\bm{\beta}}=(\hat{\beta}_1,\ldots,\hat{\beta}_{p})^{\top}$,  we define $P_{d,j}$ as the probability of selecting the covariate $\bm{X}_{j}$, that is, $P_{d,j}=P(\hat{\beta}_{j}\neq 0)$, $j\in\{1,\ldots,p\}$.
For the one-step adaptive lasso estimator $\bm{\beta}^{(1)}=(\beta_{1}^{(1)},\ldots,\beta_{p}^{(1)})^{\top}$, based on the variable selection condition \eqref{condition2}, $P_{d,j}$  
does not have an  explicit form. However,  in the  Supplementary Material S1,  we show that $P_{d,j}$ can be approximated by $P_{d,j}^{\ast}$, where
\begin{multline}
	P_{d,j}^{\ast}=\Phi\left(\frac{-\sqrt{\lambda{\rm E}(D_{0,ii})/\left[{\rm E}(D_{0,ii} x_{ij}^{2}){\rm E}(D_{0,ii})-\{{\rm E}(D_{0,ii} x_{ij})\}^2\right]} +{\beta}_{j0}}{\sqrt{\{{\rm E}(\widetilde{\mathbf{X}}^\top\mathbf{D}_{0}\widetilde{\mathbf{X}})\}^{-1}_{j+1,j+1}}}   \right)\\
	+\Phi\left(\frac{-\sqrt{\lambda{\rm E}(D_{0,ii})/\left[{\rm E}(D_{0,ii} x_{ij}^{2}){\rm E}(D_{0,ii})-\{{\rm E}(D_{0,ii} x_{ij})\}^2\right]}-{\beta}_{j0}}{\sqrt{\{{\rm E}(\widetilde{\mathbf{X}}^\top\mathbf{D}_{0}\widetilde{\mathbf{X}})\}^{-1}_{j+1,j+1}}}   \right).
	\label{detectionprob}
\end{multline}
Intuitively, in the derivation of the selection probability, we  can omit the terms of (S2) and (S3) in the Supplementary Material S1, and simplify the calculation using asymptotic theory. Then we can relax the orthogonality assumption required in \cite{shi2017weak}.  We require the following mild assumption to ensure \eqref{detectionprob}  is valid. 
\begin{assumption}
	For each $i\in\{1,\ldots,n\}$ and $j\in\{1,\ldots,p\}$, $P(D_{0,ii}>0)=1$, ${\rm E}(D_{0,ii})<\infty$, ${\rm E}(D_{0,ii} x_{ij}^{2})<\infty$, and ${\rm E}(\widetilde{\mathbf{X}}^\top\mathbf{D}_{0}\widetilde{\mathbf{X}})$  is positive definite.
	\label{assumpt1}
\end{assumption}
The condition  $P(D_{0,ii}>0)=1$ implies that the conditional log-likelihood function of $y_i$ given $\mathbf{x}_{i}$, $\ell_{i}\{\mu_{i}(\bm{\gamma})\}$, is a concave function of $\mu_{i}(\bm{\gamma})$. This is a necessary condition for  the uniqueness of the maximum likelihood estimator $\bm \gamma^{(0)}$. In addition, according to the Cauchy--Schwarz inequality, this also ensures that ${\rm E}(D_{0,ii} x_{ij}^{2}){\rm E}(D_{0,ii})-\{{\rm E}(D_{0,ii} x_{ij})\}^2>0$. The conditions of ${\rm E}(D_{0,ii})<\infty$ and ${\rm E}(D_{0,ii} x_{ij}^{2})<\infty$ guarantee that all  expectations of random variables  in \eqref{detectionprob} are bounded for finite $n$. The positive-definite condition of ${\rm E}(\widetilde{\mathbf{X}}^\top\mathbf{D}_{0}\widetilde{\mathbf{X}})$ is a necessary condition for the asymptotic normality of the maximum likelihood estimator $\bm \gamma^{(0)}$, and ensures $\{{\rm E}(\widetilde{\mathbf{X}}^\top\mathbf{D}_{0}\widetilde{\mathbf{X}})\}^{-1}_{j+1,j+1}>0$.

For a deeper understanding of $P_{d,j}^{\ast}$,  we first study the asymptotic  properties of $P_{d,j}^{\ast}$. When ${\beta}_{j0}=0$, 
\[
P_{d,j}^{\ast}=2\Phi\left(\frac{-\sqrt{n\lambda{\rm E}(D_{0,ii})/\left[{\rm E}(D_{0,ii} x_{ij}^{2}){\rm E}(D_{0,ii})-\{{\rm E}(D_{0,ii} x_{ij})\}^2\right]}}{\sqrt{\{{\rm E}(\widetilde{\mathbf{X}}^\top\mathbf{D}_{0}\widetilde{\mathbf{X}})/n\}^{-1}_{j+1,j+1}}}   \right).
\]
Under Assumption  \ref{assumpt1},  $[{\rm E}(D_{0,ii} x_{ij}^{2}){\rm E}(D_{0,ii})-\{{\rm E}(D_{0,ii} x_{ij})\}^2]/{\rm E}(D_{0,ii})$ and  $\{{\rm E}(\widetilde{\mathbf{X}}^\top\mathbf{D}_{0}\widetilde{\mathbf{X}})/n\}^{-1}_{j+1,j+1}$ are both positive and bounded. If $n\lambda\rightarrow \infty$,  then $P_{d,j}^{\ast}\rightarrow 0$.

When ${\beta}_{j0}\neq 0$, 
\begin{align*}
	P_{d,j}^{\ast}=&\Phi\left(\frac{-\sqrt{n}\left[\sqrt{\lambda{\rm E}(D_{0,ii})/\left[{\rm E}(D_{0,ii} x_{ij}^{2}){\rm E}(D_{0,ii})-\{{\rm E}(D_{0,ii} x_{ij})\}^2\right]}-{\beta}_{j0}\right]}{\sqrt{\{{\rm E}(\widetilde{\mathbf{X}}^\top\mathbf{D}_{0}\widetilde{\mathbf{X}})/n\}^{-1}_{j+1,j+1}}}\right)\\
	& +\Phi\left(\frac{-\sqrt{n}\left[\sqrt{\lambda{\rm E}(D_{0,ii})/\left[{\rm E}(D_{0,ii} x_{ij}^{2}){\rm E}(D_{0,ii})-\{{\rm E}(D_{0,ii} x_{ij})\}^2\right]}+{\beta}_{j0}\right]}{\sqrt{\{{\rm E}(\widetilde{\mathbf{X}}^\top\mathbf{D}_{0}\widetilde{\mathbf{X}})/n\}^{-1}_{j+1,j+1}}}\right).
\end{align*}
If $\sqrt{n}\lambda\rightarrow 0$, then $P_{d,j}^{\ast}\rightarrow 1$ under  Assumption \ref{assumpt1}. 

These asymptotic  properties of $P_{d,j}^{\ast}$ are consistent with the conclusion that the one-step adaptive lasso estimator enjoys  model selection consistency if  $\lambda$ satisfies the conditions of $\sqrt{n}\lambda\rightarrow 0$ and $n\lambda\rightarrow \infty$.

In the following, we study the finite-sample properties of $P_{d,j}^{\ast}$. To illustrate, we  first consider three special cases, where the likelihood-based model is taken as a linear regression model, a logistic regression model, and a Poisson regression model, respectively. 

{\bf Case One: Linear regression model}

We first illustrate the simplest case under the  linear regression model setting. Let $y_{i}=\alpha_0+\mathbf{x}_i^\top \bm{\beta}_0+\varepsilon_i$, where $\varepsilon_i\stackrel{i.i.d.}{\sim} \mathcal{N}(0,\sigma^2)$;
then, $D_{0,ii}=1/\sigma^2$.  If we assume ${\rm corr}(x_{ij},x_{ik})=0$ for any $k$, $k\neq j$, then 
\[
P_{d,j}^{\ast}=\Phi\left( \frac{\beta_{j0}-\sqrt{\lambda}\sigma}{\sigma/\sqrt{n}} \right)+\Phi\left( \frac{-\beta_{j0}-\sqrt{\lambda}\sigma}{\sigma/\sqrt{n}} \right).
\]
Note that if the tuning parameter $\lambda$ is replaced by  $\lambda_{Shi}=\lambda \sigma^{2}$, then $P_{d,j}^{\ast}$ has the same form as that in   \cite{shi2017weak}, where the covariate matrix is  assumed to be orthogonal. In this case,  $P_{d,j}^{\ast}$ does not depend on $\bm{\gamma}_{0}^{-j}$, where $\bm{\gamma}_{0}^{-j}$ stands for the components in  $\bm{\gamma}_{0}$ other than ${\beta}_{j0}$. In addition, given any values in $P_{d,j}^{\ast}$ except ${\beta}_{j0}$, $P_{d,j}^{\ast}$ is a symmetric function of $\beta_{j0}$ and increases with $|\beta_{j0}|$. Thus,  both $P_{d,j}^{\ast}$ and $|\beta_{j0}|$ can be used to measure the signal strength of  $\bm{X}_{j}$, as shown in \cite{shi2017weak}.

\begin{figure}[!t]
	\centerline{\includegraphics[width=20em,angle=270]{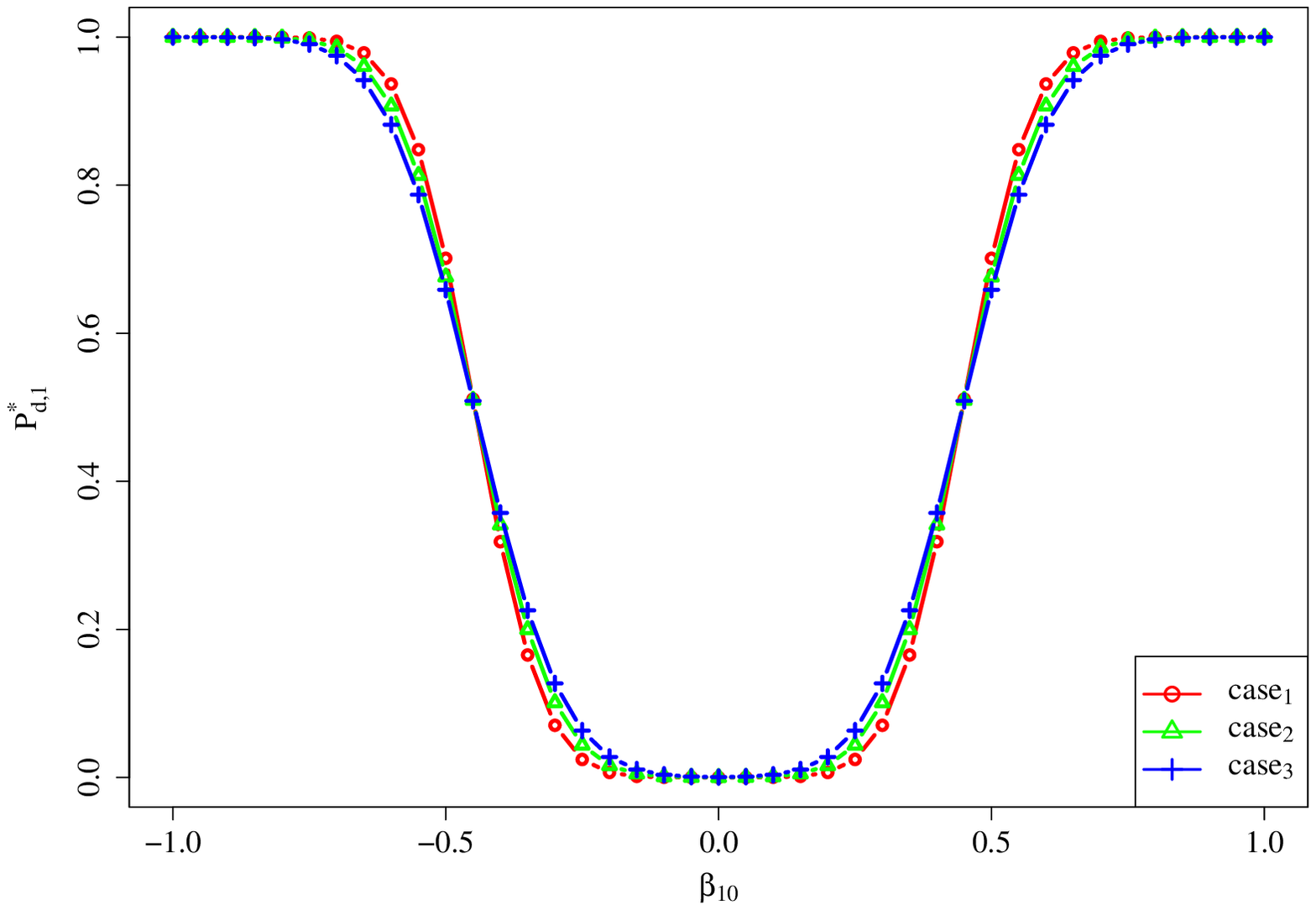}}
	\caption{ The plots for $P_{d,1}^{\ast}$ as $\beta_{10}$ varies under three different cases in linear regression models. In case 1, the correlation structure of $ \mathbf{X}$ is taken to be the independence correlation structure; in case 2, the correlation structure of $ \mathbf{X}$ is taken to be the AR(1) correlation structure with  $\rho=0.5$; in case 3, the correlation structure of $ \mathbf{X}$ is taken to be the exchangeable correlation structure with $\rho=0.5$. In all cases, $n=100$, $p=5$, $\lambda=0.2$, $\sigma=1$, and $\beta_{10}$ varies between $-1$ and $1$, with a step size of $0.05$.}
	\label{figure:linear}
\end{figure}

However, if ${\rm corr}(x_{ij},x_{ik})\neq 0$;  for some $k$, $k\neq j$, then 
\[
P_{d,j}^{\ast}=\Phi\left( \frac{\beta_{j0}-\sqrt{\lambda}\sigma}{\sigma\Big/\left[\sqrt{n}\sqrt{\{{\rm corr}(\tilde{ \mathbf{X}})\}^{-1}_{j+1,j+1}}\right]   } \right)+\Phi\left( \frac{-\beta_{j0}-\sqrt{\lambda}\sigma}{\sigma\Big/\left[\sqrt{n}\sqrt{\{{\rm corr}(\tilde{ \mathbf{X}})\}^{-1}_{j+1,j+1}} \right]  } \right).
\]
Thus, $P_{d,j}^{\ast}$ also depends on  the correlations between covariates. 
Given any values in $P_{d,j}^{\ast}$ except ${\beta}_{j0}$, $P_{d,j}^{\ast}$ is still a symmetric function of $\beta_{j0}$ and  an increasing function of $|\beta_{j0}|$. 
However, under different correlation structures of $\tilde{ \mathbf{X}}$, the shape of $P_{d,j}^{\ast}$ can vary with the value of $\beta_{j0}$. Therefore, both the value of $|\beta_{j0}|$ and the correlation structure of $\tilde{ \mathbf{X}}$  influence the signal strength of $\bm{X}_j$, as illustrated in  Figure \ref{figure:linear}.

{\bf Case Two: Logistic regression model}

Under the logistic regression model setting, 
\[
{\rm E}(y_{i}|\mathbf{x}_{i})=p_i=\frac{\exp(\alpha_{0}+\mathbf{x}_{i}^\top\bm{\beta}_{0} )}{1+\exp(\alpha_{0}+\mathbf{x}_{i}^\top\bm{\beta}_{0} )}.
\]
We obtain that  in \eqref{detectionprob}, $D_{0,ii}=p_{i}(1-p_{i})$ and $\mathbf{D}_{0}={\rm diag}\{p_{1}(1-p_{1}),\ldots,p_{n}(1-p_{n})\}$.
Thus, $P_{d,j}^{\ast}$ not only depends on ${\beta}_{j0}$, but also depends on $\bm{\gamma}_{0}^{-j}$, the coefficients of the other covariates. 
This is a fundamental difference between logistic regression models and linear regression models in terms of selection probability. In contrast to linear regression models, 
$\mathbf{x}_{i}$ influences $P_{d,j}^{\ast}$ through  the matrix ${\rm E}[\widetilde{\mathbf{X}}^\top{\rm diag}\{p_{1}(1-p_{1}),\ldots,p_{n}(1-p_{n})\}\widetilde{\mathbf{X}}]$,  rather than through the correlation matrix of $\tilde{ \mathbf{X}}$, in logistic regression models. 
In addition, in the Supplementary Material S2.1,  we show that $P_{d,j}^{\ast}$ is not necessarily a symmetric function of $\beta_{j0}$, given other values in $P_{d,j}^{\ast}$. 
Thus, $|\beta_{j0}|$ cannot be used to measure the signal strength of  $\bm{X}_{j}$ instead of $P_{d,j}^{\ast}$, which differs from \cite{shi2017weak}.

\begin{figure}[t!]
	\centerline{\includegraphics[width=20em,angle=270]{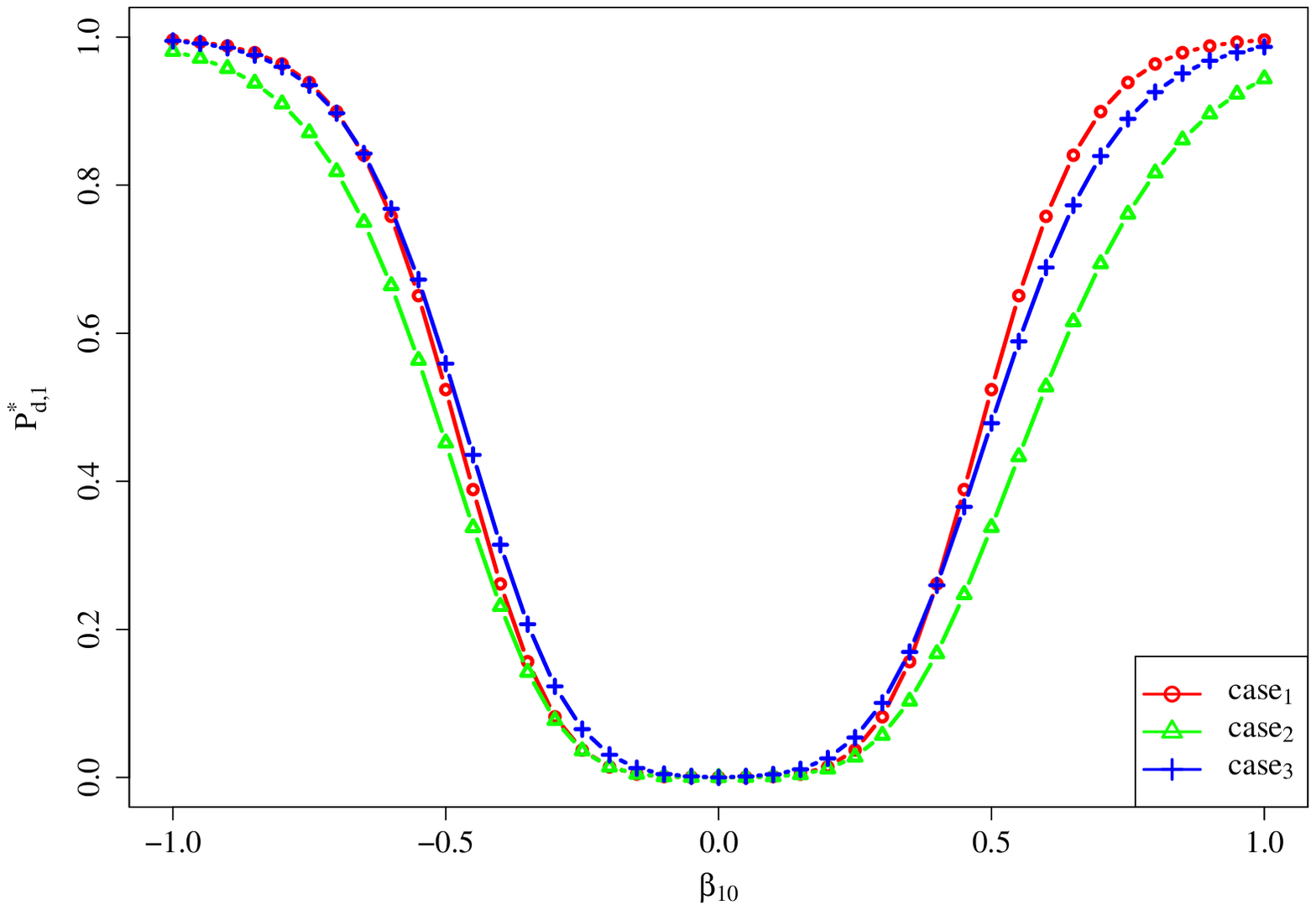}}
	\caption{\small The plots for $P_{d,1}^{\ast}$ as $\beta_{10}$ varies under three different cases in logistic regression models. In case 1, $\bm{X}_1$ and $\bm{X}_2$ both follow the standard normal distribution, and $\bm{X}_1$ and $\bm{X}_2$ are independent; in case 2, $\bm{X}_1$ and $\bm{X}_2$ both follow the centralized exponential distribution with mean zero and variance one, and $\bm{X}_1$ and $\bm{X}_2$ are independent; in case 3, $\bm{X}_1$ and $\bm{X}_2$ both follow the standard normal distribution, and $\bm{X}_1$ and $\bm{X}_2$ have the correlation of $0.5$. In all cases, $n=300$, $\bm \gamma_0=(0.3,\beta_{10},0.2)'$, $\lambda=0.05$, and $\beta_{10}$ varies between $-1$ and $1$, with a step size of $0.05$.}
	\label{figure:logistic}
\end{figure}

In addition, for the logistic regression model, the range of  $\bm{\gamma}_{0}$ is bounded so that  $p_i$ can  satisfy the condition  $0<c_1<p_i<c_2<1$, where $c_1$ and $c_2$ are some positive constants. 
We show that, given any values in $P_{d,j}^{\ast}$ except $\beta_{j0}$,   $P_{d,j}^{\ast}$ is an increasing function of $\beta_{j0}$ if $0<\beta_{j0}<c_3$, and $P_{d,j}^{\ast}$ is a decreasing function of $\beta_{j0}$ if $-c_4<\beta_{j0}<0$, where $c_3$ and $c_4$ are some bounded positive constants depending on $c_1$ and $c_2$. Proofs of the above findings are provided in the  Supplementary Material S2.2. We also illustrate these properties in  Figure \ref{figure:logistic}. Note that in this case, the response variable has  two categories. However, it can be easily extended to the case where there are more than two categories. 

\newpage
{\bf Case Three: Poisson regression model}

Under the Poisson regression model setting, 
\[
P(y_{i}=y|\mathbf{x}_{i})=\frac{\lambda_{i}^{y}}{y!}\exp(-\lambda_{i}),
\]
where $\lambda_{i}={\rm E}(y_{i}|\mathbf{x}_{i})=\exp(\alpha_{0}+\mathbf{x}_{i}^\top\bm{\beta}_{0})$. Then,  in \eqref{detectionprob}, $D_{0,ii}=\lambda_{i}$ and $\mathbf{D}_{0}={\rm diag}\{ \lambda_1,\ldots,\lambda_{n} \}$.
We obtain similar conclusions to those for logistic regression models, except that $P_{d,j}^{\ast}$ is influenced by $\mathbf{x}_{i}$ through  the matrix ${\rm E}[\widetilde{\mathbf{X}}^\top{\rm diag}\{ \lambda_1,\ldots,\lambda_{n} \}\widetilde{\mathbf{X}}]$. 
Note that under Assumption 1, the range of $\bm{\gamma}_{0}$ is bounded.  Given any other values in $P_{d,j}^{\ast}$ except $\beta_{j0}$,   $P_{d,j}^{\ast}$ is an increasing function of $\beta_{j0}$ if $0<\beta_{j0}<c_5$, and $P_{d,j}^{\ast}$ is a decreasing function of $\beta_{j0}$ if $-c_6<\beta_{j0}<0$, where $c_5$ and $c_6$ are some bounded positive constants. The proof for this finding is provided in the Supplementary Material S2.2.  Figure \ref{figure:poisson} illustrates $P_{d,j}^{\ast}$. 

\begin{figure}[t!]
	\centerline{\includegraphics[width=20em,angle=270]{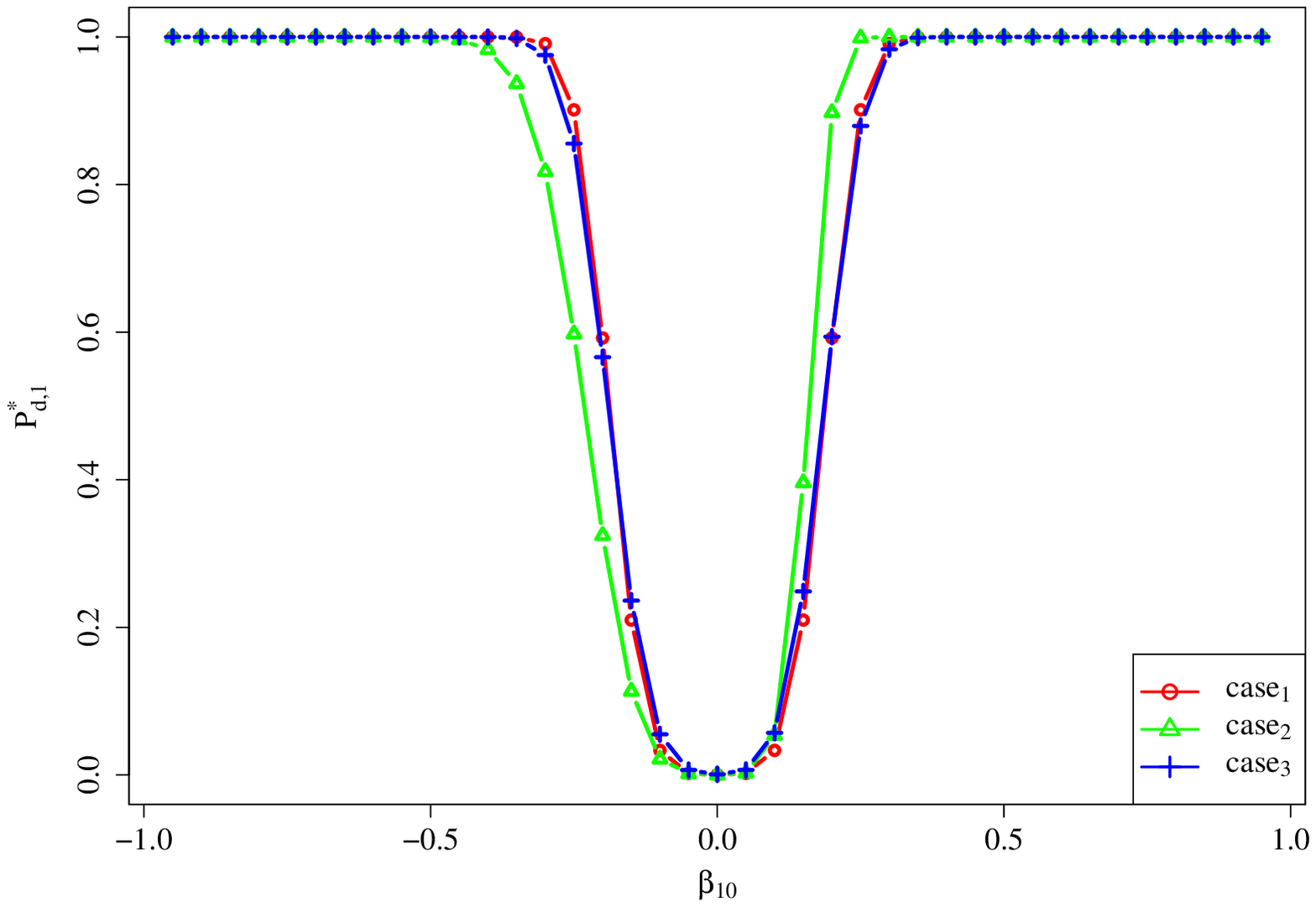}}
	\caption{\small The plots for $P_{d,1}^{\ast}$ as $\beta_{10}$ varies under three different cases in Poisson regression models.  In case 1, $\bm{X}_1$ and $\bm{X}_2$ both follow the standard normal distribution, and $\bm{X}_1$ and $\bm{X}_2$ are independent; in case 2, $\bm{X}_1$ and $\bm{X}_2$ both follow the centralized exponential distribution with mean $0$ and variance $1$, and $\bm{X}_1$ and $\bm{X}_2$ are independent; in case 3, $\bm{X}_1$ and $\bm{X}_2$ both follow the standard normal distribution, and $\bm{X}_1$ and $\bm{X}_2$ have the correlation of $0.5$. In all cases, $n=300$, $\bm \gamma_0=(0.3,\beta_{10},0.2)'$, $\lambda=0.05$, $\beta_{10}$ varies between $-0.95$ and $0.95$, with a step size of $0.05$.}
	\label{figure:poisson}
\end{figure}

The finite-sample properties of $P_{d,j}^\ast$ under other likelihood-based models can be  analyzed similarly. In general, $P_{d,j}^{\ast}$ is a comprehensive indicator. It shows how the selection probability of $\bm{X}_j$ is influenced by $\bm{\gamma}_{0}$, $\mathbf{x}_{i}$, $n$, and $\lambda$ in finite samples.  
Given  other values in $P_{d,j}^{\ast}$ except $\beta_{j0}$, $P_{d,j}^{\ast}$ is not necessarily a symmetric function of $\beta_{j0}$ or an increasing function of $|\beta_{j0}|$.

Based on the above analysis, we propose using  $P_{d,j}^{\ast}$ to measure the signal strength levels directly, rather than using $|\beta_{j0}|$.  Intuitively, if $P_{d,j}^{\ast}$ is close to one, then the variable $\bm{X}_{j}$ is defined to be a strong signal; if  $P_{d,j}^{\ast}$ is close to zero, then the variable $\bm{X}_{j}$ is defined to be a noise variable; if $P_{d,j}^{\ast}$ lies between the strong and noise levels, then the variable $\bm{X}_{j}$ is defined to be a weak signal. Specifically, we introduce two threshold values,  $\delta^{s}$ and $\delta^{w}$. Then the three levels of signal strength can be defined as
\begin{equation}
	\begin{cases}
		\bm{X}_{j}\in \mathbf{X}^{(S)}, & \text{if} \quad P_{d,j}^{\ast}>\delta^{s};\\
		\bm{X}_{j}\in \mathbf{X}^{(W)}, &\text{if} \quad \delta^{w}<P_{d,j}^{\ast}\leq \delta^{s};\\
		\bm{X}_{j}\in \mathbf{X}^{(N)}, & \text{if} \quad  P_{d,j}^{\ast}\leq\delta^{w},\\
	\end{cases}
	\label{sigleveldefi1}
\end{equation}
where $0<\tau^{w}\leq\delta^{w}<\delta^{s}\leq \tau^{s}\leq 1$, $\tau^{w}=\min_{j}P_{d,j}^{\ast}$, and $\tau^{s}=\max_{j}P_{d,j}^{\ast}$. Obviously, it is easier to select a stronger signal using the variable selection process than it is to select a weaker signal.

\subsection{Weak signal identification}
\label{ss:siganlidentify}
In this section, we show how to identify weak signals. Based on  the analysis in Section \ref{ss:weakdef},  the approximated selection probability $P_{d,j}^\ast$ depends on the true parameter $\bm{\gamma}_{0}$ and the distribution of $\mathbf{x}_{i}$, but they are always unknown in practice. In the following, we estimate $P_{d,j}^\ast$  by plugging in the  maximum likelihood estimator $\bm{\gamma}^{(0)}$ and  the empirical mean of  the random variables in (\ref{detectionprob}). That is, 
\begin{align}
	\hat{P}_{d,j}^\ast=&\Phi\left(\frac{-\sqrt{\left(n\lambda\sum_{i=1}^{n}D_{ii}^{(0)}\right)/\left\{\sum_{i=1}^{n}D_{ii}^{(0)}x_{ij}^{2}\sum_{i=1}^{n}D_{ii}^{(0)}-(\sum_{i=1}^{n}D_{ii}^{(0)}x_{ij})^2\right\}} +{\beta}_{j}^{(0)}}{\sqrt{(\widetilde{\mathbf{X}}^\top\mathbf{D}^{(0)}\widetilde{\mathbf{X}})^{-1}_{j+1,j+1}}}\right)\\
	&	+\Phi\left(\frac{-\sqrt{\left(n\lambda\sum_{i=1}^{n}D_{ii}^{(0)}\right)/\left\{\sum_{i=1}^{n}D_{ii}^{(0)}x_{ij}^{2}\sum_{i=1}^{n}D_{ii}^{(0)}-(\sum_{i=1}^{n}D_{ii}^{(0)}x_{ij})^2\right\}}-{\beta}_{j}^{(0)}}{\sqrt{(\widetilde{\mathbf{X}}^\top\mathbf{D}^{(0)}\widetilde{\mathbf{X}})^{-1}_{j+1,j+1}}}\right).
	\label{estdetectionprob}
\end{align}

In practice, we identify the signal strength level of $\bm{X}_j$ based on $\hat{P}_{d,j}^{\ast}$, and  introduce two threshold values  $\delta_1$ and $\delta_2$. We denote the identified subsets of strong signals, weak signals, and noise variables as $\hat{\mathbf{S}}^{(S)}$, $\hat{\mathbf{S}}^{(W)}$, and $\hat{\mathbf{S}}^{(N)}$, respectively:
\begin{equation}
	\begin{cases}
		\hat{\mathbf{S}}^{(S)}=\{j:\hat{P}_{d,j}^{\ast}>\delta_{1}\};\\
		\hat{\mathbf{S}}^{(W)}=\{j:\delta_{2}<\hat{P}_{d,j}^{\ast}\leq \delta_{1}\};\\
		\hat{\mathbf{S}}^{(N)}=\{j:\hat{P}_{d,j}^{\ast}\leq \delta_{2}\}.
	\end{cases}
	\label{sigleveldefi2}
\end{equation}

The selections of $\delta_{1}$ and $\delta_{2}$ are crucial to determining the signal type. The threshold value $\delta_{1}$  is selected to ensure that we can identify strong signals when the  selection probabilities of signals are high.  Assume $\alpha$ is a significance level,  and we choose  $\delta_{1}$ to be larger than $1-\alpha$, so  that the identified strong signals are strong.
The threshold value $\delta_{2}$ is selected to control the false positive rate of selecting variable $\bm{X}_j$. Denote the false positive rate as $\tau$.  Then $\tau$ can be defined as
\begin{equation}
	\tau=P(j\notin\hat{\mathbf{S}}^{(N)}\mid \beta_{j0}=0,\bm{\gamma}_{0}^{-j})=P(\hat{P}_{d,j}^{\ast}>\delta_{2}\mid \beta_{j0}=0,\bm{\gamma}_{0}^{-j}).
	\label{eq:falsepositiverate}
\end{equation}
Thus,  $\delta_{2}$ can be estimated based on \eqref{eq:falsepositiverate}. Because the value of $\bm \gamma_0$ is unknown in practice,  we estimate it using the one-step adaptive lasso estimator $\bm \gamma^{(1)}$. Furthermore,  to make  the estimated value of the false positive rate  equal  to $\tau$ based on the observed data, we take the value of $\delta_{2}$ as the $100(1-\tau)\%$ quantile of $ \{\hat{P}_{d,j}^\ast: \beta_j^{(1)}=0,j=1,\ldots,p\}$.
Because we intend to recover weak signals given  finite samples, $\tau$ is chosen to be larger than zero. However, the value of  $\tau$ cannot be too large, because there is a trade-off between recovering weak signals and including noise variables. In practice, if we want to recover more weak signals, we can choose a larger $\tau$; if we want to make the false positive rate lower, we can choose a smaller $\tau$. In the simulation studies, we perform a sensitivity analysis for the choice of $\delta_1$ and $\tau$.

\section{Weak Signal Inference}
\label{sec:weakinfer}
In this section, we propose a two-step inference procedure for  constructing  confidence intervals for the regression coefficients. The  procedure consists of two parts: if a covariate is identified as a strong signal, then its confidence interval is constructed based on the asymptotic theory for the nonzero one-step adaptive lasso estimator \citep{zou2008one}; if a covariate is identified as a weak signal or a noise variable, then we provide a confidence interval based on the following inference  theory for the maximum likelihood estimator.

Similarly to the theory in  \cite{zou2008one}, we can obtain the asymptotic distribution of the one-step adaptive lasso estimator. Without loss of generality, assume $\mathscr{A}_{n}=\{1,\ldots, s\}$, where $s$ is the number of nonzero elements in $\bm{\beta}^{(1)}$. 
Define $\mathscr{B}_{n}=\{k: \gamma_{k}^{(1)}\neq 0, k=1,\ldots,p+1\}$, then $\mathscr{B}_{n}=\{1,\ldots, s+1\}$.
Although the one-step adaptive lasso estimator $\bm{\beta}_{\mathscr{A}_{n}}^{(1)}$  is biased, owing to the shrinkage effect in finite samples, we can construct a de-biased confidence interval for the true coefficient based on the estimated bias and  covariance matrix for $\bm{\beta}_{\mathscr{A}_{n}}^{(1)}$, as shown in Theorem \ref{theo:1}. The proof of Theorem \ref{theo:1} is given in the Supplementary Material S3.
\begin{thm}
	Denote $\mathbf{X}^\top\mathbf{D}^{\dagger(0)}\mathbf{X}$ and $\widetilde{\mathbf{X}}^\top \mathbf{D}^{(0)}\widetilde{\mathbf{X}}/n$ as $\mathbf{Z}^{(0)}$ and  $\mathbf{I}^{(0)}$, respectively. The estimators of the bias  and the  covariance matrix of $\bm{\beta}_{\mathscr{A}_{n}}^{(1)}$ are given by
	\[
	\widehat{\mathrm{bias}}(\bm{\beta}_{\mathscr{A}_{n}}^{(1)})=-\left\{\frac{1}{n}\mathbf{Z}_{\mathscr{A}_{n}}^{(0)}+\Sigma_{\lambda}(\bm{\beta}_{\mathscr{A}_{n}}^{(0)},\bm{\beta}_{\mathscr{A}_{n}}^{(1)})\right\}^{-1}\left(\frac{\lambda}{|\beta_{1}^{(0)}|}\mathrm{sgn}(\beta_{1}^{(1)}),\ldots,  \frac{\lambda}{|\beta_{s}^{(0)}|}\mathrm{sgn}(\beta_{s}^{(1)})\right)^\top,
	\]
	and
	\begin{multline*}
		\widehat{\mathrm{cov}}(\bm{\beta}_{\mathscr{A}_{n}}^{(1)})
		=\frac{1}{n^{3}}\left\{\frac{1}{n}\mathbf{Z}_{\mathscr{A}_{n}}^{(0)}+\Sigma_{\lambda}(\bm{\beta}_{\mathscr{A}_{n}}^{(0)},\bm{\beta}_{\mathscr{A}_{n}}^{(1)})\right\}^{-1}\mathbf{Z}_{\mathscr{A}_{n}}^{(0)} \{(\mathbf{I}^{(0)}_{\mathscr{B}_{n}})^{-1}\}_{\mathscr{A}_{n}} \mathbf{Z}_{\mathscr{A}_{n}}^{(0)}   \\\times \left\{\frac{1}{n}\mathbf{Z}_{\mathscr{A}_{n}}^{(0)}+\Sigma_{\lambda}(\bm{\beta}_{\mathscr{A}_{n}}^{(0)},\bm{\beta}_{\mathscr{A}_{n}}^{(1)})\right\}^{-1},
	\end{multline*}
	respectively, where $\Sigma_{\lambda}(\bm{\beta}_{\mathscr{A}_{n}}^{(0)},\bm{\beta}_{\mathscr{A}_{n}}^{(1)})={\rm diag}\{\lambda/(|\beta_{1}^{(0)}||\beta_{1}^{(1)}|),\ldots,\lambda/(|\beta_{s}^{(0)}||\beta_{s}^{(1)}|)\}$,  $\mathbf{Z}_{\mathscr{A}_{n}}^{(0)}$ is the sub-matrix of $\mathbf{Z}^{(0)}$ corresponding to $\bm{\beta}^{(0)}_{\mathscr{A}_{n}}$,  and  $\mathbf{I}^{(0)}_{\mathscr{B}_{n}}$  is the sub-matrix of $\mathbf{I}^{(0)}$ corresponding to $\bm{\gamma}^{(0)}_{\mathscr{B}_{n}}$.
	\label{theo:1}
\end{thm}

Based on Theorem \ref{theo:1}, if the covariate $\bm{X}_{j}$ is identified as  a strong signal, then the $100(1-\alpha)\%$ confidence interval for $\beta_{j0}$ can be constructed as 
\begin{equation}
	(\beta_{j}^{(1)}-\hat{b}_{j}-z_{\alpha/2}\hat{\sigma}_{j},\beta_{j}^{(1)}-\hat{b}_{j}+z_{\alpha/2}\hat{\sigma}_{j}),
	\label{interval1}
\end{equation}
where  $\hat{b}_{j}$ is the corresponding component  of $\widehat{\mathrm{bias}}(\bm{\beta}_{\mathscr{A}_{n}}^{(1)})$ and $\hat{\sigma}_{j}$ is the positive square root of the corresponding diagonal  component of $\widehat{\mathrm{cov}}(\bm{\beta}_{\mathscr{A}_{n}}^{(1)})$.

If the covariate $\bm{X}_{j}$ is identified  as a weak signal or a noise variable,  then the $100(1-\alpha)\%$ confidence interval for $\beta_{j0}$ can be constructed as 
\begin{equation}
	(\beta_{j}^{(0)}-z_{\alpha/2}\sigma_{j}^{(0)},\beta_{j}^{(0)}+z_{\alpha/2}\sigma_{j}^{(0)}),
	\label{interval2}
\end{equation}
where $\sigma_{j}^{(0)}$ is the positive square root of the corresponding diagonal  component of $\widehat{\mathrm{cov}}(\bm{\gamma}^{(0)})=(\widetilde{\mathbf{X}}^\top\mathbf{D}^{(0)}\widetilde{\mathbf{X}})^{-1}$. 
\begin{remark}
	Note that  \cite{shi2017weak}  did not construct confidence intervals for the noise variables, whereas we do.  As shown in Figure \ref{figure:coverprobpart1} in the simulation studies, this improves the coverage probabilities for the noise variables and weak signals. Using the  two-step inference method based on  \cite{shi2017weak}, the coverage probabilities for the noise variables tend to be lower than  $1-\alpha$, and the coverage probabilities for weak signals tend to be higher than $1-\alpha$. This is because one will construct confidence intervals for the noise variables only when the noise variables are misidentified as weak signals or strong signals, in which case the estimated values of  the coefficients tend to be far from the true values, leading to lower coverage probabilities; one will not construct confidence intervals for the weak signals when the weak signals are misidentified as noise variables, making the coverage probabilities of the confidence intervals higher. To solve these problems, we propose constructing confidence intervals for the identified noise variables as well. As a result,   the coverage probabilities of the confidence intervals become closer to $1-\alpha$.
\end{remark}

In summary, our proposed confidence interval for $\beta_{j0}$ can be written as
\begin{multline}
	(\beta_{j}^{(1)}-\hat{b}_{j}-z_{\alpha/2}\hat{\sigma}_{j},\beta_{j}^{(1)}-\hat{b}_{j}+z_{\alpha/2}\hat{\sigma}_{j})\mathrm{I}\{j\in\hat{\mathbf{S}}^{(S)} \}\\
	+(\beta_{j}^{(0)}-z_{\alpha/2}\sigma_{j}^{(0)},\beta_{j}^{(0)}+z_{\alpha/2}\sigma_{j}^{(0)})\mathrm{I}\{j\in\hat{\mathbf{S}}^{(W)}\cup\hat{\mathbf{S}}^{(N)} \},
	\label{interval}
\end{multline}
which combines both (\ref{interval1}) and (\ref{interval2}).

\section{Simulation Studies}
\label{sec:simu}

In this section, we conduct simulation studies to evaluate the finite-sample performance of the proposed signal identification criterion and  two-step inference procedure. Consider the following logistic regression model:
\[
P(y_{i}=1\mid \mathbf{x}_{i})=\frac{\exp(\alpha_{0}+\mathbf{x}_{i}^\top\bm{\beta}_{0})}{1+\exp(\alpha_{0}+\mathbf{x}_{i}^\top\bm{\beta}_{0})},\quad i=1,\ldots,n.
\]
We generate the covariate vector $\mathbf{x}_{i}=(x_{i1},\ldots,x_{ip})^\top$ from  a multivariate normal distribution with mean zero and covariance matrix $\mathbf{R}({\rho}) {\sigma }^{2}$, where $\mathbf{R}({\rho})$ is a correlation matrix with the AR(1)  correlation structure and  ${\sigma }^{2}=1$.  All the generated covariates are standardized by subtracting their sample means and dividing by their sample standard deviations. 
For each setting, we choose $n=350$ or $550$, $p=25$ or $35$, $\rho=0$, $0.2$, or  $0.5$, and $\alpha_{0}=0.5$. The regression coefficient vector $\bm{\beta}_{0}$ is set to $(1,1,0.5,\theta,\underbrace{0,\ldots,0}_{p-4})^\top$, which consists of two large coefficients $1$, one moderate size coefficient $0.5$, one varying coefficient $\theta$, and $(p-4)$ zero coefficients. The coefficient $\theta$ ranges from zero to one, with a step size of $0.05$. In each simulation setting, we repeat the simulations $500$ times. The implementation details of the one-step adaptive lasso estimators are given in the Supplementary Material S4.

Figure 	\ref{figure:selectindep} displays the results for different types of selection probability for $\bm X_4$ when $\rho=0$.  In  Figure \ref{figure:selectindep},  the approximated selection probability  based on (\ref{detectionprob}) is close to the empirical selection probability, indicating a small approximation error from the  approximated selection probability. In addition, both the empirical selection probability and the approximated selection probability increase with $\theta$, implying that a larger value of $\theta$ leads to a stronger signal strength. This observation supports the result in Section \ref{ss:weakdef}. Although the median of the estimated selection probabilities is not too close to the empirical selection probability when $\theta$ is small,   the estimated selection probability still increases with the signal strength. We can still use the estimated selection probability to identify the signal strength level.  The simulation results for the correlated covariates are provided in Figures S1 and S2 of the Supplementary Material S5, and the approximated selection probability  is similar to the empirical selection probability. In addition, the empirical selection probability,  approximated selection probability, and estimated selection probability, in general, increase  with the value of $\theta$.  Thus,  we can also identify the signal strength level based on the value of $\theta$.

\begin{figure}[t!]
	\centerline{\includegraphics[width=40em,angle=0]{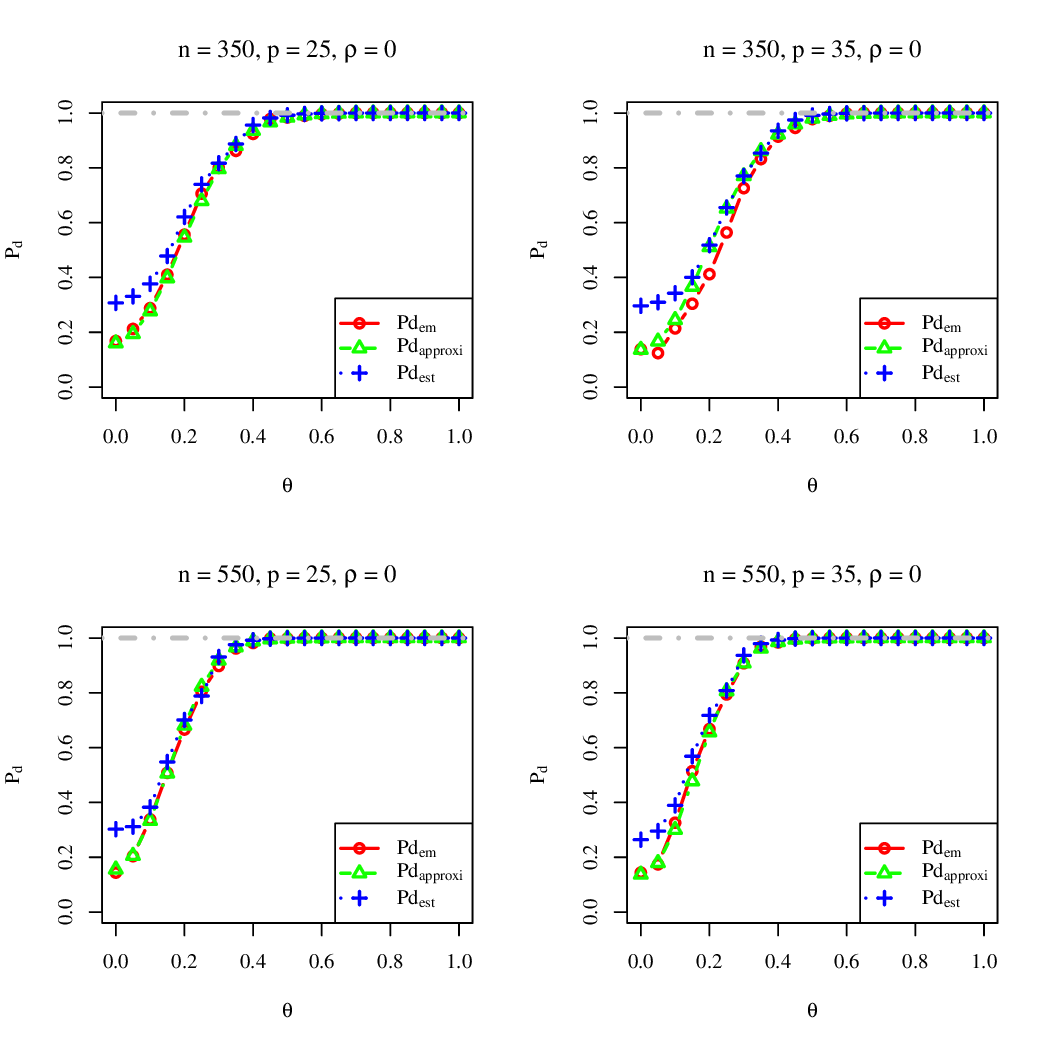}}
	\caption{\small Different types of selection probability for $\bm X_4$ when $\rho=0$. ${\rm Pd_{em}}$:  empirical selection probability, which is equal to the empirical probability of $\{\theta^{(1)}\neq 0\}$ based on $500$ Monte Carlo samples; ${\rm Pd_{approxi}}$:   approximated selection probability based on  (3.1), where the expectations in  (3.1) are calculated by using the function \texttt{cubintegrate} in \textsf{R}; ${\rm Pd_{est}}$:  median of estimated selection probabilities based on (3.3) for $500$ Monte Carlo samples.}
	\label{figure:selectindep}
\end{figure}

\begin{figure}[t!]
	\centerline{\includegraphics[width=40em,angle=0]{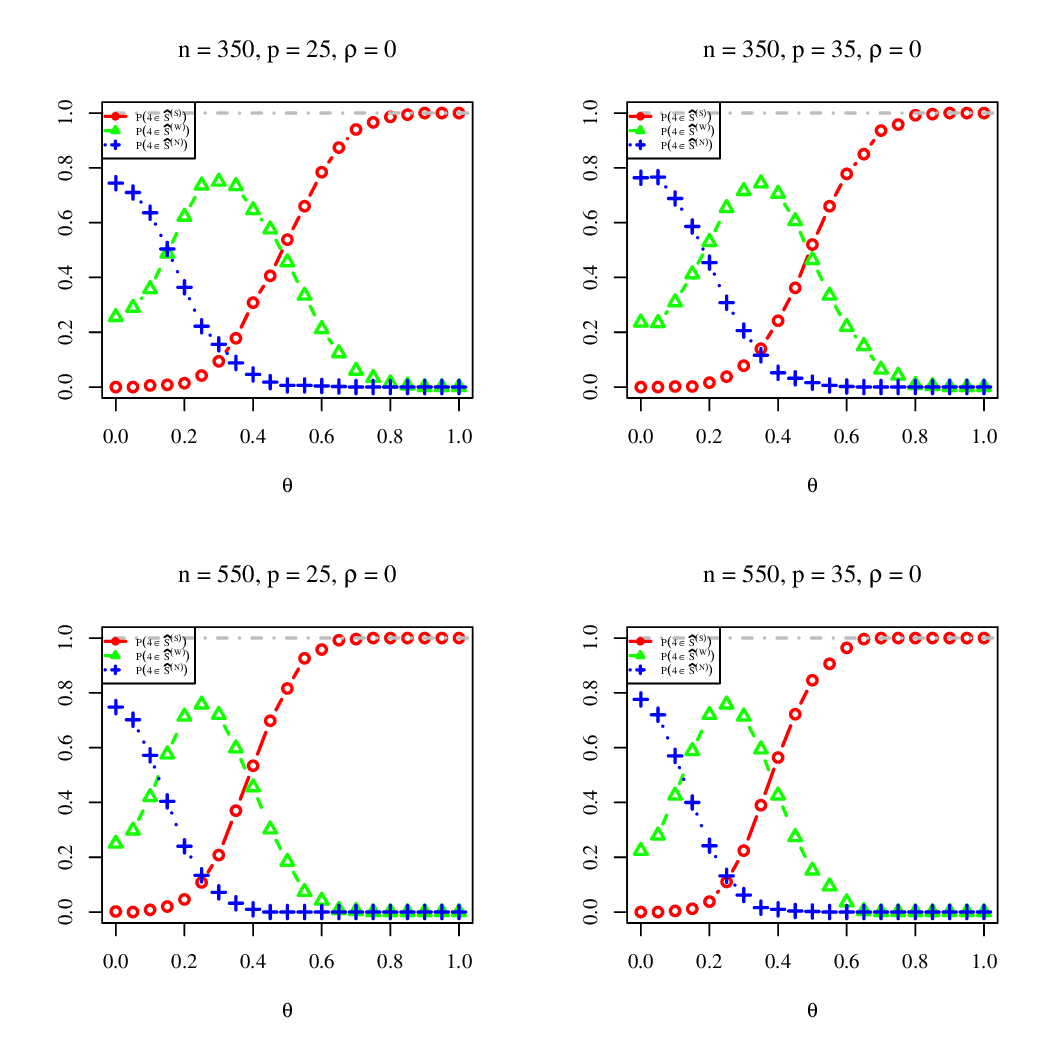}}
	\caption{\small Empirical probabilities of assigning the covariate $\bm X_4$ to different signal categories  when $\rho=0$.}
	\label{figure:signalidentifyindep}
\end{figure}

We then identify whether a covariate is a strong signal,  weak signal, or noise variable based on the criterion in (\ref{sigleveldefi2}).  For illustration, we choose $\delta_{1}$ to be $0.99$ and $\tau$ to be $0.1$.  Figure \ref{figure:signalidentifyindep}  represents the empirical probabilities of assigning the covariate $\bm X_4$ to different signal categories as $\theta$ varies and $\rho=0$.
Figure  \ref{figure:signalidentifyindep}  shows that when $\theta$ is close to zero, $\bm X_4$ is more  likely to be identified as a noise variable; when $\theta$ is far away from zero and one, the empirical probability of $\bm X_4$ being identified as a weak signal is highest; as $\theta$ becomes larger, the empirical probability of $\bm X_4$ being identified as a strong signal becomes more dominant, and gradually increases to one.  The results for the correlated covariates are given in Figures S3 and S4 of the Supplementary Material S5, and we have similar findings. Therefore, our proposed signal identification criterion (\ref{sigleveldefi2}) performs well in practice. 

After identifying the signal strength levels, we construct the $95\%$ confidence intervals based on the  proposed two-step inference procedure. We also compare our method with the  two-step inference method based on \cite{shi2017weak}, which does not construct confidence intervals for the identified noise variables.
In addition, we construct confidence intervals based on the asymptotic theory for the one-step adaptive lasso estimator, as shown in (\ref{interval1}),  the maximum likelihood estimation method, as shown in \eqref{interval2}, the perturbation method \citep{minnier2011perturbation}, the estimating equation-based method \citep{neykov2018unified}, the standard bootstrap method \citep{efron1994introduction}, the smoothed bootstrap method \citep{efron2014estimation}, the de-biased lasso method \citep{javanmard2014confidence,van2014asymptotically,zhang2014confidence}, and two different types of bootstrap de-biased lasso methods \citep{dezeure2017high}. The number of bootstrap resampling is set to  $4000$  for all  bootstrap methods, and the resampling number is set to  $500$ for the perturbation method. The implementation details of  the estimating equation-based method  and the two types of bootstrap de-biased lasso methods can be found in the Supplementary Material S4. For the method based on the asymptotic theory for the one-step adaptive lasso estimator, if a variable is not selected, then we do not construct a confidence interval for it, because the asymptotic normality is established only for the selected variables. 

\begin{figure}[t!]
	\centerline{\includegraphics[width=25em,angle=360]{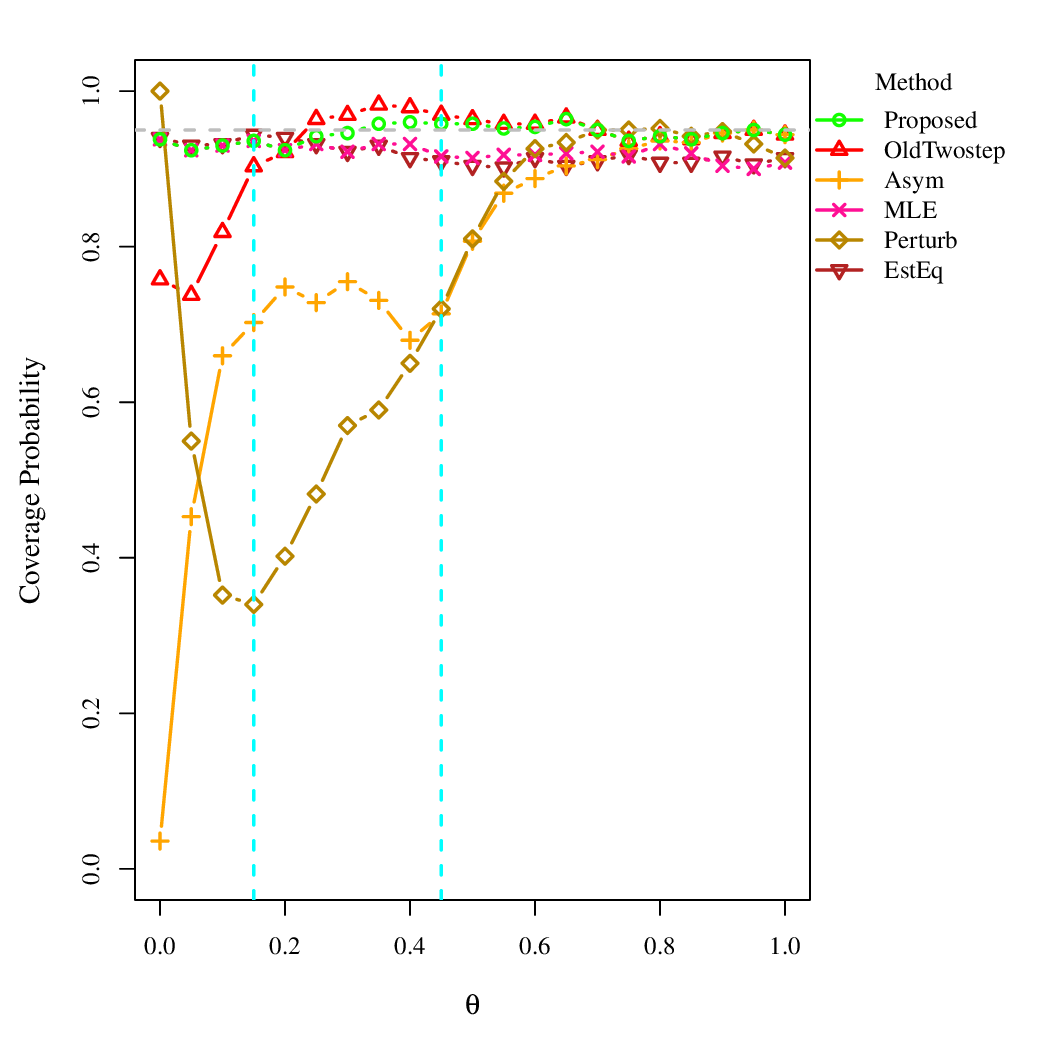}}
	\caption{\small Coverage probabilities of the $95\%$ confidence intervals  when $(n,p,\rho)=(350,25,0)$. Proposed: the proposed two-step inference method; OldTwostep: the  two-step inference method based on  \cite{shi2017weak}, which does not construct confidence intervals for identified noise variables; Asym: the method based on the asymptotic theory using the one-step adaptive lasso estimator; MLE: the maximum likelihood estimation method; Perturb: the perturbation method; EstEq: the estimating equation-based method. }
	\label{figure:coverprobpart1}
\end{figure}

\begin{figure}[t!]
	\centerline{\includegraphics[width=25em,angle=360]{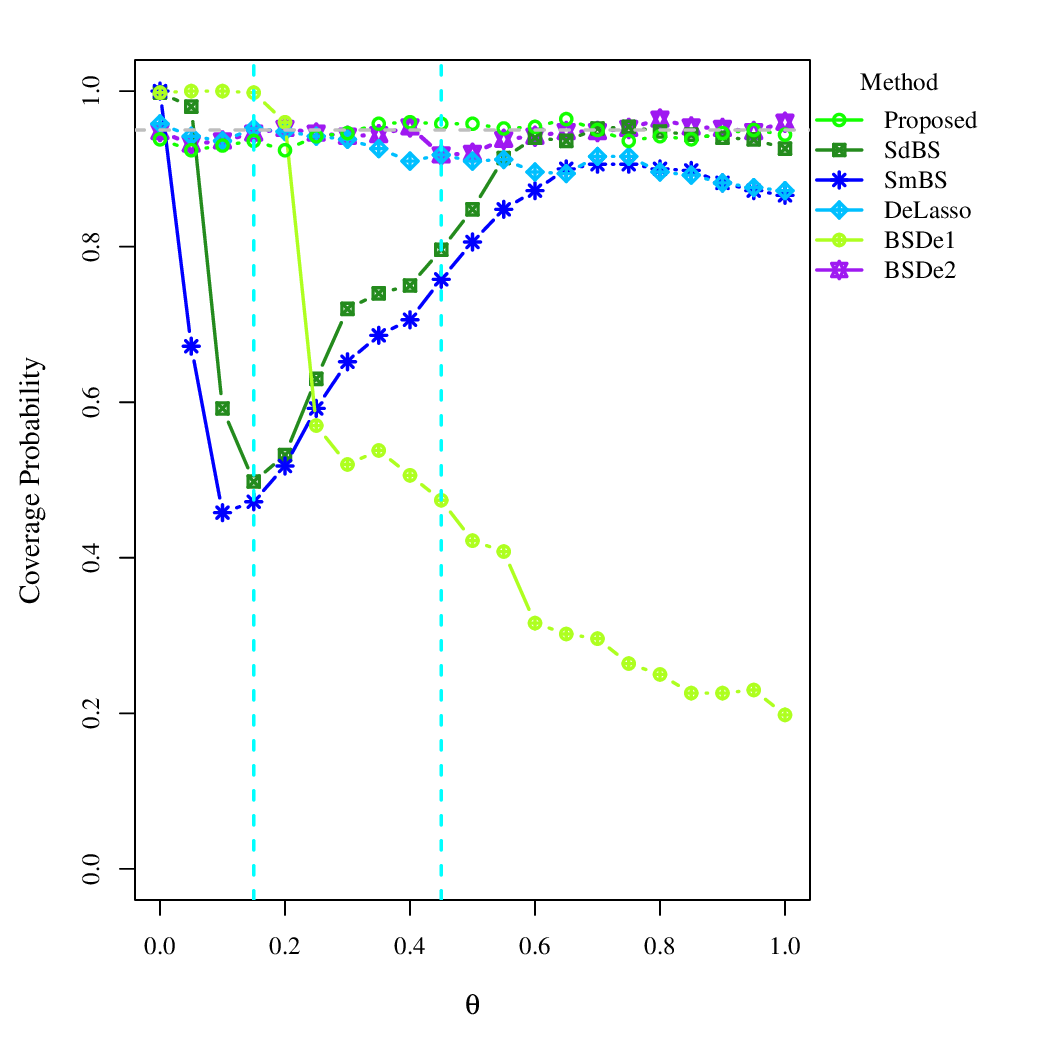}}
	\caption{\small Coverage probabilities of the $95\%$ confidence intervals  when $(n,p,\rho)=(350,25,0)$. Proposed: the proposed two-step inference method; SdBS: the standard bootstrap method; SmBS: the smoothed bootstrap method; DeLasso:  the de-biased lasso method; BSDe1: the type-I bootstrap de-biased lasso method; BSDe2: the type-II  bootstrap de-biased lasso method.}
	\label{figure:coverprobpart2}
\end{figure}

Figures \ref{figure:coverprobpart1}  and \ref{figure:coverprobpart2} provide coverage probabilities of the $95\%$ confidence intervals as $\theta$ varies and $(n,p,\rho)=(350,25,0)$. In Figures \ref{figure:coverprobpart1}  and \ref{figure:coverprobpart2}, the  vertical line on the left  shows whether $\bm X_4$ is more likely to be identified as a noise variable or  a weak signal, and the  vertical line on the right distinguishes whether $\bm X_4$ is more likely to be identified as a weak signal or  a strong signal. The threshold values are obtained from Figure \ref{figure:signalidentifyindep}. Comparing the proposed two-step inference method with the   two-step inference method based on  \cite{shi2017weak}, when $\theta$ is small, the former outperforms the latter. When $\theta$ is close to zero,  the coverage probability of the asymptotic method is too low and  close to zero, while the perturbation method, standard bootstrap method, smoothed bootstrap method, and  type-I bootstrap de-biased lasso method provide over-coverage confidence intervals, with coverage probabilities approximating to one.   When the signal is weak, the asymptotic method, perturbation method, standard bootstrap method, smoothed bootstrap method, and type-I bootstrap de-biased lasso method all perform poorly, and  their  coverage probabilities  are much lower than $95\%$. In addition, the coverage probability of the estimating equation-based method is slightly lower than $95\%$.  When the signal is stronger, the performance of the  maximum likelihood estimation method,  estimating equation-based method, de-biased lasso method, and  type-I bootstrap de-biased lasso method also become worse.
However, the coverage probabilities of the $95\%$ confidence intervals for the proposed method and the type-II bootstrap de-biased  lasso method  are close to $95\%$ under all signal strength levels of $\theta$.

\begin{figure}[t!]
	\centerline{\includegraphics[width=25em,angle=360]{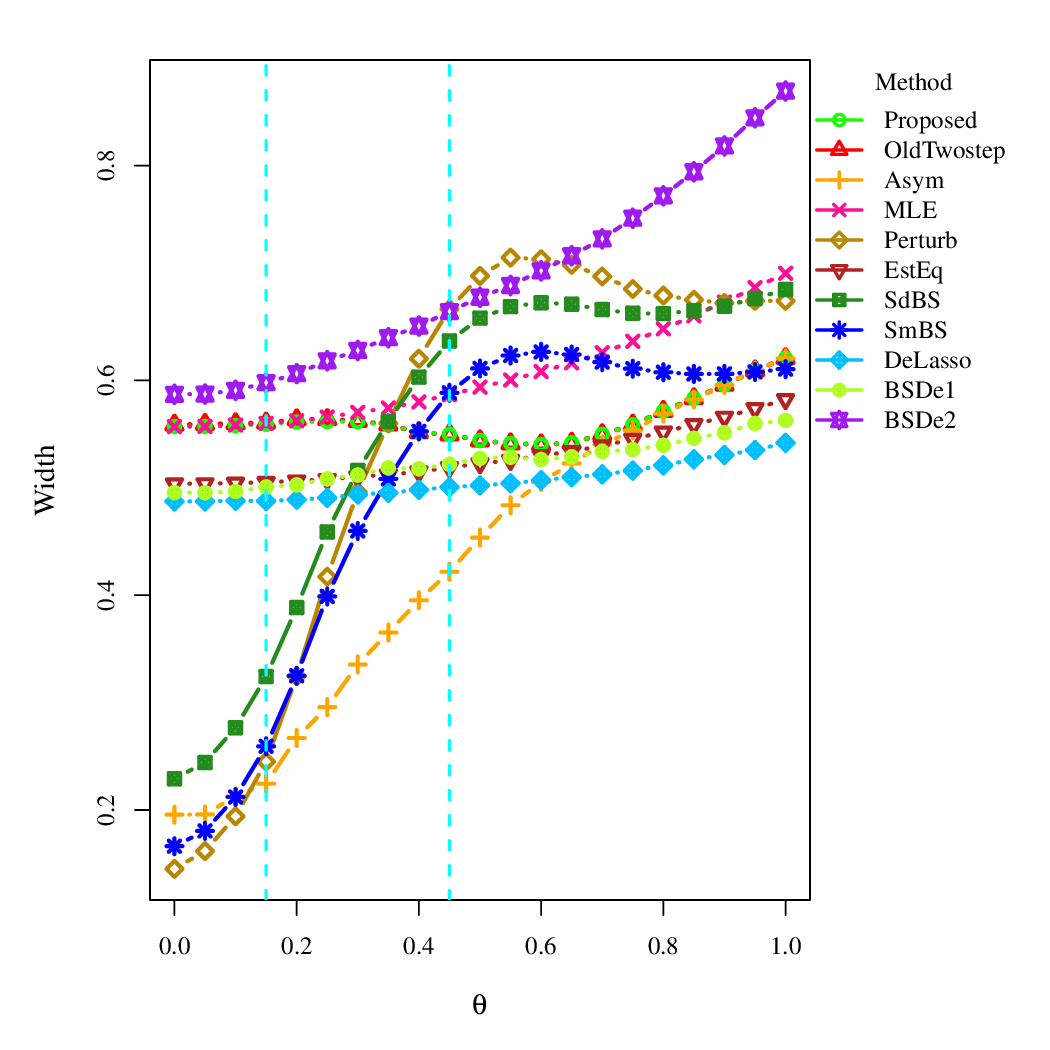}}
	\caption{\small Average widths of the $95\%$  confidence intervals  when $(n,p,\rho)=(350,25,0)$.   Proposed: the proposed two-step inference method; OldTwostep: the  two-step inference method based on  \cite{shi2017weak}, which does not construct confidence intervals for identified noise variables; Asym: the method based on the asymptotic theory using the one-step adaptive lasso estimator; MLE: the maximum likelihood estimation method; Perturb: the perturbation method; EstEq: the estimating equation-based method; SdBS: the standard bootstrap method; SmBS: the smoothed bootstrap method; DeLasso:  the de-biased lasso method; BSDe1: the type-I bootstrap de-biased lasso method; BSDe2: the type-II  bootstrap de-biased lasso method.}
	\label{figure:coverwidth}
\end{figure}

Figure \ref{figure:coverwidth} provides the average widths of  the $95\%$ confidence intervals as $\theta$ varies and $(n,p,\rho)=(350,25,0)$. Note that the widths of  the confidence intervals for the two types of two-step inference methods are both very close, while their coverage probabilities are not similar when $\theta$ is small. The width of the confidence interval using the proposed method is between those of the  maximum likelihood estimation method and the asymptotic method. This is not surprising, because the proposed method combines the strengths of these two methods.  Although the confidence intervals based on the asymptotic method, perturbation method, standard bootstrap method, and smoothed bootstrap method are narrow when $\theta$ is close to zero, the coverage probabilities are not accurate,  because they are  either too small or too large. When the signal is strong, the widths of the confidence intervals for the perturbation method, standard bootstrap method, and smoothed bootstrap method are, in general, larger than that for the proposed method.  Although the estimating equation-based method, de-biased lasso method, and type-I bootstrap de-biased lasso method have shorter confidence intervals than that of the proposed method, their coverage probabilities of the confidence intervals decrease as the signal becomes stronger. Overall, the confidence interval for the type-II bootstrap de-biased lasso method is wider than that of the proposed method.

The coverage probabilities and average widths of  the $95\%$ confidence intervals under all  simulation settings are  summarized in Tables S1--S4 of the Supplementary Material S5.
For each simulation setting, we select three different values of $\theta$, under which $\bm X_4$ is identified as  a noise variable,  weak signal, and  strong signal, respectively. In summary, the findings from the  simulation setting of $(n,p,\rho)=(350,25,0)$ still hold under other simulation settings when $\rho=0$. By comparison, the average widths of the confidence intervals for all methods decrease with the sample size and increase with the correlations between the covariates.  When $\bm X_4$ is not a strong signal, regardless of the correlations among covariates, the confidence intervals for the asymptotic method have relatively low coverage probabilities. When $\bm X_4$ is a strong signal, if $\rho$ is  $0$ or $0.2$,  the asymptotic method provides accurate confidence intervals, but if $\rho$ increases to $0.5$, the performance of the  asymptotic method deteriorates. However, the coverage probabilities of the confidence intervals for the proposed method are still close to $95\%$ under all simulation settings.

In order to see whether the performance of the proposed method  is sensitive to the choice of the threshold values $\delta_1$ and $\tau$, we also consider other combinations of  threshold values. For example, when $(n,p,\rho)=(350,25,0)$, we set $\tau$  as $0.1$ and  choose $\delta_{1}$ to be $0.96, 0.97, 0.98$, or $0.99$, which is larger than $1-\alpha=0.95$. The empirical probabilities of assigning the covariate $\bm X_4$ to different signal categories are shown in Figure S5 of the Supplementary Material S5. As the value of $\delta_{1}$ becomes larger and the value of $\theta$ is fixed, the empirical probability of identifying $\bm X_4$ as a weak signal becomes larger, and that of identifying $\bm X_4$ as a strong signal becomes smaller if $\theta$ is not sufficiently large. Furthermore,  the empirical probability of identifying $\bm X_4$ as a noise variable does not change.  This is because of the proposed signal identification criterion. Figures S6--S7 in the Supplementary Material S5 show the corresponding coverage probabilities and average widths of the $95\%$ confidence intervals for the proposed two-step inference method. As shown, the coverage probability becomes larger as $\delta_{1}$ increases and $\theta$ is between $0.6$ and $0.75$, and the average width becomes larger as $\delta_{1}$ increases and $\theta$ is between $0.15$ and $0.75$. This is not surprising because when $\delta_{1}$ increases, the probability of using the maximum likelihood method to construct the confidence intervals becomes larger. As  shown in Figures 	\ref{figure:coverprobpart2} and \ref{figure:coverwidth}, when $\theta$ is not too large, the coverage probability and average width of the confidence interval based on the maximum likelihood method is higher than that based on the asymptotic method. However, as $\delta_{1}$ varies, the changes of the coverage probability and average width are not large. 

We also consider another situation where $\delta_{1}$ is set to $0.99$ and $\tau$ is chosen to be $0.05, 0.1, 0.15$, or $0.2$. Figure S8 in the Supplementary Material S5 shows the empirical probabilities of assigning the covariate $\bm X_4$ to different signal categories in this situation. Here, we find that as $\tau$ increases, the empirical probability of identifying $\bm X_4$ as a weak signal is larger, and that of identifying $\bm X_4$ as a noise variable is smaller if $\theta$ is not too large. The empirical probability of identifying $\bm X_4$ as a strong signal remains the same. This is consistent with the proposed signal selection criterion. However, because  the proposed two-step inference method uses the same confidence interval construction method for the identified noise variables and weak signals, the confidence interval does not change with the value of $\tau$, as shown in Figures  S9--S10 of the Supplementary Material S5.

We also  examine whether the performance of the proposed method is sensitive to the total number of weak signals. We reset the  regression coefficient vector $\bm{\beta}_{0}$ to be\\ $(1,1,0.5,\theta,\underbrace{0.3,\ldots,0.3}_{q},\underbrace{0,\ldots,0}_{p-q-4})^\top$, where $q$ is taken to be $0, 1, 2$, or  $3$.  For illustration, let $(n,p,\rho)=(350,25,0)$,  $\delta_1$  be $0.99$, and $\tau$ be $0.1$.  Based on the signal identification criterion, all the $q$ covariates corresponding to the coefficient $0.3$ are weak signals if $\theta$ ranges from zero to one. If the covariate $\bm X_4$ is identified as a weak signal, then the total number of weak signals is $q+1$; otherwise  it is $q$. The empirical probabilities of assigning the covariate $\bm X_4$ to different signal categories are shown in Figure S11 of the Supplementary Material S5, which are not sensitive to the value of $q$. 	Figures S12--S13 in the Supplementary Material S5 respectively  show the coverage probabilities and average widths  of the $95\%$ confidence intervals for the proposed two-step inference method,  showing that when $\theta$ is small, the average width increases with the value of $q$, while the coverage probability does not change monotonously with the value of  $q$. In addition, as $q$ varies, the variations of average width and coverage probability are not large. Thus, the performance of the proposed method is quite robust to the total number of weak signals.

\section{Real-Data Application}
\label{sec:realdata}
To  illustrate the performance of the proposed method, we apply it to a data set in the Practice Fusion diabetes study, which was provided by Kaggle as part of the ``Practice Fusion Diabetes Classification" challenge \citep{kaggle2012diabetes}. The data set consists of de-identified electronic medical records for over $10,000$ patients.
There are a total of 9948 patients in the training data,  including a binary variable indicating whether  a patient is diagnosed with Type 2 diabetes mellitus (T2DM),  or not. In this analysis, we aim to determine the most important risk factors for the incidence of  T2DM, which can  be used to identify patients with a high risk of  T2DM. 

We first extract $119$ predictors  from the predictors selected by the first-place winner in the Kaggle competition  by removing some highly correlated predictors (details can be found in \url{https://www.kaggle.com/c/pf2012-diabetes/overview/winners}).  These predictors can be divided into six categories: basic information, transcript records,  diagnosis information, medication information, lab result, and smoking status. Detailed information about these predictors can be found in Table  S5 in the Supplementary Material S6.
One outlying patient is also removed owing to  inaccurate information on the predictors.  All the predictors are standardized beforehand.  We adopt the following logistic regression model to fit the data set:
\[
P(y_i=1\mid \mathbf{x}_i)=\frac{\exp\Big(\alpha+\sum\limits_{j=1}^{p}x_{ij}\beta_{j}\Big)}{1+\exp\Big(\alpha+\sum\limits_{j=1}^{p}x_{ij}\beta_{j}\Big)},\quad i=1,\ldots,n,
\]
where $p=119$ and $n=9947$.

We first obtain the one-step adaptive lasso estimates of the  regression coefficients following the tuning parameter selection procedure given in the  Supplementary Material S4.
We then identify whether a predictor is a strong signal,  weak signal, or  noise variable based on criterion (\ref{sigleveldefi2}).  Here, we choose $\delta_{1}$ to be $0.99$ and $\tau$ to be $0.1$.  From all the predictors, we identify $18$ strong signals, $32$ weak signals, and $69$ noise variables. The $18$ strong signals are  all selected by the one-step adaptive lasso estimator, indicating  consistency between it and our method  for strong signal selection.  Among the $32$ weak signals, $24$ are also selected by the one-step adaptive lasso estimator, while the other eight predictors are only identified by our method. These eight additional predictors include  (1) the number of times being diagnosed with herpes zoster, hypercholesterolemia, hypertensive heart disease, respiratory infection, sleep apnea, and joint pain, respectively; (2) the number of transcripts for  cardiovascular disease; and (3) the number of diagnoses per weighted year. The relationships between these eight predictors and diabetes have also been studied  by other researchers. For example, \cite{papagianni2018herpes} reviewed studies on associations between herpes zoster and diabetes mellitus, and found that herpes zoster and T2DM were likely to coexist for the same patient. 

\begin{figure}[t!]
	\centerline{\includegraphics[width=30em,angle=360]{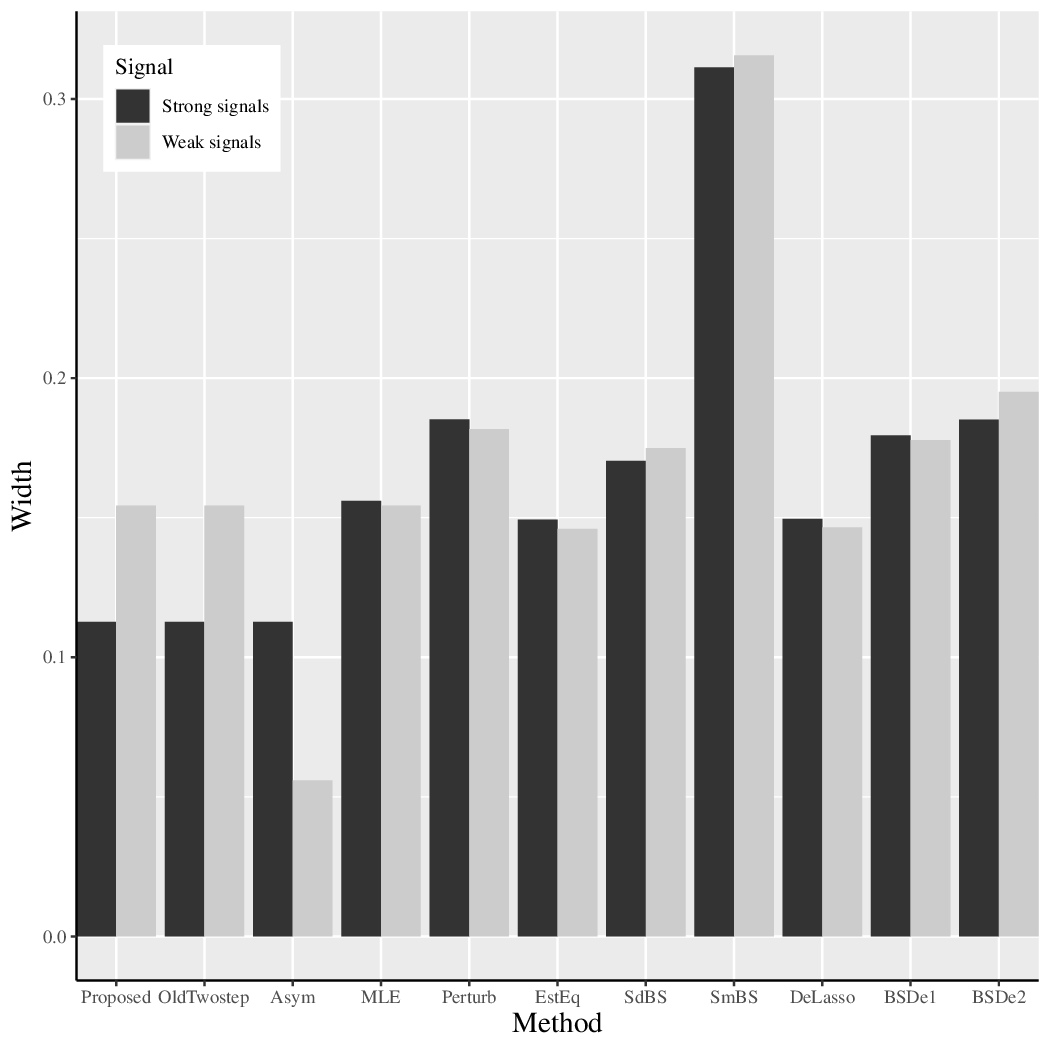}}
	\caption{\small The average widths of the $95\%$ confidence intervals for the diabetes data set.  Note that the asymptotic method does not construct confidence intervals for all the weak signals, the result for the weak signals is the average width of the confidence intervals for the weak signals, which are also selected by the asymptotic method. For the meanings of the notation, see Figures \ref{figure:coverprobpart1}  and \ref{figure:coverprobpart2}.} 
	\label{figure:widthrealdata}
\end{figure}

Next, we construct the $95\%$ confidence intervals using our two-step inference method,  together with all other comparison methods in Section \ref{sec:simu}.
Figure \ref{figure:widthrealdata} shows the average widths of the confidence intervals for the strong and weak signals.  For both, the widths of the confidence intervals for the two types of  two-step inference methods are the same.  For strong signals, the proposed method and the asymptotic method provide the shortest confidence intervals. For weak signals, the widths of the  confidence intervals based on the proposed method are smaller than those based on the perturbation method,  standard bootstrap method, smoothed bootstrap method, and two types of bootstrap  de-biased lasso methods.

\section{Conclusion}
\label{sec:discussion}
We have proposed a new unified approach for weak signal identification and inference in penalized likelihood models, including the special case when the responses are categorical. To identify weak signals, we propose using the estimated selection probability of each covariate as a measure of the signal strength, and  develop a signal identification criterion based directly on the estimated selection probability. To construct confidence intervals for the regression coefficients, we propose a  two-step inference procedure. Extensive simulation studies and a real-data application show that the proposed  signal identification method and two-step inference procedure outperform several existing methods in finite samples. 

The proposed method can  be extended  to a high-dimensional setting where $p$ is not fixed. One possible way is to use the de-biased lasso estimator as an initial estimator for the one-step adaptive lasso estimator, and then leverage the asymptotic properties of the de-biased lasso estimator to derive the selection probability. We can also use a penalized method to estimate the inverse of the information matrix, such as the CLIME estimator  \citep{cai2011constrained}.
In addition, our signal identification and inference framework can  be extended to longitudinal data.
For longitudinal data, we can replace the negative log-likelihood function with the generalized estimating function in the estimation. 
Finally, in the fields of causal inference and econometrics, there is a popular ``weak instrument''  problem \citep{chao2005consistent,burgess2011bias,choi2018weak}, which can be considered  a  weak signal problem. This is worth further development using our approach.

\vskip 14pt
\noindent {\large\bf Supplementary Material}

The online Supplementary Material contains six sections. Section S1 derives the approximated selection probability. Section S2 provide an additional detailed analysis of the approximated selection probability in finite samples.  Section S3 contains a proof for Theorem 1. Section S4 presents the implementation details of several methods. Sections S5 and S6 provide  additional simulation results and  information related to the real-data application, respectively.
\par
\vskip 14pt
\noindent {\large\bf Acknowledgments}

 This work was partially supported by the National Science Foundation of the United States (DMS-1821198, DMS-1952406), National Natural Science Foundation of China (11671096, 11731011, 12071087), and Natural
Sciences and Engineering Research Council of Canada  (RGPIN-2019-07052, DGECR-2019-00453, RGPAS-2019-00093).
 
\par

\markboth{\hfill{\footnotesize\rm ZHANG ET AL.} \hfill}
{\hfill {\footnotesize\rm WEAK SIGNAL IDENTIFICATION AND INFERENCE} \hfill}

\bibhang=1.7pc
\bibsep=2pt
\fontsize{9}{14pt plus.8pt minus .6pt}\selectfont
\renewcommand\bibname{\large \bf References}
\expandafter\ifx\csname
natexlab\endcsname\relax\def\natexlab#1{#1}\fi
\expandafter\ifx\csname url\endcsname\relax
  \def\url#1{\texttt{#1}}\fi
\expandafter\ifx\csname urlprefix\endcsname\relax\def\urlprefix{URL}\fi


\bibliography{weaksignal}
\bibliographystyle{apalike}

\vskip .65cm
\noindent
Yuexia Zhang\\
Department of Management Science and Statistics, The University of Texas at San Antonio, TX 78249, USA.\\
\noindent
E-mail: yuexia.zhang@utsa.edu\\
\noindent
Peibei Shi\\
Meta, Menlo Park, CA 94025, United States.\\
\noindent
E-mail: pshi@meta.com\\
\noindent
Zhongyi Zhu\\
Department of Statistics, Fudan University, Shanghai  200433, China.\\
\noindent
E-mail: zhuzy@fudan.edu.cn\\
\noindent
Linbo Wang\\
Department of Statistical Sciences, University of Toronto, ON M5S 3G3, Canada.\\
\noindent
E-mail: linbo.wang@utoronto.ca\\
\noindent
Annie Qu\\
Department of Statistics, University of California, Irvine, CA 92697, United States.\\
\noindent
E-mail: aqu2@uci.edu

\clearpage

\renewcommand{\baselinestretch}{2}

\fontsize{12}{14pt plus.8pt minus .6pt}\selectfont \vspace{0.8pc}

\centerline{\large\bf  SUPPLEMENTARY MATERIALS FOR}
\vspace{2pt}
\centerline{\large\bf ``WEAK SIGNAL IDENTIFICATION AND INFERENCE}
\vspace{2pt}
\centerline{\large\bf IN PENALIZED LIKELIHOOD MODELS }
\vspace{2pt}
\centerline{\large\bf  FOR CATEGORICAL RESPONSES''}
\vspace{.25cm} 
\centerline{Yuexia Zhang, Peibei Shi, Zhongyi Zhu, Linbo Wang and Annie Qu} 
\vspace{.4cm}
\centerline{\it
	The University of Texas at San Antonio, Meta, Fudan University} 
\centerline{\it
	University of Toronto and University of California, Irvine} 
\vspace{.55cm} 

\fontsize{9}{11.5pt plus.8pt minus
	.6pt}\selectfont

\begin{quotation}
	\noindent {\it Abstract:}\\
The online Supplementary Material contains six sections. Section S1 derives the approximated selection probability. Section S2 provide an additional detailed analysis of the approximated selection probability in finite samples.  Section S3 contains a proof for Theorem 1. Section S4 presents the implementation details of several methods. Sections S5 and S6 provide  additional simulation results and  information related to the real-data application, respectively.
	
	\par
\end{quotation}\par

\setcounter{section}{0}
\setcounter{equation}{0}
\setcounter{figure}{0}
\setcounter{table}{0}
\def\theequation{S\arabic{section}.\arabic{equation}}
\def\thesection{S\arabic{section}}

\renewcommand{\thefigure}{S\arabic{figure}}
\renewcommand{\thetable}{S\arabic{table}}
\renewcommand{\theassumption}{S\arabic{assumption}}

\fontsize{12}{14pt plus.8pt minus .6pt}\selectfont

\newpage

\section{Derivation of the Approximated Selection Probability \label{s:deridetectdegree}}

In Section 2 of the main paper, we have obtained the following condition for selecting the covariate $\bm{X}_{j}$, $j\in\{1,\ldots,p\}$:
\begin{multline*}
	\left|\sum_{i=1}^{n}\Big(\sum_{s=1}^{n}d_{is}^{(0)}x_{sj}\Big)^2(\beta_{j}^{(0)})^2+\sum_{k\neq j}\sum_{i=1}^{n}\Big(\sum_{s=1}^{n}d_{is}^{(0)}x_{sk}\Big)\Big(\sum_{s=1}^{n}d_{is}^{(0)}x_{sj}\Big)\beta_{j}^{(0)}(\beta_{k}^{(0)}-\beta_{k}^{(1)})\right|\\
	>n\lambda.
\end{multline*}
It is equivalent to
\begin{equation}
	\begin{aligned}
		&\Bigg|\frac{\sum\limits_{i=1}^{n}\Big(\sum\limits_{s=1}^{n}d_{is}^{(0)}x_{sj}\Big)^2(\beta_{j}^{(0)})^2}{n}
		+\frac{\sum\limits_{k\neq j}\sum\limits_{i=1}^{n}\Big(\sum\limits_{s=1}^{n}d_{is}^{(0)}x_{sk}\Big)\Big(\sum\limits_{s=1}^{n}d_{is}^{(0)}x_{sj}\Big)\beta_{j}^{(0)}(\beta_{k}^{(0)}-\beta_{k0}+\beta_{k0})}{n}\\
		&-\frac{\sum\limits_{k\neq j}\sum\limits_{i=1}^{n}\Big(\sum\limits_{s=1}^{n}d_{is}^{(0)}x_{sk}\Big)\Big(\sum\limits_{s=1}^{n}d_{is}^{(0)}x_{sj}\Big)\beta_{j}^{(0)}(\beta_{k}^{(1)}-\beta_{k0}+\beta_{k0})}{n}\Bigg|\\
		=&\Bigg|\frac{\sum\limits_{i=1}^{n}\Big(\sum\limits_{s=1}^{n}d_{is}^{(0)}x_{sj}\Big)^2(\beta_{j}^{(0)})^2}{n}+\frac{\sum\limits_{k\neq j}\sum\limits_{i=1}^{n}\Big(\sum\limits_{s=1}^{n}d_{is}^{(0)}x_{sk}\Big)\Big(\sum\limits_{s=1}^{n}d_{is}^{(0)}x_{sj}\Big)\beta_{j}^{(0)}(\beta_{k}^{(0)}-\beta_{k0})}{n}\\
		&-\frac{\sum\limits_{k\neq j}\sum\limits_{i=1}^{n}\Big(\sum\limits_{s=1}^{n}d_{is}^{(0)}x_{sk}\Big)\Big(\sum\limits_{s=1}^{n}d_{is}^{(0)}x_{sj}\Big)\beta_{j}^{(0)}(\beta_{k}^{(1)}-\beta_{k0})}{n}\Bigg|\\
		>& \lambda.
	\end{aligned}
	\label{eq:condition}
\end{equation}

We consider the following three formulas respectively,
\begin{equation*}
	\frac{\sum\limits_{i=1}^{n}\Big(\sum\limits_{s=1}^{n}d_{is}^{(0)}x_{sj}\Big)^2(\beta_{j}^{(0)})^2}{n},
\end{equation*}
\begin{equation}
	\frac{\sum\limits_{k\neq j}\sum\limits_{i=1}^{n}\Big(\sum\limits_{s=1}^{n}d_{is}^{(0)}x_{sk}\Big)\Big(\sum\limits_{s=1}^{n}d_{is}^{(0)}x_{sj}\Big)\beta_{j}^{(0)}(\beta_{k}^{(0)}-\beta_{k0})}{n},
	\label{eq:conditionpart2}
\end{equation}
and 
\begin{equation}
	\frac{\sum\limits_{k\neq j}\sum\limits_{i=1}^{n}\Big(\sum\limits_{s=1}^{n}d_{is}^{(0)}x_{sk}\Big)\Big(\sum\limits_{s=1}^{n}d_{is}^{(0)}x_{sj}\Big)\beta_{j}^{(0)}(\beta_{k}^{(1)}-\beta_{k0})}{n}.
	\label{eq:conditionpart3}
\end{equation}

Since $d_{is}^{(0)}$ is the $(i,s)$th element of $\mathbf{D}^{\star(0)}$, $\mathbf{D}^{\star(0)}=(\mathbf{D}^{(0)})^{1/2}-(\mathbf{D}^{(0)})^{1/2}\bm{1}\\ \times(\bm{1}^\top\mathbf{D}^{(0)}\bm{1})^{-1}\bm{1}^{\top}\mathbf{D}^{(0)}$ and $\mathbf{D}^{(0)}$ is an $n\times n$ diagonal matrix with the $(i,i)$th element  $D_{ii}^{(0)}$, then by calculation, 
\[
\frac{\sum\limits_{i=1}^{n}\Big(\sum\limits_{s=1}^{n}d_{is}^{(0)}x_{sj}\Big)^2}{n}=\frac{\sum\limits_{i=1}^{n}D_{ii}^{(0)}x_{ij}^{2}}{n}-\frac{\Big(\frac{\sum\limits_{i=1}^{n}D_{ii}^{(0)}x_{ij}}{n}\Big)^{2}}{\frac{\sum\limits_{i=1}^{n}D_{ii}^{(0)}}{n}}.
\]
Since  $(\mathbf{x}_{1},y_{1}),\ldots,(\mathbf{x}_{n},y_{n})$ are independent and identically distributed random vectors, $D_{ii}(\bm{\gamma})$ is a continuous function of $\bm{\gamma}$ and the maximum likelihood estimator  $\bm{\gamma}^{(0)}\stackrel{P}{\rightarrow}\bm{\gamma}_{0}$ under some regularity conditions, then by the Law of Large Numbers and  Continuous Mapping Theorem, we have $\sum_{i=1}^{n}D_{ii}^{(0)} x_{ij}^{2}/n\stackrel{P}{\rightarrow}{\rm E}(D_{0,ii} x_{ij}^{2})$, $\sum_{i=1}^{n}D_{ii}^{(0)} x_{ij}/n\stackrel{P}{\rightarrow}{\rm E}(D_{0,ii} x_{ij})$ and $\sum_{i=1}^{n}D_{ii}^{(0)}/n\\\stackrel{P}{\rightarrow}{\rm E}(D_{0,ii})$. Then 
\begin{equation*}
	\frac{\sum\limits_{i=1}^{n}\Big(\sum\limits_{s=1}^{n}d_{is}^{(0)}x_{sj}\Big)^2(\beta_{j}^{(0)})^2}{n}-\left[{\rm E}(D_{0,ii} x_{ij}^{2})-\frac{\{{\rm E}(D_{0,ii} x_{ij})\}^2}{{\rm E}(D_{0,ii})}\right](\beta_{j}^{(0)})^2\stackrel{P}{\rightarrow} 0.
\end{equation*}

By calculation, \eqref{eq:conditionpart2} equals
\[
\begin{aligned}
	&\sum\limits_{k\neq j}\left(\frac{\sum\limits_{i=1}^{n}x_{ik}D_{ii}^{(0)}x_{ij}}{n}-\frac{\sum\limits_{i=1}^{n}\sum\limits_{s=1}^{n}x_{ik}D_{ii}^{(0)}D_{ss}^{(0)}x_{sj}}{n\sum\limits_{i=1}^{n}D_{ii}^{(0)}}\right)\beta_{j}^{(0)}(\beta_{k}^{(0)}-\beta_{k0})\\
	=&\sum\limits_{k\neq j}\left(\frac{\sum\limits_{i=1}^{n}x_{ik}D_{ii}^{(0)}x_{ij}}{n}-\frac{\frac{\sum\limits_{i=1}^{n}x_{ik}D_{ii}^{(0)}}{n}\frac{\sum\limits_{s=1}^{n}D_{ss}^{(0)}x_{sj}}{n}}{\frac{\sum\limits_{i=1}^{n}D_{ii}^{(0)}}{n}}\right)\beta_{j}^{(0)}\frac{\sqrt{n}(\beta_{k}^{(0)}-\beta_{k0})}{\sqrt{n}}.
\end{aligned}
\]
Because of the same reason as before, $\sum_{i=1}^{n}x_{ik}D_{ii}^{(0)}x_{ij}/n\stackrel{P}{\rightarrow}{\rm E}(x_{ik}D_{0,ii}x_{ij})$, $\sum_{i=1}^{n}x_{ik}D_{ii}^{(0)}/n \stackrel{P}{\rightarrow}{\rm E}(x_{ik}D_{0,ii})$, $\sum_{s=1}^{n}D_{ss}^{(0)}x_{sj}/n\stackrel{P}{\rightarrow}{\rm E}(D_{0,ss}x_{sj})$ and $\sum_{i=1}^{n}D_{ii}^{(0)}/n\stackrel{P}{\rightarrow}{\rm E}(D_{0,ii})$.  By the Central Limit Theorem, $\sqrt{n}(\beta_{k}^{(0)}-\beta_{k0})\stackrel{D}{\rightarrow}\mathcal{N}(0,\{\mathbf{I}^{-1}(\bm{\gamma}_{0})\}_{k+1,k+1})$, where $\mathbf{I}(\bm{\gamma}_{0})={\rm E}(\widetilde{\mathbf{X}}^\top\mathbf{D}_{0}\widetilde{\mathbf{X}})/n$. Then $\sqrt{n}(\beta_{k}^{(0)}-\beta_{k0})=O_p(1)$. Furthermore, since $\beta_{j}^{(0)}\stackrel{P}{\rightarrow}\beta_{j0}$ and  the number of covariates $p$ is finite, then according to the  Slutsky's Theorem, \eqref{eq:conditionpart2} is $O_p(1/\sqrt{n})$.

Based on the oracle properties of $\bm{\beta}^{(1)}$, if $\beta_{k0}=0$, then $P(\beta_{k}^{(1)}=0)\rightarrow 1$. Therefore, similar to the previous proof,
\begin{equation}
	\begin{aligned}
		&\frac{\sum\limits_{i=1}^{n}\Big(\sum\limits_{s=1}^{n}d_{is}^{(0)}x_{sk}\Big)\Big(\sum\limits_{s=1}^{n}d_{is}^{(0)}x_{sj}\Big)\beta_{j}^{(0)}(\beta_{k}^{(1)}-\beta_{k0})}{n}\\
		=&\left(\frac{\sum\limits_{i=1}^{n}x_{ik}D_{ii}^{(0)}x_{ij}}{n}-\frac{\frac{\sum\limits_{i=1}^{n}x_{ik}D_{ii}^{(0)}}{n}\frac{\sum\limits_{s=1}^{n}D_{ss}^{(0)}x_{sj}}{n}}{\frac{\sum\limits_{i=1}^{n}D_{ii}^{(0)}}{n}}\right)\beta_{j}^{(0)}(\beta_{k}^{(1)}-\beta_{k0})\stackrel{P}{\rightarrow}0.
	\end{aligned}
	\label{eq:conditionpart21}
\end{equation}
If $\beta_{k0}\neq 0$, then $\sqrt{n}(\beta_{k}^{(1)}-\beta_{k0})\stackrel{D}{\rightarrow}\mathcal{N}(0,[\mathbf{I}^{-1}\{(\bm{\gamma}_{0})_{\mathscr{A}}\}]_{\bm{X}_{k}})$, where $\mathbf{I}\{(\bm{\gamma}_{0})_{\mathscr{A}}\}$ is the Fisher information matrix knowing $(\bm{\gamma}_{0})_{\mathscr{A}^{c}}=\bm{0}$ and  $[\mathbf{I}^{-1}\{(\bm{\gamma}_{0})_{\mathscr{A}}\}]_{\bm{X}_{k}}$ is an element of the matrix $\mathbf{I}^{-1}\{(\bm{\gamma}_{0})_{\mathscr{A}}\}$ corresponding to $\bm{X}_{k}$. Therefore, $\sqrt{n}(\beta_{k}^{(1)}-\beta_{k0})=O_p(1)$. Furthermore, 
\begin{equation}
	\begin{aligned}
		&\frac{\sum\limits_{i=1}^{n}\Big(\sum\limits_{s=1}^{n}d_{is}^{(0)}x_{sk}\Big)\Big(\sum\limits_{s=1}^{n}d_{is}^{(0)}x_{sj}\Big)\beta_{j}^{(0)}(\beta_{k}^{(1)}-\beta_{k0})}{n}\\
		=&\left(\frac{\sum\limits_{i=1}^{n}x_{ik}D_{ii}^{(0)}x_{ij}}{n}-\frac{\frac{\sum\limits_{i=1}^{n}x_{ik}D_{ii}^{(0)}}{n}\frac{\sum\limits_{s=1}^{n}D_{ss}^{(0)}x_{sj}}{n}}{\frac{\sum\limits_{i=1}^{n}D_{ii}^{(0)}}{n}}\right)\beta_{j}^{(0)}\frac{\sqrt{n}(\beta_{k}^{(1)}-\beta_{k0})}{\sqrt{n}}=O_{p}\left(\frac{1}{\sqrt{n}}\right).
	\end{aligned}
	\label{eq:conditionpart22}
\end{equation}
According to  \eqref{eq:conditionpart21} and  \eqref{eq:conditionpart22}, \eqref{eq:conditionpart3} is also $O_p(1/\sqrt{n})$.

In summary, the  condition for selecting the covariate $\bm{X}_{j}$  becomes
\[
\left|\left[{\rm E}(D_{0,ii} x_{ij}^{2})-\frac{\{{\rm E}(D_{0,ii} x_{ij})\}^2}{{\rm E}(D_{0,ii})}\right](\beta_{j}^{(0)})^2+o_p(1)\right|>\lambda.
\]
Furthermore, 
\begin{equation}
	P(\beta_{j}^{(1)}\neq 0)\approx P\left(\left[{\rm E}(D_{0,ii} x_{ij}^{2})-\frac{\{{\rm E}(D_{0,ii} x_{ij})\}^2}{{\rm E}(D_{0,ii})}\right](\beta_{j}^{(0)})^2>\lambda \right).
	\label{eq:detectprob2}
\end{equation}
By the Central Limit Theorem, $\sqrt{n}(\beta_{j}^{(0)}-\beta_{j0})\stackrel{D}{\rightarrow}\mathcal{N}(0,\{\mathbf{I}^{-1}(\bm{\gamma}_{0})\}_{j+1,j+1})$ and $\mathbf{I}(\bm{\gamma}_{0})={\rm E}(\widetilde{\mathbf{X}}^\top\mathbf{D}_{0}\widetilde{\mathbf{X}})/n$. Therefore, the right hand side of  \eqref{eq:detectprob2} can be approximated by
\begin{multline}
	P_{d,j}^{\ast}=\Phi\left(\frac{-\sqrt{\frac{\lambda{\rm E}(D_{0,ii})}{{\rm E}(D_{0,ii} x_{ij}^{2}){\rm E}(D_{0,ii})-\{{\rm E}(D_{0,ii} x_{ij})\}^2}} +{\beta}_{j0}}{\sqrt{\{{\rm E}(\widetilde{\mathbf{X}}^\top\mathbf{D}_{0}\widetilde{\mathbf{X}})\}^{-1}_{j+1,j+1}}}   \right)\\
	+\Phi\left(\frac{-\sqrt{\frac{\lambda{\rm E}(D_{0,ii})}{{\rm E}(D_{0,ii} x_{ij}^{2}){\rm E}(D_{0,ii})-\{{\rm E}(D_{0,ii} x_{ij})\}^2}}-{\beta}_{j0}}{\sqrt{\{{\rm E}(\widetilde{\mathbf{X}}^\top\mathbf{D}_{0}\widetilde{\mathbf{X}})\}^{-1}_{j+1,j+1}}}   \right).
	\label{truepdsecondmethod}
\end{multline}


\section{Additional Detailed Analysis of the Approximated Selection Probability in Finite Samples \label{s:propselpro}}
In this selection, we provide an additional detailed analysis of finite-sample properties of the approximated selection probability $P_{d,j}^\ast$ and provide some plots  to illustrate the finite-sample properties of $P_{d,j}^\ast$ under three different kinds of likelihood-based models.
\subsection{Symmetry of the approximated selection probability\label{ss:sympro}}
In order to study given any values in $P_{d,j}^{\ast}$ except $\beta_{j0}$, whether $P_{d,j}^\ast$ is a symmetric function of $\beta_{j0}$ or not, we need to study for any $\beta_{j0}\neq 0$, whether  $P_{d,j}^\ast(\beta_{j0})$ is equal to  $P_{d,j}^\ast(-\beta_{j0})$. According to (\ref{truepdsecondmethod}),
\[
\begin{aligned}
	P_{d,j}^\ast(\beta_{j0})=&\Phi\left(\frac{-\sqrt{\frac{\lambda{\rm E}\{D_{0,ii}(\beta_{j0},\bm{\gamma}_{0}^{-j})\}}{{\rm E}\{D_{0,ii}(\beta_{j0},\bm{\gamma}_{0}^{-j}) x_{ij}^{2}\}{\rm E}\{D_{0,ii}(\beta_{j0},\bm{\gamma}_{0}^{-j})\}-[{\rm E}\{D_{0,ii}(\beta_{j0},\bm{\gamma}_{0}^{-j}) x_{ij}\}]^2}} +{\beta}_{j0}}{\sqrt{\big[{\rm E}\{\widetilde{\mathbf{X}}^\top\mathbf{D}_{0}(\beta_{j0},\bm{\gamma}_{0}^{-j})\widetilde{\mathbf{X}}\}\big]^{-1}_{j+1,j+1}}}   \right)\\
	&+\Phi\left(\frac{-\sqrt{\frac{\lambda{\rm E}\{D_{0,ii}(\beta_{j0},\bm{\gamma}_{0}^{-j})\}}{{\rm E}\{D_{0,ii}(\beta_{j0},\bm{\gamma}_{0}^{-j}) x_{ij}^{2}\}{\rm E}\{D_{0,ii}(\beta_{j0},\bm{\gamma}_{0}^{-j})\}-[{\rm E}\{D_{0,ii}(\beta_{j0},\bm{\gamma}_{0}^{-j}) x_{ij}\}]^2}} -{\beta}_{j0}}{\sqrt{\big[{\rm E}\{\widetilde{\mathbf{X}}^\top\mathbf{D}_{0}(\beta_{j0},\bm{\gamma}_{0}^{-j})\widetilde{\mathbf{X}}\}\big]^{-1}_{j+1,j+1}}}   \right)
\end{aligned}
\]
and 
\[
\begin{aligned}
	P_{d,j}^\ast(-\beta_{j0})=&\Phi\left(\frac{-\sqrt{\frac{\lambda{\rm E}\{D_{0,ii}(-\beta_{j0},\bm{\gamma}_{0}^{-j})\}}{{\rm E}\{D_{0,ii}(-\beta_{j0},\bm{\gamma}_{0}^{-j}) x_{ij}^{2}\}{\rm E}\{D_{0,ii}(-\beta_{j0},\bm{\gamma}_{0}^{-j})\}-[{\rm E}\{D_{0,ii}(-\beta_{j0},\bm{\gamma}_{0}^{-j}) x_{ij}\}]^2}}-{\beta}_{j0}}{\sqrt{\big[{\rm E}\{\widetilde{\mathbf{X}}^\top\mathbf{D}_{0}(-\beta_{j0},\bm{\gamma}_{0}^{-j})\widetilde{\mathbf{X}}\}\big]^{-1}_{j+1,j+1}}}   \right)\\
	&+\Phi\left(\frac{-\sqrt{\frac{\lambda{\rm E}\{D_{0,ii}(-\beta_{j0},\bm{\gamma}_{0}^{-j})\}}{{\rm E}\{D_{0,ii}(-\beta_{j0},\bm{\gamma}_{0}^{-j}) x_{ij}^{2}\}{\rm E}\{D_{0,ii}(-\beta_{j0},\bm{\gamma}_{0}^{-j})\}-[{\rm E}\{D_{0,ii}(-\beta_{j0},\bm{\gamma}_{0}^{-j}) x_{ij}\}]^2}}+{\beta}_{j0}}{\sqrt{\big[{\rm E}\{\widetilde{\mathbf{X}}^\top\mathbf{D}_{0}(-\beta_{j0},\bm{\gamma}_{0}^{-j})\widetilde{\mathbf{X}}\}\big]^{-1}_{j+1,j+1}}}   \right).
\end{aligned}
\]
Since $D_{0,ii}(\beta_{j0},\bm{\gamma}_{0}^{-j})=-\partial^{2}\ell_{i}\{\mu_{i}(\beta_{j0},\bm{\gamma}_{0}^{-j})\}/\partial \mu_{i}^{2}$ with $\mu_{i}(\beta_{j0},\bm{\gamma}_{0}^{-j})=\alpha_{0}+\sum_{k\neq j}x_{ik}\beta_{k0}+x_{ij}\beta_{j0}$, and $D_{0,ii}(-\beta_{j0},\bm{\gamma}_{0}^{-j})=-\partial^{2}\ell_{i}\{\mu_{i}(-\beta_{j0},\bm{\gamma}_{0}^{-j})\}/\partial \mu_{i}^{2}$ with $\mu_{i}(-\beta_{j0},\bm{\gamma}_{0}^{-j})=\alpha_{0}+\sum_{k\neq j}x_{ik}\beta_{k0}-x_{ij}\beta_{j0}$,
then one of the sufficient conditions for $P_{d,j}^\ast(\beta_{j0})=P_{d,j}^\ast(-\beta_{j0})$ is that the distribution of $x_{ij}$ is symmetric about zero and $x_{ij}$ is independent of $x_{ik}$ for any $k\neq j$. Under this condition, we have ${\rm E}\{D_{0,ii}(\beta_{j0},\bm{\gamma}_{0}^{-j})\}={\rm E}\{D_{0,ii}(-\beta_{j0},\bm{\gamma}_{0}^{-j})\}$, ${\rm E}\{D_{0,ii}(\beta_{j0},\bm{\gamma}_{0}^{-j}) x_{ij}^{2}\}={\rm E}\{D_{0,ii}(-\beta_{j0},\bm{\gamma}_{0}^{-j}) x_{ij}^{2}\}$, ${\rm E}\{D_{0,ii}(\beta_{j0},\bm{\gamma}_{0}^{-j}) x_{ij}\}=-{\rm E}\{D_{0,ii}(-\beta_{j0},\bm{\gamma}_{0}^{-j}) x_{ij}\}$ and ${\rm E}\{\widetilde{\mathbf{X}}^\top\mathbf{D}_{0}(\beta_{j0},\bm{\gamma}_{0}^{-j})\widetilde{\mathbf{X}}\}={\rm E}\{\widetilde{\mathbf{X}}^\top\mathbf{D}_{0}(-\beta_{j0},\bm{\gamma}_{0}^{-j})\widetilde{\mathbf{X}}\}$. Furthermore, $P_{d,j}^\ast(\beta_{j0})=P_{d,j}^\ast(-\beta_{j0})$. 

However, this sufficient condition may not be satisfied in practice and it is easy to find a case where $P_{d,j}^\ast(\beta_{j0})\neq  P_{d,j}^\ast(-\beta_{j0})$. So given any values in $P_{d,j}^{\ast}$ except $\beta_{j0}$, $P_{d,j}^\ast$ is not necessarily a symmetric function of $\beta_{j0}$.

\subsection{Monotonicity of the approximated selection probability\label{ss:monotopro}}
In order to study the monotonicity of the approximated selection probability, we need to study the first order derivative of $P_{d,j}^\ast$ with respect to $\beta_{j0}$.  By calculation,
\[
\frac{\partial P_{d,j}^\ast}{\partial \beta_{j0}}=\frac{1}{f_{2j}}\phi\left(\frac{-\sqrt{f_{1j}}-\beta_{j0}}{\sqrt{f_{2j}}} \right)\delta(\beta_{j0}),
\]
where 
\[
f_{1j}=\frac{\lambda{\rm E}(D_{0,ii})}{{\rm E}(D_{0,ii} x_{ij}^{2}){\rm E}(D_{0,ii})-\{{\rm E}(D_{0,ii} x_{ij})\}^2},
\]
\[
f_{2j}=\{{\rm E}(\widetilde{\mathbf{X}}^\top\mathbf{D}_{0}\widetilde{\mathbf{X}})\}^{-1}_{j+1,j+1},
\]
and 
\begin{multline*}
	\delta(\beta_{j0})\\
	=\bigg[\Big\{-\frac{1}{2}(f_{1j})^{-\frac{1}{2}}\frac{\partial f_{1j}}{\partial \beta_{j0}}+1\Big\}\sqrt{f_{2j}}-\frac{1}{2}(f_{2j})^{-\frac{1}{2}}(-\sqrt{f_{1j}}+\beta_{j0})\frac{\partial f_{2j}}{\partial \beta_{j0}}\bigg]\exp\Big(\frac{2\sqrt{f_{1j}}\beta_{j0}}{f_{2j}}\Big)\\
	+\Big\{-\frac{1}{2}(f_{1j})^{-\frac{1}{2}}\frac{\partial f_{1j}}{\partial \beta_{j0}}-1\Big\}\sqrt{f_{2j}}+\frac{1}{2}(f_{2j})^{-\frac{1}{2}}(\sqrt{f_{1j}}+\beta_{j0})\frac{\partial f_{2j}}{\partial \beta_{j0}},
\end{multline*}
with
\[
\begin{aligned}
	&\frac{\partial f_{1j}}{\partial \beta_{j0}}\\
	=&\frac{\lambda\frac{\partial {\rm E}(D_{0,ii})}{\partial \beta_{j0}}\left[{\rm E}(D_{0,ii}x_{ij}^{2}){\rm E}(D_{0,ii})-\{{\rm E}(D_{0,ii}x_{ij})\}^2\right]   }{\big[{\rm E}(D_{0,ii}x_{ij}^{2}){\rm E}(D_{0,ii})-\{{\rm E}(D_{0,ii}x_{ij})\}^2\big]^2}\\
	&-\frac{\lambda{\rm E}(D_{0,ii})\left\{ \frac{\partial {\rm E}(D_{0,ii}x_{ij}^{2})}{\partial\beta_{j0}}{\rm E}(D_{0,ii})+{\rm E}(D_{0,ii}x_{ij}^{2})\frac{\partial {\rm E}(D_{0,ii})}{\partial \beta_{j0}}-2{\rm E}(D_{0,ii}x_{ij})\frac{\partial {\rm E}(D_{0,ii}x_{ij})}{\partial \beta_{j0}} \right\}  }{\left[{\rm E}(D_{0,ii}x_{ij}^{2}){\rm E}(D_{0,ii})-\{{\rm E}(D_{0,ii}x_{ij})\}^2\right]^2},
\end{aligned}
\]
\[
\frac{\partial f_{2j}}{\partial \beta_{j0}}=\left[\{{\rm E}(\widetilde{\mathbf{X}}^\top\mathbf{D}_{0}\widetilde{\mathbf{X}})\}^{-1}\big\{\rm E(\widetilde{\mathbf{X}}^\top\mathbf{M}_{0} \widetilde{\mathbf{X}})\big\}\{{\rm E}(\widetilde{\mathbf{X}}^\top\mathbf{D}_{0}\widetilde{\mathbf{X}})\}^{-1}\right]_{j+1,j+1},
\]
and 
\[
\mathbf{M}_{0}={\rm diag}\left\{\frac{\partial^{3}\ell_{1}\{\mu_{1}(\bm{\gamma}_{0})\}}{\partial \mu_{1}^{3}}x_{1j},\ldots,\frac{\partial^{3}\ell_{n}\{\mu_{n}(\bm{\gamma}_{0})\}}{\partial \mu_{n}^{3}}x_{nj} \right\}.
\]

To simplify the proof, we first consider the case where $(\mathbf{x}_{i},y_{i})$ follows a logistic regression model, that is,
\[
{\rm E}(y_{i}|\mathbf{x}_{i})=p_i=\frac{\exp( \alpha_{0}+\mathbf{x}_{i}^\top\bm{\beta}_{0}  )}{1+\exp(\alpha_{0}+\mathbf{x}_{i}^\top\bm{\beta}_{0})}.
\]
By calculation, $D_{0,ii}=p_{i}(1-p_{i})$ and $\mathbf{D}_{0}=\mathrm{diag}\{p_{1}(1-p_{1}),\ldots,p_{n}(1-p_{n})\}$.
Assume $p=2$, $x_{i1}$ and $x_{i2}$ are independent, ${\rm E}(x_{ij})=0$ and ${\rm Var}(x_{ij})=1$, $j=1,2$. Denote $\exp(\alpha_{0}+x_{ik}\beta_{k0})$ as $t_{k}$, $k\neq j$. It is easy to show that $\delta(0)=0$ and 
\begin{multline*}
	\frac{\partial\delta(\beta_{j0})}{\partial\beta_{j0}}\bigg|_{\beta_{j0}=0}\\
	=\sqrt{\frac{\lambda}{n}}\times\frac{2\left[{\rm E}\left\{\frac{t_{k}(1-t_{k})x_{ik}}{(1+t_{k})^3}\right\} {\rm E}\left\{\frac{t_{k}}{(1+t_{k})^2}\right\}-{\rm E}\left\{\frac{t_{k}x_{ik}}{(1+t_{k})^2}\right\}{\rm E}\left\{\frac{t_{k}(1-t_{k})}{(1+t_{k})^3}\right\} \right]^2}{ \left[{\rm E}\left\{\frac{t_{k}}{(1+t_{k})^2}\right\} \right]^3\left[{\rm E}\left\{\frac{t_{k}x_{ik}^2}{(1+t_{k})^2}\right\}{\rm E}\left\{\frac{t_{k}}{(1+t_{k})^2}\right\}-\left[{\rm E}\left\{\frac{t_{k}x_{ik}}{(1+t_{k})^2}\right\}\right]^2\right] }+2\sqrt{n\lambda}>0.
\end{multline*}
Therefore, 
\[
\frac{\partial P_{d,j}^\ast}{\partial \beta_{j0}}\bigg|_{\beta_{j0}=0}=0\quad  \mbox{and} \quad \frac{\partial^2P_{d,j}^\ast}{\partial \beta_{j0}^2}\bigg|_{\beta_{j0}=0}>0.
\]
It means that $P_{d,j}^\ast$ obtains a minimum value at $\beta_{j0}=0$.
Furthermore, there exists two positive constant $c_1$  and $c_2$ such that  $\delta(\beta_{j0})\geq 0$ for any $\beta_{j0}\in [0,c_1]$ and $\delta(\beta_{j0})\leq 0$ for any $\beta_{j0}\in [-c_2,0]$. Thus, $\partial P_{d,j}^\ast/\partial \beta_{j0}\geq 0$ for any $\beta_{j0}\in [0,c_1]$ and $\partial P_{d,j}^\ast/\partial \beta_{j0}\leq 0$ for any $\beta_{j0}\in [-c_2,0]$. In other words, $P_{d,j}^{\ast}$ is an increasing function of $\beta_{j0}$ if $0<\beta_{j0}<c_1$ and $P_{d,j}^{\ast}$ is a decreasing function of $\beta_{j0}$ if $-c_2<\beta_{j0}<0$.

Second, we consider the case  where $(\mathbf{x}_{i},y_{i})$ follows a Poisson regression model, that is,
\[
P(y_{i}=y|\mathbf{x}_{i})=\frac{\lambda_{i}^{y}}{y!}\exp(-\lambda_{i}),
\]
where $\lambda_{i}={\rm E}(y_{i}|\mathbf{x}_{i})=\exp(\alpha_{0}+\mathbf{x}_{i}^\top\bm{\beta}_{0})$. By calculation, $D_{0,ii}=\lambda_{i}$ and $\mathbf{D}_{0}=\mathrm{diag}\{\lambda_{1},\ldots,\lambda_{n}\}$.
Assume $p=2$, $x_{i1}$ and $x_{i2}$ are independent, ${\rm E}(x_{ij})=0$ and ${\rm Var}(x_{ij})=1$, $j=1,2$. Denote $\exp(\alpha_{0}+x_{ik}\beta_{k0})$ as $t_{k}$, $k\neq j$. Then 
\[
\frac{\partial P_{d,j}^\ast}{\partial \beta_{j0}}=\frac{n\lambda}{f_{1j}}\phi\left(-\sqrt{n\lambda}-\beta_{j0}\sqrt{ \frac{n\lambda}{f_{1j}} }\right)\delta(\beta_{j0}),
\]
with 
\[
\delta(\beta_{j0})=\left(\sqrt{\frac{f_{1j}}{n\lambda}}-\frac{\beta_{j0}}{2\sqrt{n\lambda f_{1j}}}\frac{\partial f_{1j}}{\beta_{j0}}\right)\left\{\exp\left(\frac{2\beta_{j0}n\lambda}{\sqrt{f_{1j}}}\right)-1  \right\},
\]
\[
f_{1j}=\frac{ \lambda{\rm E}\left\{\exp(x_{ij}\beta_{j0})\right\}}{ {\rm E}(t_{k})\left[ {\rm E}\left\{\exp(x_{ij}\beta_{j0})x_{ij}^{2}\right\} {\rm E}\left\{\exp(x_{ij}\beta_{j0})\right\}-\left[{\rm E}\left\{\exp(x_{ij}\beta_{j0})x_{ij}\right\}\right]^2 \right]},
\]
and
\[
\begin{aligned}
	\frac{\partial f_{1j}}{\partial \beta_{j0}}=&\frac{2\lambda{\rm E}\left\{\exp(x_{ij}\beta_{j0})\}{\rm E}\{\exp(x_{ij}\beta_{j0})x_{ij}\right\}{\rm E}\left\{\exp(x_{ij}\beta_{j0})x_{ij}^{2}\right\}}{ {\rm E}(t_{k})\left[ {\rm E}\left\{\exp(x_{ij}\beta_{j0})x_{ij}^{2}\right\} {\rm E}\left\{\exp(x_{ij}\beta_{j0})\right\}-\left[{\rm E}\left\{\exp(x_{ij}\beta_{j0})x_{ij}\right\}\right]^2 \right]^2}\\
	&-\frac{\lambda\left[{\rm E}\left\{\exp(x_{ij}\beta_{j0})x_{ij}\right\}\right]^{3}+\lambda\left[{\rm E}\left\{\exp(x_{ij}\beta_{j0})\right\}\right]^{2} {\rm E}\left\{\exp(x_{ij}\beta_{j0})x_{ij}^{3}\right\} }{ {\rm E}(t_{k})\left[ {\rm E}\left\{\exp(x_{ij}\beta_{j0})x_{ij}^{2}\right\} {\rm E}\left\{\exp(x_{ij}\beta_{j0})\right\}-\left[{\rm E}\left\{\exp(x_{ij}\beta_{j0})x_{ij}\right\}\right]^2 \right]^2}.
\end{aligned}
\]

In particular, if $x_{ij}$ follows the standard normal distribution, then 
\[
\begin{aligned}
	\frac{\partial P_{d,j}^\ast}{\partial \beta_{j0}}=&n{\rm E}(t_{k})\exp(\beta_{j0}^2/2)\phi\left[-\sqrt{n\lambda}-\beta_{j0}\sqrt{n{\rm E}(t_{k})\exp(\beta_{j0}^2/2)}\right]\\
	&\times  \left\{\sqrt{ \frac{1}{n{\rm E}(t_{k})\exp(\beta_{j0}^2/2)}+\frac{\beta_{j0}^2}{2\sqrt{n{\rm E}(t_{k})\exp(\beta_{j0}^2/2)}} } \right \}\\
	&\times \left[\exp\left\{2\beta_{j0}n\sqrt{\lambda{\rm E}(t_{k})\exp(\beta_{j0}^2/2)}\right\}-1 \right].
\end{aligned}
\]
Obviously, $\partial P_{d,j}^\ast/\partial \beta_{j0}>0$ if $\beta_{j0}>0$, $\partial P_{d,j}^\ast/\partial \beta_{j0}=0$ if $\beta_{j0}=0$ and $\partial P_{d,j}^\ast/\partial \beta_{j0}<0$ if $\beta_{j0}<0$.
Thus,  $P_{d,j}^\ast$ is an increasing function of $\beta_{j0}$ if $\beta_{j0}>0$ and $P_{d,j}^\ast$ is a decreasing function of $\beta_{j0}$ if $\beta_{j0}<0$.

\section{Proof for Theorem 1 \label{s:prooftheo1}}
According to (2.4) in the main paper, the objective function about $\bm{\beta}$ for the one-step adaptive lasso estimator is 
$$
Q(\bm{\beta})=\frac{1}{2n}(\bm{\beta}-\bm{\beta}^{(0)})^\top\mathbf{X}^\top\mathbf{D}^{\dagger(0)}\mathbf{X}(\bm{\beta}-\bm{\beta}^{(0)})+\sum_{j=1}^{p}\lambda \frac{|\beta_{j}|}{|\beta_{j}^{(0)}|}.
$$
For $\beta_{j}\approx \beta_{j}^{(1)}$, $Q(\bm{\beta})$ can be approximated by
\begin{align*}
	&\frac{1}{2n}(\bm{\beta}-\bm{\beta}^{(0)})^\top\mathbf{X}^\top\mathbf{D}^{\dagger(0)}\mathbf{X}(\bm{\beta}-\bm{\beta}^{(0)})+\sum_{j=1}^{p}\lambda \frac{|\beta_{j}^{(1)}|}{|\beta_{j}^{(0)}|}+\frac{1}{2}\sum_{j=1}^{p}\frac{\lambda}{|\beta_{j}^{(0)}||\beta_{j}^{(1)}|}\{\beta_{j}^{2}-(\beta_{j}^{(1)})^2\}\\
	=&L(\bm{\beta})+\sum_{j=1}^{p}\lambda \frac{|\beta_{j}^{(1)}|}{|\beta_{j}^{(0)}|}+\frac{1}{2}\sum_{j=1}^{p}\frac{\lambda}{|\beta_{j}^{(0)}||\beta_{j}^{(1)}|}\{\beta_{j}^{2}-(\beta_{j}^{(1)})^2\},
\end{align*}
where $L(\bm{\beta})=(\bm{\beta}-\bm{\beta}^{(0)})^\top\mathbf{X}^\top\mathbf{D}^{\dagger(0)}\mathbf{X}(\bm{\beta}-\bm{\beta}^{(0)})/(2n)$.

It can be shown easily that there exists  a $\bm{\beta}_{\mathscr{A}}^{(1)}$ that is a $\sqrt{n}$-consistent local minimizer of $Q\{( \bm{\beta}_{\mathscr{A}}^\top, \bm{0}_{\mathscr{A}^{c}}^\top )^\top \}$ and satisfies the following condition:
$$
\frac{\partial Q(\bm{\beta})}{\partial \beta_{j}}\Bigg|_{\bm{\beta}=\left(\begin{smallmatrix}
		\bm{\beta}_{\mathscr{A}}^{(1)}\\ \bm{0}_{\mathscr{A}^{c}}
	\end{smallmatrix}\right)}=0 \quad {\rm for} \quad j=1,\ldots, q,
$$
where $\mathscr{A}=\{j:\beta_{j0}\neq 0,j=1,\ldots,p\}$ and $\mathscr{A}^c=\{j:\beta_{j0}=0,j=1,\ldots,p\}$. Without loss of generality, assume $\mathscr{A}=\{1,\ldots, q\}$ and $q\leq p$.

Note that $\bm{\beta}_{\mathscr{A}}^{(1)}$ is a consistent estimator, then 
\begin{equation}
	\begin{aligned}
		&\frac{\partial L(\bm{\beta})}{\partial \beta_{j}}\Bigg|_{\bm{\beta}=\left(\begin{smallmatrix}
				\bm{\beta}_{\mathscr{A}}^{(1)}\\ \bm{0}_{\mathscr{A}^{c}}
			\end{smallmatrix}\right)}+\frac{\lambda}{|\beta_{j}^{(0)}||\beta_{j}^{(1)}|}\beta_{j}^{(1)}\\
		=&\frac{\partial L(\bm{\beta})}{\partial \beta_{j}}\Bigg|_{\bm{\beta}=\left(\begin{smallmatrix}
				\bm{\beta}_{\mathscr{A}}^{(1)}\\ \bm{0}_{\mathscr{A}^{c}}
			\end{smallmatrix}\right)}+\frac{\lambda}{|\beta_{j}^{(0)}|}{\rm sgn}(\beta_{j}^{(1)})\\
		=&\frac{\partial L(\bm{\beta_{0}})}{\partial \beta_{j}}+\sum_{\ell=1}^{q}\left\{\frac{\partial^{2}L(\bm{\beta}_{0})}{\partial \beta_{j}\partial \beta_{\ell}}+o_{p}(1)  \right\}(\beta_{\ell}^{(1)}-\beta_{\ell 0})\\
		&+\frac{\lambda}{|\beta_{j}^{(0)}|}{\rm sgn}(\beta_{j0})+\frac{\lambda}{|\beta_{j}^{(0)}||\beta_{j}^{(1)}|}(\beta_{j}^{(1)}-\beta_{j0})=0.
	\end{aligned}
	\label{eq:deri}
\end{equation}
Denote $\mathbf{X}^\top\mathbf{D}^{\dagger(0)}\mathbf{X}$ as $\mathbf{Z}^{(0)}$, then according to (\ref{eq:deri}),
\begin{equation}
	\begin{aligned}
		&\sqrt{n}\left\{\frac{1}{n}\mathbf{Z}^{(0)}_{\mathscr{A}}+ \Sigma_{\lambda}(\bm{\beta}_{\mathscr{A}}^{(0)},\bm{\beta}_{\mathscr{A}}^{(1)})\right\}\\
		&\times \left[\bm{\beta}_{\mathscr{A}}^{(1)}- \bm{\beta}_{0,\mathscr{A}}+\left\{\frac{1}{n}\mathbf{Z}^{(0)}_{\mathscr{A}}+ \Sigma_{\lambda}(\bm{\beta}_{\mathscr{A}}^{(0)},\bm{\beta}_{\mathscr{A}}^{(1)})\right\}^{-1}\bm{b}(\bm{\beta}_{0,\mathscr{A}},\bm{\beta}_{\mathscr{A}}^{(0)})\right]\\
		=&-\sqrt{n}\frac{\partial L(\bm{\beta}_{0})}{\partial \bm{\beta}_{\mathscr{A}}}=\frac{1}{\sqrt{n}}\mathbf{Z}_{\mathscr{A}}^{(0)}(\bm{\beta}_{\mathscr{A}}^{(0)}-\bm{\beta}_{0,\mathscr{A}}),
	\end{aligned}
	\label{eq:taylorexpan}
\end{equation}
where $\Sigma_{\lambda}(\bm{\beta}_{\mathscr{A}}^{(0)},\bm{\beta}_{\mathscr{A}}^{(1)})={\rm diag}\{\lambda/(|\beta_{1}^{(0)}||\beta_{1}^{(1)}|),\ldots,\lambda/(|\beta_{q}^{(0)}||\beta_{q}^{(1)}|)\}$ and $\bm{b}(\bm{\beta}_{0,\mathscr{A}},\bm{\beta}_{\mathscr{A}}^{(0)})\\=(\lambda\times {\rm sgn}(\beta_{10})/|\beta_{1}^{(0)}|,\ldots, \lambda\times {\rm sgn}(\beta_{q0})/|\beta_{q}^{(0)}| )^\top$.
According to the Central Limit Theorem,  $\sqrt{n}(\bm{\beta}_{\mathscr{A}}^{(0)}-\bm{\beta}_{0,\mathscr{A}})\stackrel{D}{\rightarrow} \mathcal{N}(\bm{0}, \{(\mathbf{I}_{0,\mathscr{B}})^{-1}\}_{\mathscr{A}})$, where $\mathscr{B}=\{k:\gamma_{k0}\neq 0,k=1,\ldots,p+1\}$.  Furthermore, according to the Slutsky's Theorem,  the asymptotic bias of $\bm{\beta}_{\mathscr{A}}^{(1)}$ is 
$$
\mathrm{bias}(\bm{\beta}_{\mathscr{A}}^{(1)})=-\left\{\frac{1}{n}\mathbf{Z}_{0,\mathscr{A}}+\Sigma_{\lambda}(\bm{\beta}_{0,\mathscr{A}},\bm{\beta}_{0,\mathscr{A}})\right\}^{-1}\bm{b}(\bm{\beta}_{0,\mathscr{A}},\bm{\beta}_{0,\mathscr{A}}),
$$
where $\mathbf{Z}_{0}={\rm E}(\mathbf{X}^\top\mathbf{D}^{\dagger}_{0}\mathbf{X})$. The asymptotic covariance matrix of $\bm{\beta}_{\mathscr{A}}^{(1)}$ is
\begin{multline*}
	\mathrm{cov}(\bm{\beta}_{\mathscr{A}}^{(1)})=\frac{1}{n^{3}}\left\{\frac{1}{n}\mathbf{Z}_{0,\mathscr{A}}+\Sigma_{\lambda}(\bm{\beta}_{0,\mathscr{A}},\bm{\beta}_{0,\mathscr{A}})\right\}^{-1}\mathbf{Z}_{0,\mathscr{A}} \{(\mathbf{I}_{0,\mathscr{B}})^{-1}\}_{\mathscr{A}}\mathbf{Z}_{0,\mathscr{A}}\\ \times \left\{\frac{1}{n}\mathbf{Z}_{0,\mathscr{A}}+\Sigma_{\lambda}(\bm{\beta}_{0,\mathscr{A}},\bm{\beta}_{0,\mathscr{A}})\right\}^{-1}.
\end{multline*}
If $\lambda\rightarrow 0$ as $n$ goes to infinity, then $\mathrm{bias}(\bm{\beta}_{\mathscr{A}}^{(1)})\rightarrow \bm{0}$ and $n\mathrm{cov}(\bm{\beta}_{\mathscr{A}}^{(1)})\rightarrow  \{(\mathbf{I}_{0,\mathscr{B}})^{-1}\}_{\mathscr{A}}$.

If $n$ is finite, then the bias of $\bm{\beta}_{\mathscr{A}}^{(1)}$ can not be ignored and ${\mathscr{A}}_{n}$ is not necessarily equal to ${\mathscr{A}}$. Without loss of generality, assume $\mathscr{A}_{n}=\{j: \beta_{j}^{(1)}\neq 0, j=1,\ldots,p\}=\{1,\ldots, s\}$. Then $\mathscr{B}_{n}=\{k: \gamma_{k}^{(1)}\neq 0, k=1,\ldots,p+1\}=\{1,\ldots, s+1\}$. Furthermore, the estimators of bias  and  covariance matrix of $\bm{\beta}_{\mathscr{A}_{n}}^{(1)}$ are given by
\[
\widehat{\mathrm{bias}}(\bm{\beta}_{\mathscr{A}_{n}}^{(1)})=-\left\{\frac{1}{n}\mathbf{Z}_{\mathscr{A}_{n}}^{(0)}+\Sigma_{\lambda}(\bm{\beta}_{\mathscr{A}_{n}}^{(0)},\bm{\beta}_{\mathscr{A}_{n}}^{(1)})\right\}^{-1}\bm{b}(\bm{\beta}_{\mathscr{A}_{n}}^{(1)},\bm{\beta}_{\mathscr{A}_{n}}^{(0)})
\]
and
\begin{multline*}
	\widehat{\mathrm{cov}}(\bm{\beta}_{\mathscr{A}_{n}}^{(1)})=\frac{1}{n^{3}}\left\{\frac{1}{n}\mathbf{Z}_{\mathscr{A}_{n}}^{(0)}+\Sigma_{\lambda}(\bm{\beta}_{\mathscr{A}_{n}}^{(0)},\bm{\beta}_{\mathscr{A}_{n}}^{(1)})\right\}^{-1}\mathbf{Z}_{\mathscr{A}_{n}}^{(0)} \{(\mathbf{I}^{(0)}_{\mathscr{B}_{n}})^{-1}\}_{\mathscr{A}_{n}} \mathbf{Z}_{\mathscr{A}_{n}}^{(0)} \\ \times \left\{\frac{1}{n}\mathbf{Z}_{\mathscr{A}_{n}}^{(0)}+\Sigma_{\lambda}(\bm{\beta}_{\mathscr{A}_{n}}^{(0)},\bm{\beta}_{\mathscr{A}_{n}}^{(1)})\right\}^{-1},
\end{multline*}
where $\Sigma_{\lambda}(\bm{\beta}_{\mathscr{A}_{n}}^{(0)},\bm{\beta}_{\mathscr{A}_{n}}^{(1)})={\rm diag}\{\lambda/(|\beta_{1}^{(0)}||\beta_{1}^{(1)}|),\ldots,\lambda/(|\beta_{s}^{(0)}||\beta_{s}^{(1)}|)\}$ and $\bm{b}(\bm{\beta}_{\mathscr{A}_{n}}^{(1)},\bm{\beta}_{\mathscr{A}_{n}}^{(0)})=(\lambda\times {\rm sgn}(\beta_{1}^{(1)})/|\beta_{1}^{(0)}|,\ldots, \lambda\times {\rm sgn}(\beta_{s}^{(1)})/|\beta_{s}^{(0)}| )^\top$.

\section{Implementation Details of Several Methods}
In this section, we introduce the implementation details of several methods mentioned in the main paper. 
\subsection{One-step adaptive lasso estimator}
To obtain the one-step adaptive lasso estimator, we use the function \texttt{glmnet} in \textsf{R} to solve (2.5). The selection of tuning parameter $\lambda$ is important. In finite samples, if $\lambda$ is too large, the bias of the one-step adaptive lasso estimator will be large and the coverage probability of the confidence interval constructed based on the asymptotic theory for the one-step adaptive lasso estimator will be low; if $\lambda$ is too small, the number of false positives will be large and the width of the confidence interval will also be large. The Bayesian information criterion (BIC) and cross-validation (CV) method are two commonly used tuning parameter selection methods. Based on the simulation results, $\lambda$ selected  based on the Bayesian information criterion proposed by \cite{wang2007unified} is much larger than the value of $\lambda$ selected by the 5-fold cross-validation method. Denote the values of $\lambda$ selected by these two methods as $\lambda_{\rm BIC}$ and $\lambda_{\rm CV}$, respectively. We choose $\lambda$ to be $(\lambda_{\rm BIC}+\lambda_{\rm CV})/2$ as a trade-off of these two methods.

\subsection{Estimating equation-based method}
In our simulation studies and  real-data application, we compare the proposed method with an estimating equation-based method, which is proposed by \cite{neykov2018unified} and denoted as ``EstEq.''   We apply their method based on Algorithm 1 in their paper. Using the same notations as in our paper, the implementation details are as follows: 
\begin{enumerate}
	\item [Step 1:] Use the R functions {\tt gds} and {\tt cv\_gds} to get the generalized Dantzig selector of the regression coefficient $\bm{\gamma}_0=(\alpha_0,\bm{\beta}_0^\top)^\top$ in a logistic regression model and denote the estimator as $\hat{\bm{\gamma}}$. That is, solve the following optimization problem to obtain an estimate $\hat{\bm{\gamma}}$:
	\begin{flalign*}
		\hat{\bm{\gamma}}=&\arg\min \|\bm{\gamma}\|_1, \\
		& \text{subject to} \left\|\bm{t}(\bm{\gamma})\right\|= \left\|-\frac{1}{n}\sum_{i=1}^{n}\frac{\partial \ell_i(\bm{\gamma})}{\partial \bm{\gamma}}\right\|_{\infty}=\left\|-\frac{1}{n}\sum_{i=1}^{n}\{y_i-p_i(\bm{\gamma})\} \tilde{\mathbf{x}}_i\right\|_{\infty}\le \lambda,
	\end{flalign*}
	where $\ell_i(\bm{\gamma})$ is the conditional log-likelihood function of $y_i$ given $\mathbf{x}_i$ for a logistic regression model and $p_i(\bm{\gamma})=\exp(\tilde{\mathbf{x}}_i^\top\bm{\gamma} )/\{1+\exp(\tilde{\mathbf{x}}_i^\top\bm{\gamma} )\}$, $i=1,\ldots, n$.
	The tuning parameter of the  generalized Dantzig selector, $\lambda$,  is selected by the 10-fold cross-validation method. 
	\item [Step 2:] Calculate the inverse of $\mathbf{T}(\hat{\bm{\gamma}})=\partial \bm{t}(\hat{\bm{\gamma}})/\partial \bm{\gamma}^\top=\tilde{\mathbf{X}}^\top \mathbf{D}(\hat{\bm{\gamma}})\tilde{\mathbf{X}}/n$, where $\mathbf{D}(\hat{\bm{\gamma}})={\rm diag}\{p_1(\hat{\bm{\gamma}})(1-p_1(\hat{\bm{\gamma}})),\ldots,p_n(\hat{\bm{\gamma}})(1-p_n(\hat{\bm{\gamma}}))\}$. Denote the inverse of $\mathbf{T}(\hat{\bm{\gamma}})$ as $\mathbf{\Omega}$. Define  the projection direction  for the $j$th element of $\bm{\beta}_0$, $\beta_{j0}$,  as $\hat{\mathbf{v}}_j=\mathbf{\Omega}_{(j+1).}$, where $\mathbf{\Omega}_{(j+1).}$ is the $(j+1)$th row element of $\mathbf{\Omega}$. Note that in \cite{neykov2018unified}, the authors used the CLIME estimator to estimate the inverse of $\mathbf{T}(\hat{\bm{\gamma}})$. However, in our problem, we assume $n>p$ and $p$ is fixed, then the inverse of $\mathbf{T}(\hat{\bm{\gamma}})$ can be calculated directly. 
	\item [Step 3:] Use the R function {\tt uniroot} to solve the sparse projected test function and denote the estimated value of $\beta_{j0}$ as $\tilde{\beta}_j$. 
	\item [Step 4:]  Construct a two-sided $100(1-\alpha)\%$ confidence interval for $\beta_{j0}$ as 
	\[
	{\rm CI}_{j}=\left(\tilde{\beta}_j-\Phi^{-1}(1-\alpha/2)\hat{\sigma}_j/\sqrt{n},\tilde{\beta}_j+\Phi^{-1}(1-\alpha/2)\hat{\sigma}_j/\sqrt{n} \right),
	\]
	where $\hat{\sigma}_j^2=\hat{\mathbf{v}}_j^\top\tilde{\mathbf{X}}^\top\mathbf{D}(\hat{\bm{\gamma}})\tilde{\mathbf{X}}\hat{\mathbf{v}}_j/n$.
\end{enumerate}

\subsection{Two types of bootstrap de-biased lasso methods}
Motivated by the idea of \cite{dezeure2017high}, we establish two xy-paired bootstrap de-biased lasso methods, which are referred to as ``the  type-I bootstrap de-biased lasso method'' and  ``the type-II bootstrap de-biased lasso method,'' respectively.  The  bootstrap de-biased lasso method is based on the de-biased lasso method proposed by \cite{zhang2014confidence}, \cite{van2014asymptotically} and  \cite{javanmard2014confidence}.  Following  the idea of \cite{dezeure2017high}, the procedure for  the  type-I bootstrap de-biased lasso method is as follows:
\begin{enumerate}
	\item [(i)] Based on the original data points  $(\mathbf{X}_{1},Y_1), \ldots, (\mathbf{X}_{n},Y_n)$, calculate the lasso estimator and de-biased lasso estimator  of the $j$th element of $\bm{\beta}_0$, $\beta_{j0}$.  Denote them as $\hat{b}_j$ and $\hat{\beta}_j$, respectively. Calculate the standard error of the de-biased lasso estimator, $\widehat{\rm s.e.}_{j}$.
	\item [(ii)] Resample $(\mathbf{X}^\ast_{1},Y^\ast_1), \ldots, (\mathbf{X}^\ast_{n},Y^\ast_n)$ with replacement from $(\mathbf{X}_{1},Y_1), \ldots, (\mathbf{X}_{n},Y_n)$ for $B$ times. For the $k$th bootstrap sample, calculate the de-biased lasso estimator  $\hat{b}^{\ast}_{jk}$, the standard error for the de-biased lasso estimator $\widehat{\rm s.e.}^\ast_{jk}$ and $T_{jk}^\ast=(\hat{b}^{\ast}_{jk}-\hat{\beta}_{j})/\widehat{\rm s.e.}^\ast_{jk}$. Denote the $\nu$-quantile of $\{T_{j1}^\ast,\ldots, T_{jB}^\ast\}$ as $q_{j;\nu}^\ast$ . 
	\item [(iii)] Construct a two-sided $100(1-\alpha)\%$ confidence interval for $\beta_{j0}$ as 
	\[
	{\rm CI}_{j}=\left(\hat{b}_{j}-q_{j;1-\alpha/2}^\ast\widehat{\rm s.e.}_{j},\hat{b}_{j}-q_{j;\alpha/2}^\ast\widehat{\rm s.e.}_{j}  \right).
	\]
\end{enumerate}

In addition,  the procedure for  the  type-II bootstrap de-biased lasso method is as follows:
\begin{enumerate}
	\item [(i)] Resample $(\mathbf{X}^\ast_{1},Y^\ast_1), \ldots, (\mathbf{X}^\ast_{n},Y^\ast_n)$ with replacement from $(\mathbf{X}_{1},Y_1), \ldots, (\mathbf{X}_{n},Y_n)$ for $B$ times. For the $k$th bootstrap sample, calculate the de-biased lasso estimator  of the $j$th element of $\bm{\beta}_0$, $\beta_{j0}$, which is denoted as $\hat{b}^{\ast}_{jk}$. Denote  the $\nu$-quantile of $\{\hat{b}^{\ast}_{j1},\ldots, \hat{b}^{\ast}_{jB}\}$ as $q_{j;\nu}^\ast$. 
	\item [(iii)] Construct a two-sided $100(1-\alpha)\%$ confidence interval for $\beta_{j0}$ as 
	\[
	{\rm CI}_{j}=\left(q_{j;\alpha/2}^\ast,q_{j;1-\alpha/2}^\ast \right).
	\]
\end{enumerate}

\section{Additional Simulation Results\label{ss:addsimu}}
\begin{figure}[t!]
	\centerline{\includegraphics[width=40em,angle=0]{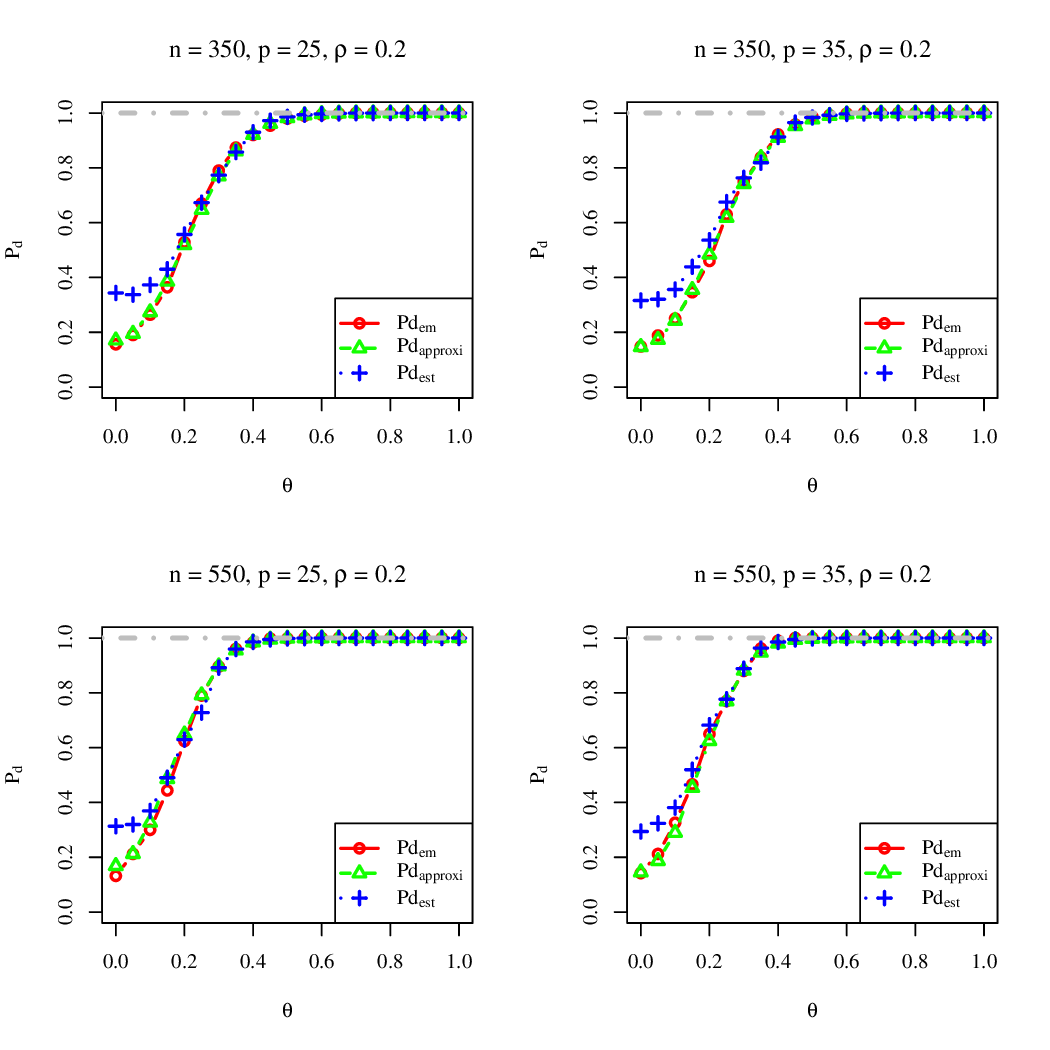}}
	\caption{\small Different types of selection probability for $\bm X_4$ when $\rho=0.2$.  ${\rm Pd_{em}}$:  empirical selection probability, which equals the empirical probability of $\{\theta^{(1)}\neq 0\}$ based on $500$ Monte Carlo samples; ${\rm Pd_{approxi}}$:   approximated selection probability based on  (3.1), where the expectations in  (3.1) are calculated by using the function \texttt{cubintegrate} in \textsf{R}; ${\rm Pd_{est}}$:  median of estimated selection probabilities based on (3.3) for $500$ Monte Carlo samples.}
	\label{figure:selectwcorr}
\end{figure}

\begin{figure}[t!]
	\centerline{\includegraphics[width=40em,angle=0]{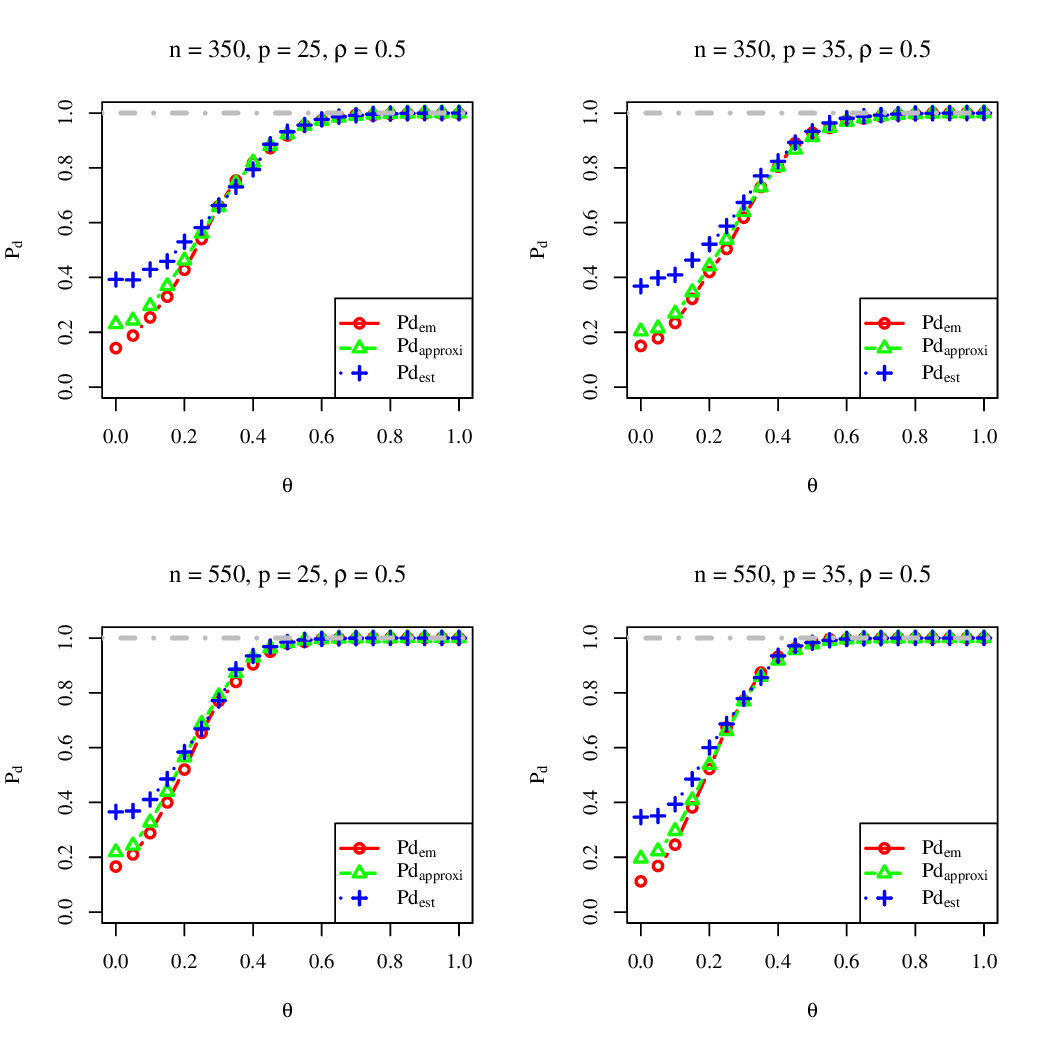}}
	\caption{\small Different types of selection probability for $\bm X_4$ when $\rho=0.5$. The meanings of notations:  see Figure	\ref{figure:selectwcorr}.}
	\label{figure:selectmcorr}
\end{figure}

\begin{figure}[t!]
	\centerline{\includegraphics[width=40em,angle=0]{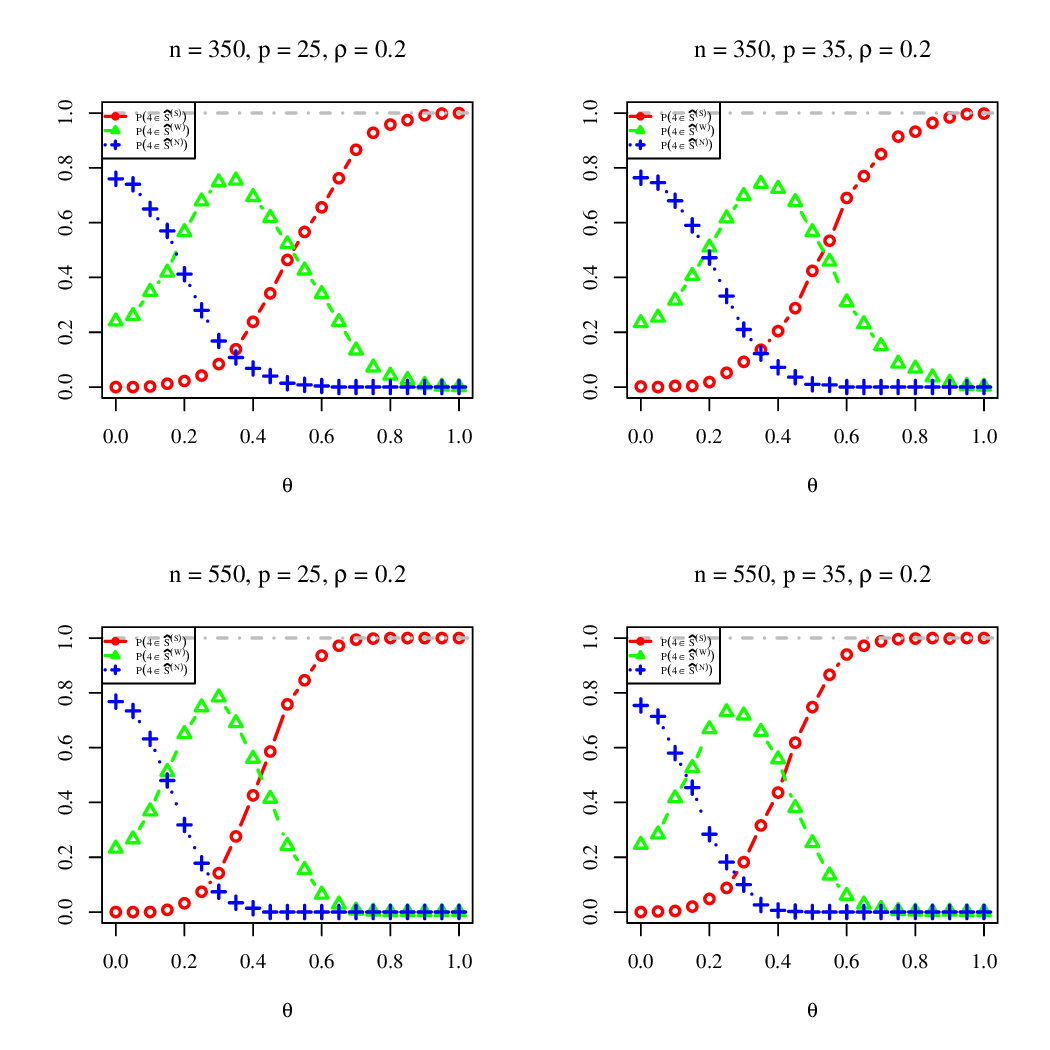}}
	\caption{\small Empirical probabilities of assigning the covariate $\bm X_4$ to different signal categories  when $\rho=0.2$.}
	\label{figure:signalidentifywcorr}
\end{figure}

\begin{figure}[t!]
	\centerline{\includegraphics[width=40em,angle=0]{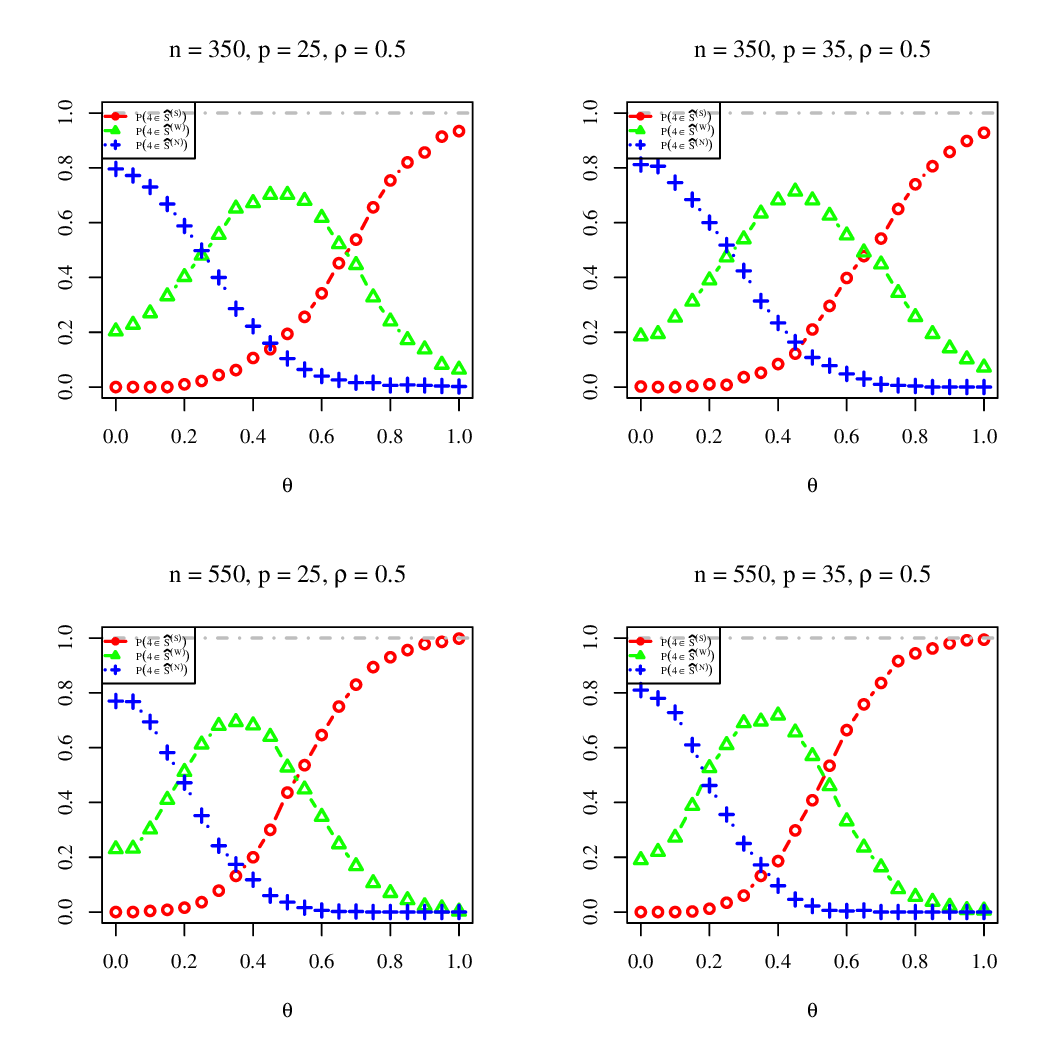}}
	\caption{\small Empirical probabilities of assigning the covariate $\bm X_4$ to different signal categories  when $\rho=0.5$.}
	\label{figure:signalidentifymcorr}
\end{figure}

\begin{figure}[t!]
	\centerline{\includegraphics[width=30em,angle=0]{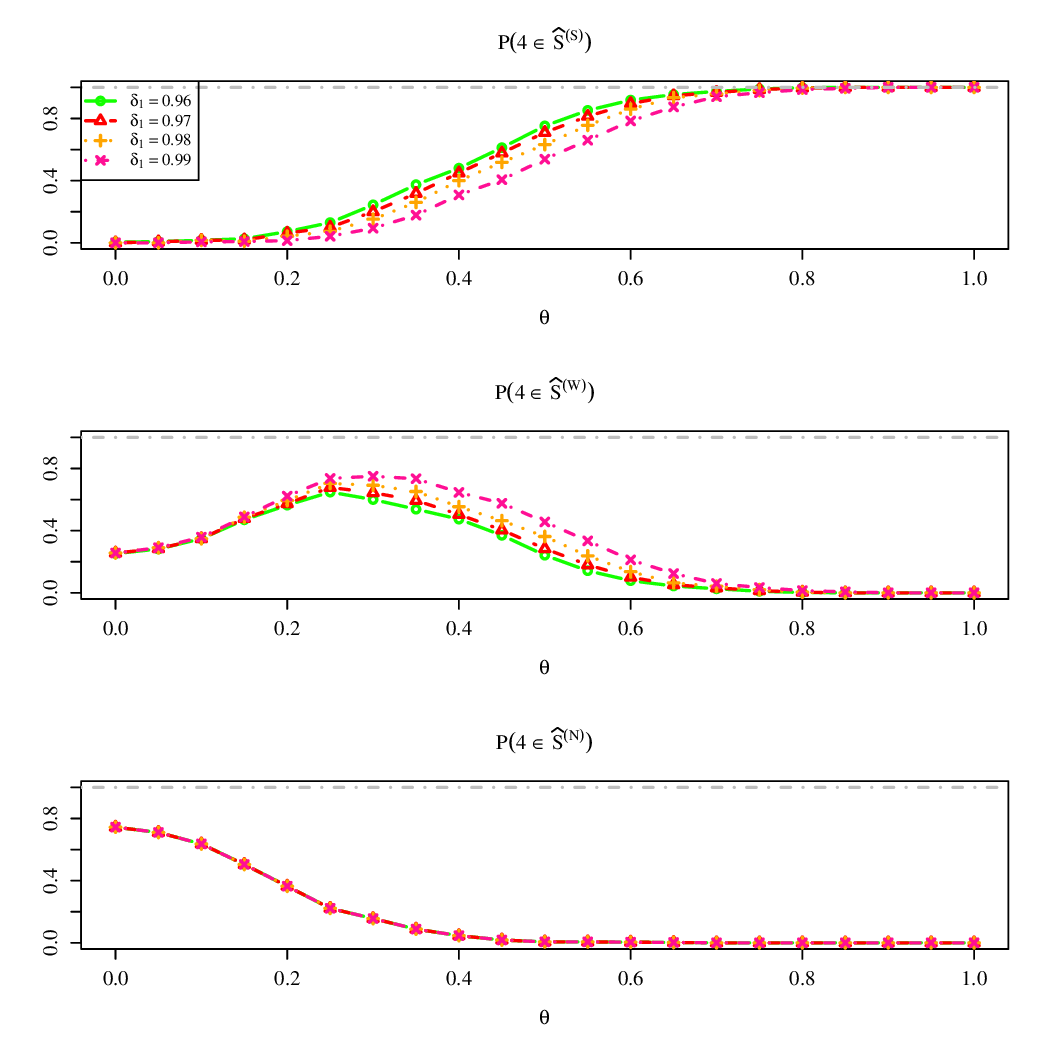}}
	\caption{\small Empirical probabilities of assigning the covariate $\bm X_4$ to different signal categories  when $(n,p,\rho)=(350,25,0)$, $\tau=0.1$ and the threshold value $\delta_1$ varies.}
	\label{figure:signalidentifyfixtau}
\end{figure}

\begin{figure}[t!]
	\centerline{\includegraphics[width=30em,angle=360]{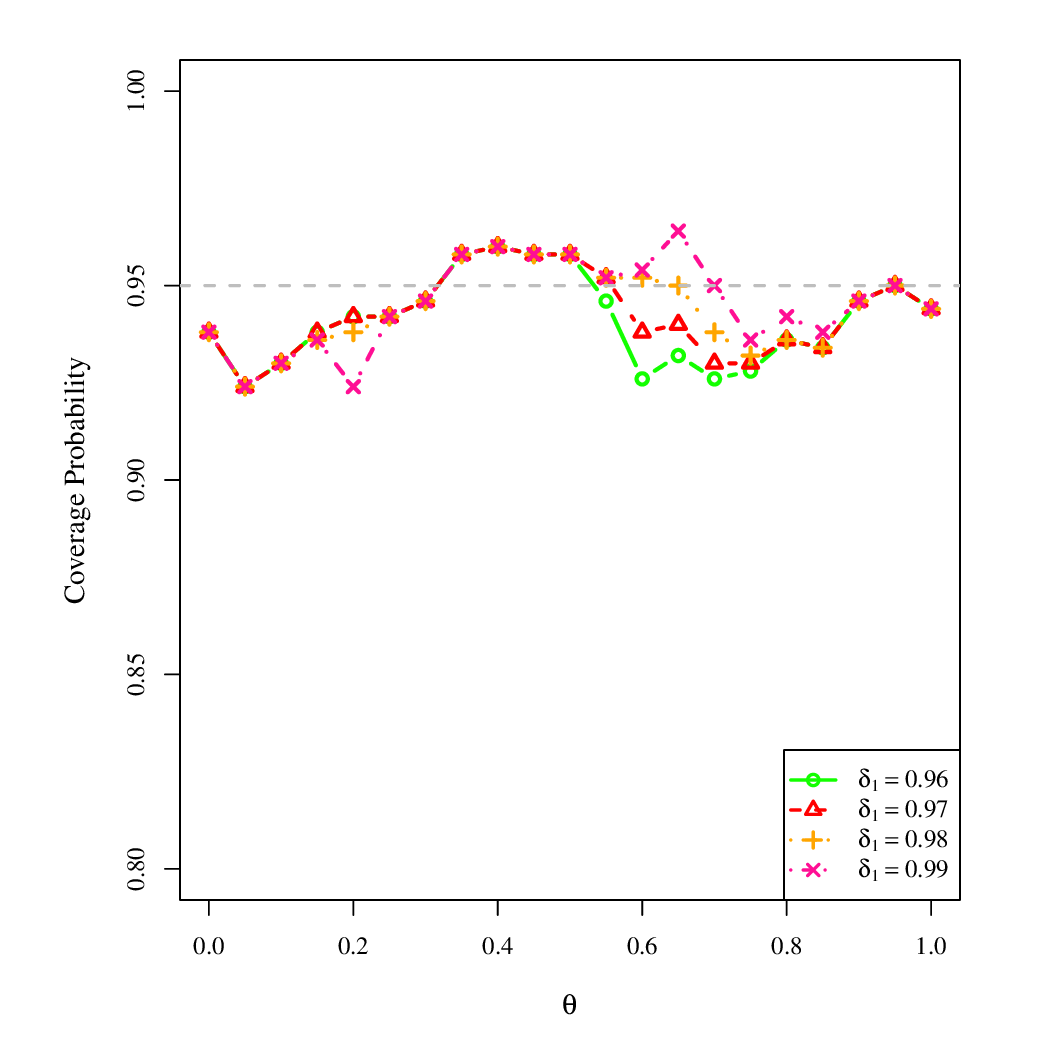}}
	\caption{\small Coverage probabilities of the $95\%$ confidence intervals for the proposed two-step inference method when $(n,p,\rho)=(350,25,0)$, $\tau=0.1$  and the threshold value $\delta_1$ varies.}
	\label{figure:coverprobfixtau}
\end{figure}

\begin{figure}[t!]
	\centerline{\includegraphics[width=30em,angle=360]{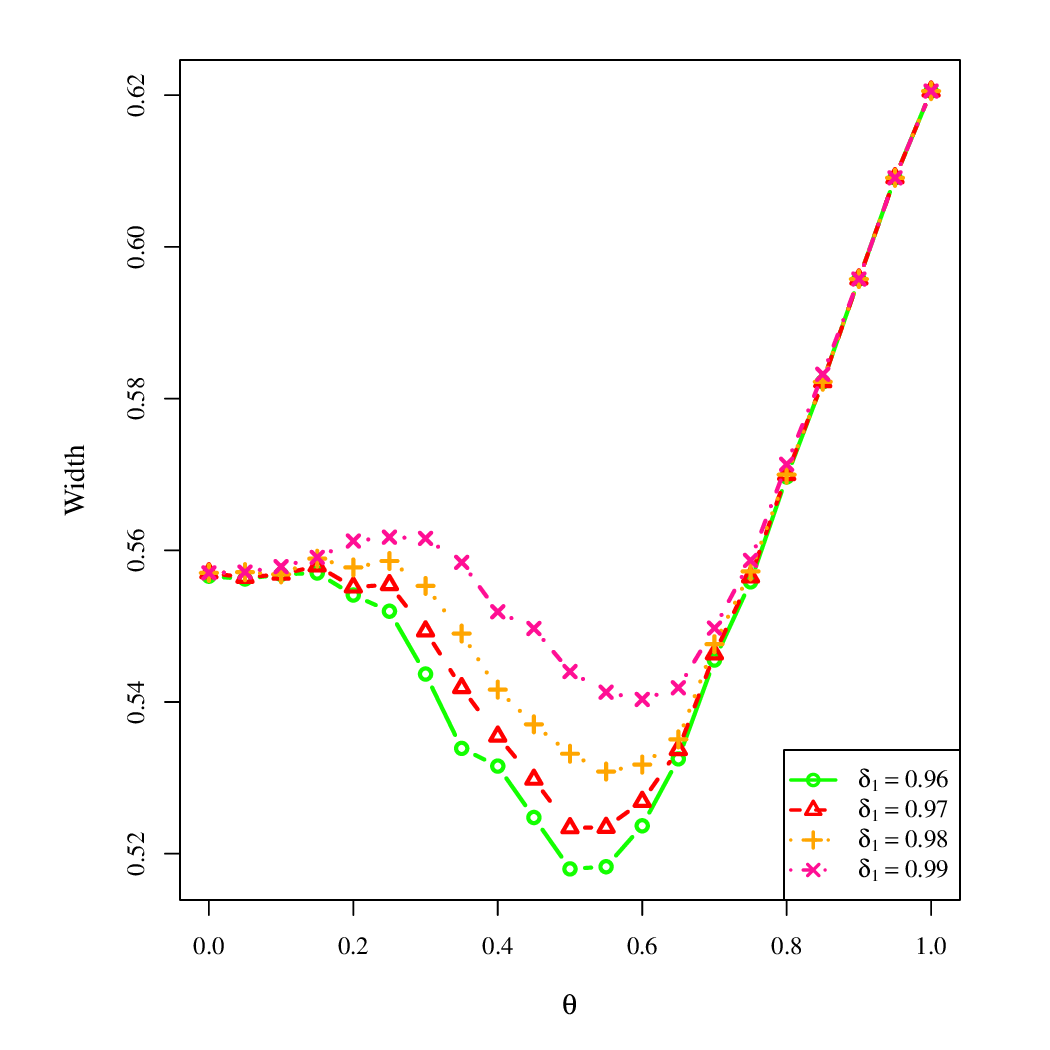}}
	\caption{\small Average widths of the $95\%$  confidence intervals for the proposed two-step inference method when $(n,p,\rho)=(350,25,0)$,  $\tau=0.1$  and the threshold value $\delta_1$ varies.}
	\label{figure:coverwidthfixtau}
\end{figure}

\begin{figure}[t!]
	\centerline{\includegraphics[width=30em,angle=0]{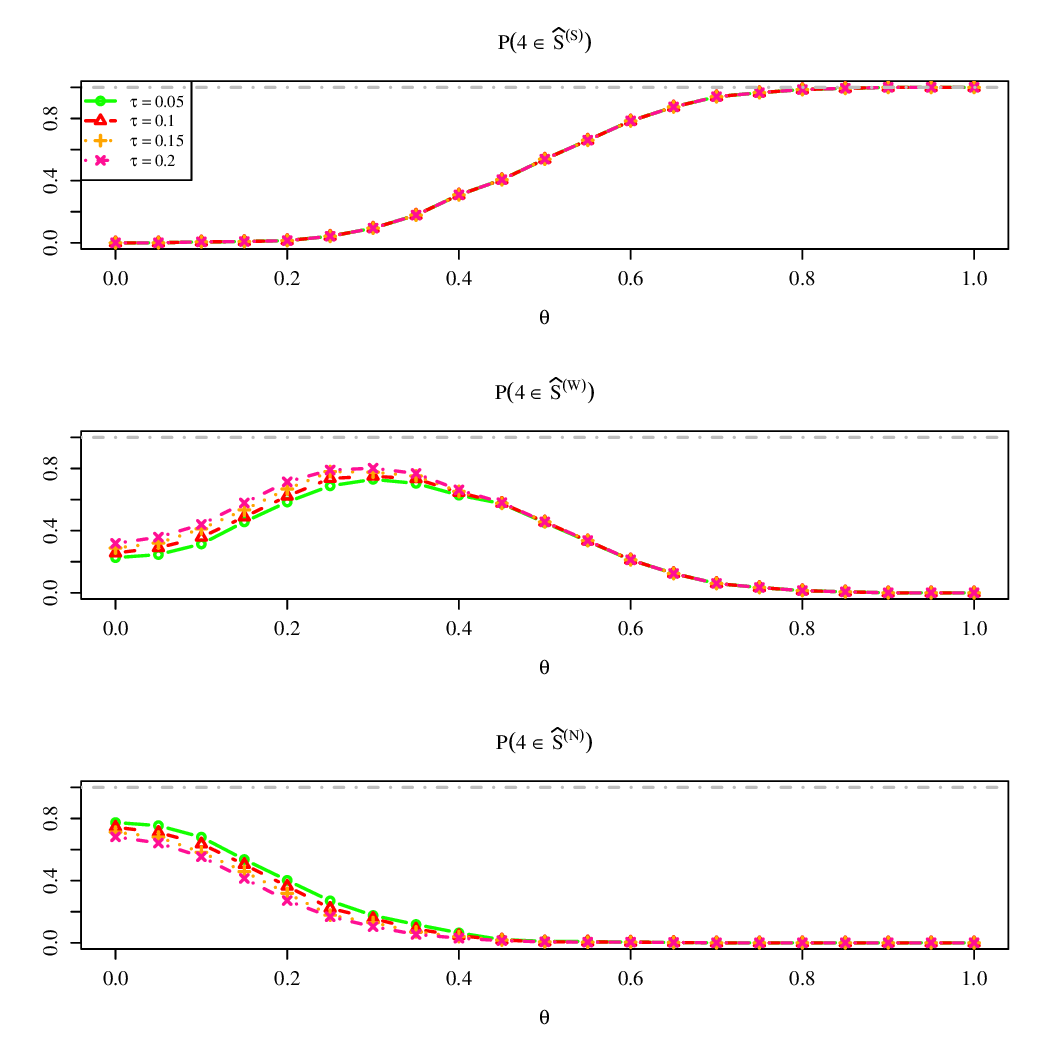}}
	\caption{\small Empirical probabilities of assigning the covariate $\bm X_4$ to different signal categories  when $(n,p,\rho)=(350,25,0)$, $\delta_1=0.99$ and the threshold value $\tau$ varies.}
	\label{figure:signalidentifyfixdelta1}
\end{figure}

\begin{figure}[t!]
	\centerline{\includegraphics[width=30em,angle=360]{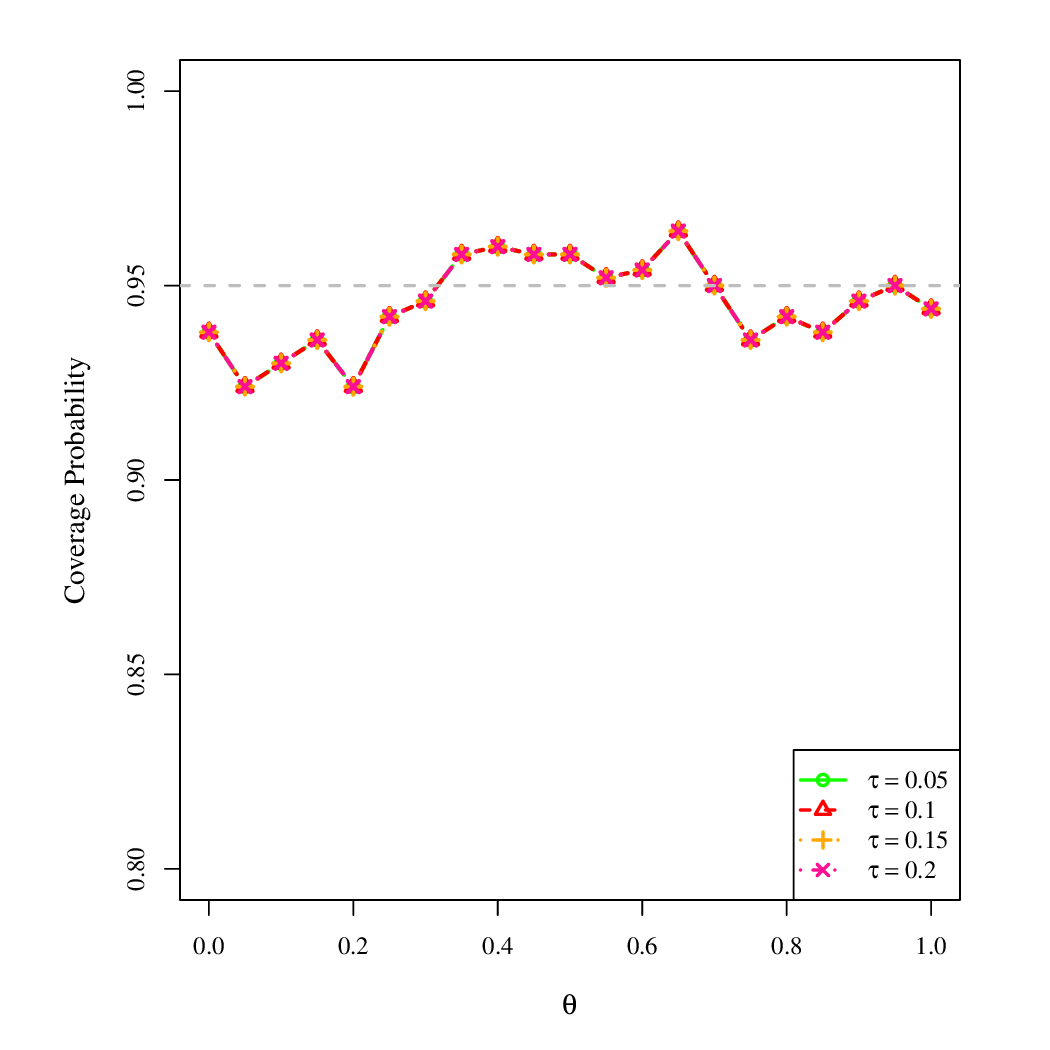}}
	\caption{\small Coverage probabilities of the $95\%$ confidence intervals for the proposed two-step inference method when $(n,p,\rho)=(350,25,0)$, $\delta_1=0.99$ and the threshold value $\tau$ varies.}
	\label{figure:coverprobfixtdelta1}
\end{figure}

\begin{figure}[t!]
	\centerline{\includegraphics[width=30em,angle=360]{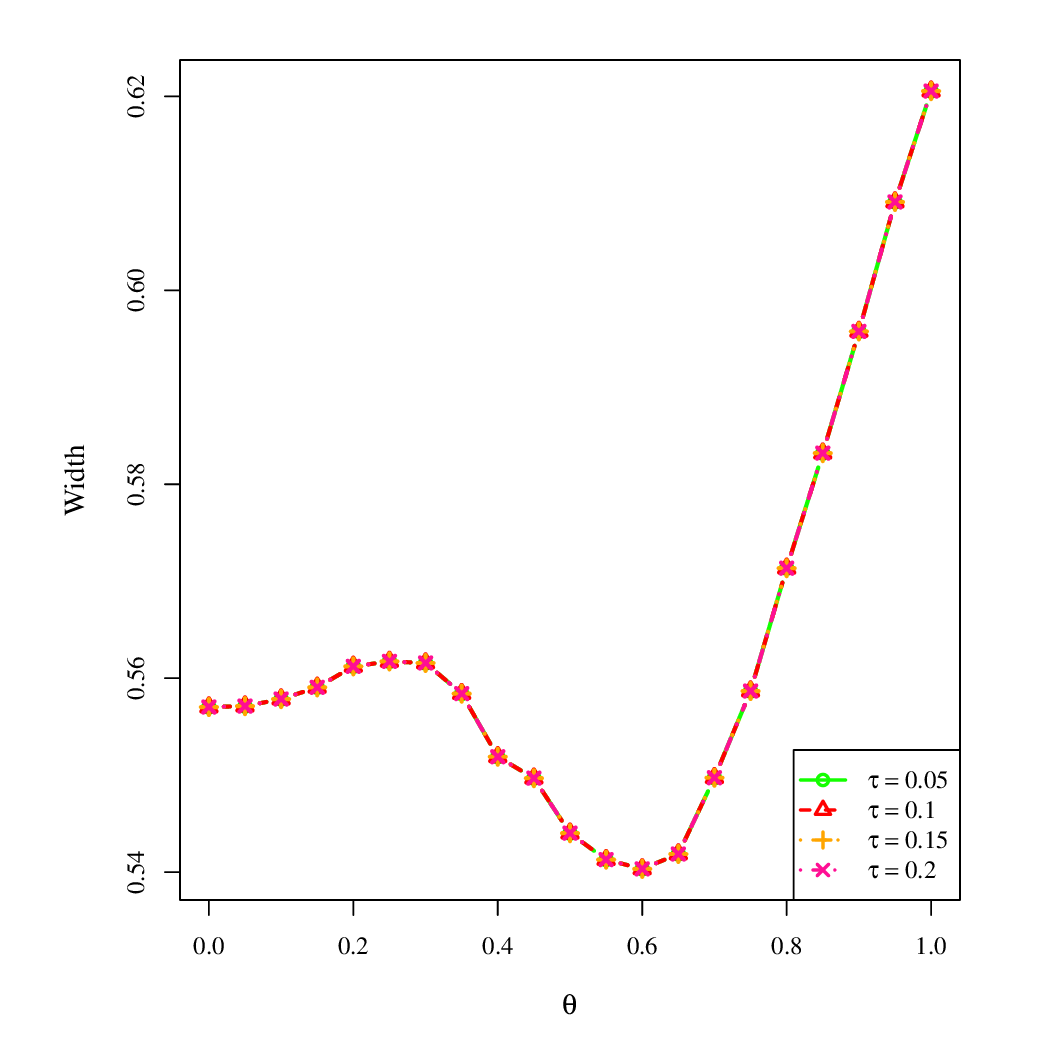}}
	\caption{\small Average widths of the $95\%$  confidence intervals for the proposed two-step inference method when $(n,p,\rho)=(350,25,0)$, $\delta_1=0.99$ and the threshold value $\tau$ varies.}
	\label{figure:coverwidthfixdelta1}
\end{figure}

\begin{figure}[t!]
	\centerline{\includegraphics[width=30em,angle=0]{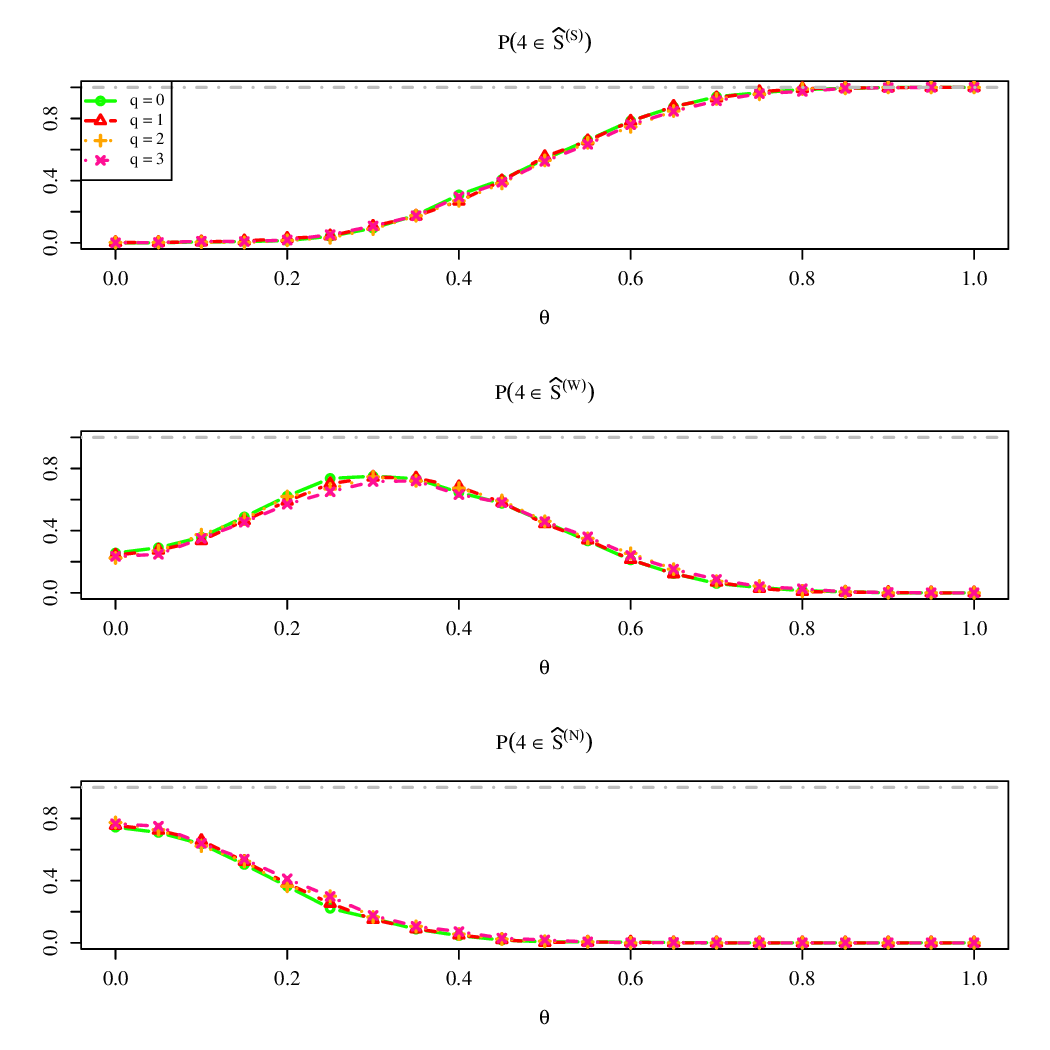}}
	\caption{\small Empirical probabilities of assigning the covariate $\bm X_4$ to different signal categories  when $(n,p,\rho)=(350,25,0)$,   $\delta_1=0.99$, $\tau=0.1$ and the total number of weak signals varies.}
	\label{figure:signalidentifychangeweak}
\end{figure}

\begin{figure}[t!]
	\centerline{\includegraphics[width=30em,angle=360]{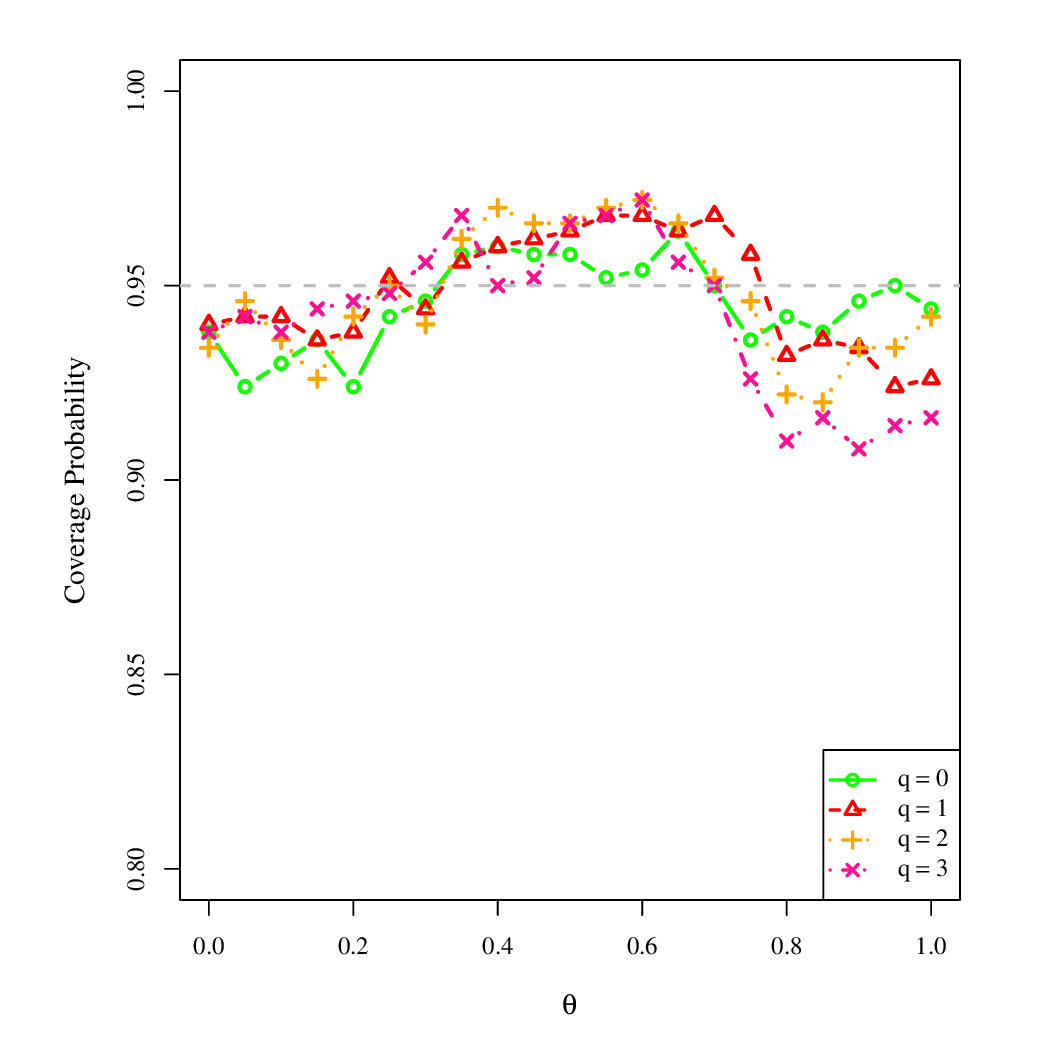}}
	\caption{\small Coverage probabilities of the $95\%$ confidence intervals for the proposed two-step inference method when $(n,p,\rho)=(350,25,0)$,   $\delta_1=0.99$, $\tau=0.1$ and the total number of weak signals varies.}
	\label{figure:coverprobchangeweak}
\end{figure}

\begin{figure}[t!]
	\centerline{\includegraphics[width=30em,angle=360]{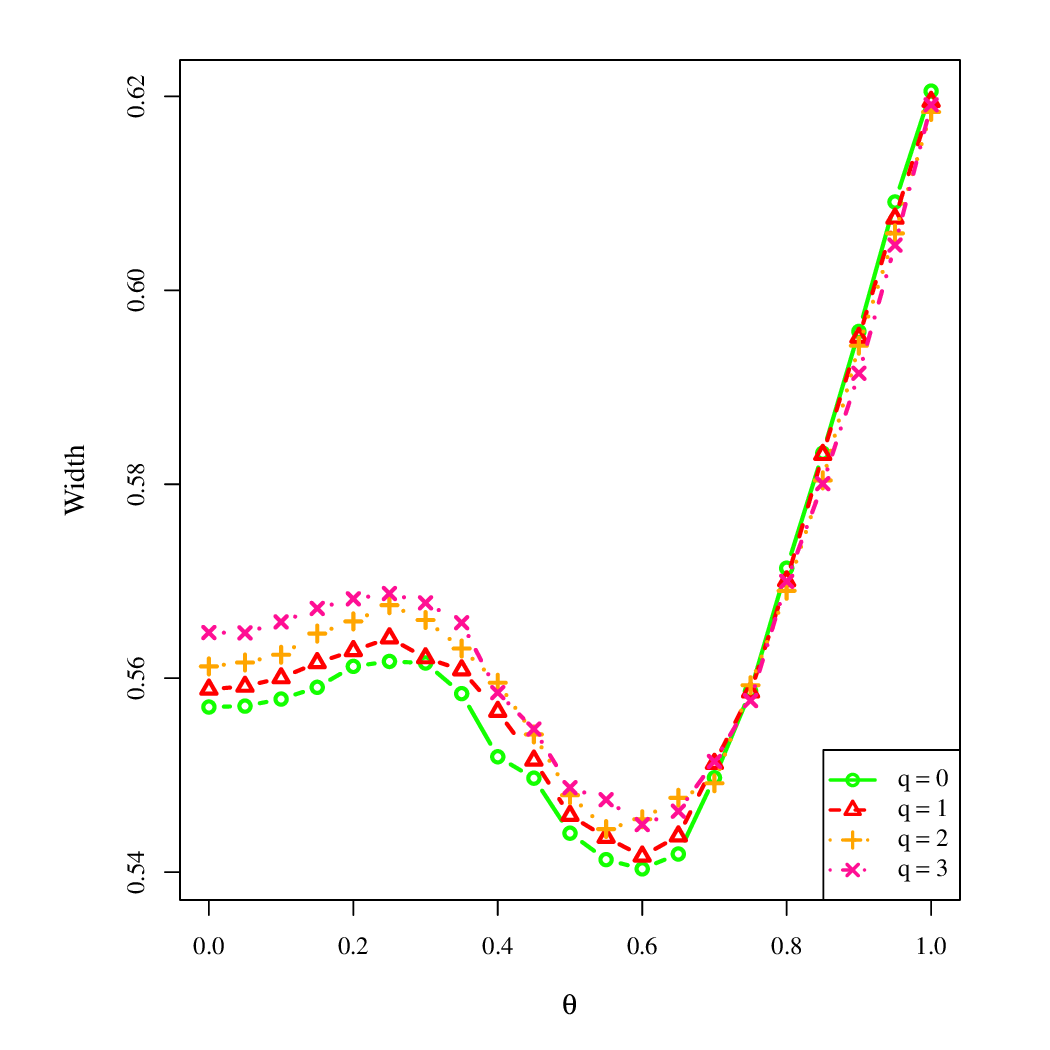}}
	\caption{\small Average widths of the $95\%$  confidence intervals for the proposed two-step inference method when $(n,p,\rho)=(350,25,0)$,   $\delta_1=0.99$, $\tau=0.1$ and the total number of weak signals varies.}
	\label{figure:coverwidthchangeweak}
\end{figure}

In this section, we present  additional simulation results under the simulation settings in Section 5. Figures \ref{figure:selectwcorr} and  \ref{figure:selectmcorr} display the results for different types of selection probability for $\bm X_4$ when $\rho=0.2$ and $0.5$, respectively.  Figures  \ref{figure:signalidentifywcorr} and   \ref{figure:signalidentifymcorr}  present the empirical probabilities of assigning the covariate $\bm X_4$ to different signal categories as the value of $\theta$ varies when $\rho= 0.2$ and $0.5$, respectively. Tables \ref{t:tableone}--\ref{t:tablefour}  show the coverage probabilities and average widths of  the $95\%$ confidence intervals under all  simulation settings. Figures 	\ref{figure:signalidentifyfixtau}--\ref{figure:coverwidthfixtau} show the simulation results for the proposed method when the threshold value $\delta_1$ varies. 
Figures 	\ref{figure:signalidentifyfixdelta1}--\ref{figure:coverwidthfixdelta1} show the simulation results for the proposed method when the threshold value $\tau$ varies.  Figures 	\ref{figure:signalidentifychangeweak}--\ref{figure:coverwidthchangeweak} show the simulation results for the proposed method when the total number of weak signals varies.

\begin{table*}[ht]
	\caption{The coverage probabilities (\%) of  the $95\%$ confidence intervals when the sample size is $n=350$.}
	\label{t:tableone}
	\begin{footnotesize}
		\begin{threeparttable}
			\begin{tabular*}{\textwidth}{@{}c@{\extracolsep{\fill}}c@{\extracolsep{\fill}}c@{\extracolsep{\fill}}c@{\extracolsep{\fill}}c@{\extracolsep{\fill}}c@{\extracolsep{\fill}}c@{\extracolsep{\fill}}c@{}}	
				\toprule
				& &\multicolumn{3}{c}{$p=25$}&\multicolumn{3}{c}{$p=35$}\\
				\cline{3-5} \cline{6-8}\\
				$\theta$ &Method&$\rho=0$&$\rho=0.2$&$\rho=0.5$&$\rho=0$&$\rho=0.2$&$\rho=0.5$ \\
				\midrule
				\multirow{11}{*}{$0$}
				&	Proposed	&	93.8	&	94.4	&	96.2	&	94.6	&	92.2	&	94.8	\\
				&	OldTwostep	&	75.8	&	76.7	&	81.4	&	77.1	&	66.9	&	72.3	\\
				&	Asym	&	3.6	&	3.8	&	12.7	&	4.3	&	1.4	&	4.0	\\
				&	MLE	&	93.8	&	94.4	&	96.2	&	94.6	&	92.2	&	94.8	\\
				&	Perturb	&	100.0	&	100.0	&	100.0	&	100.0	&	100.0	&	100.0	\\
				&EstEq &94.0 & 94.2 & 96.6 & 95.6 & 92.8 & 94.8 \\ 
				&	SdBS	&	99.8	&	100.0	&	99.8	&	99.8	&	99.8	&	99.0	\\
				&	SmBS	&	100.0	&	100.0	&	99.8	&	100.0	&	100.0	&	100.0	\\
				&	DeLasso	&	95.8	&	96.0	&	98.2	&	96.4	&	95.2	&	96.4	\\
				& BSDe1& 99.8 & 100.0 & 99.8 & 100.0 & 100.0 & 100.0 \\
				& BSDe2& 94.8 & 94.4 & 96.2 & 95.4 & 91.8 & 94.4 \\[1em]
				
				\multirow{11}{*}{$0.3$}
				&	Proposed	&	94.6	&	95.2	&	92.8	&	95.2	&	96.4	&	94.6	\\
				&	OldTwostep	&	96.9	&	96.6	&	92.0	&	98.0	&	96.7	&	92.4	\\
				&	Asym	&	75.5	&	71.6	&	61.5	&	65.8	&	69.6	&	69.3	\\
				&	MLE	&	92.2	&	93.4	&	92.6	&	92.4	&	92.0	&	93.6	\\
				&	Perturb	&	57.0	&	55.0	&	52.0	&	38.8	&	49.0	&	44.0	\\
				&EstEq & 92.2 & 92.6 & 93.8 & 92.6 & 91.6 & 94.2 \\ 
				&	SdBS	&	72.0	&	69.6	&	62.8	&	53.0	&	61.0	&	53.4	\\
				&	SmBS	&	65.2	&	64.6	&	59.8	&	39.8	&	49.4	&	47.8	\\
				&	DeLasso	&	93.8	&	94.0	&	92.8	&	93.0	&	93.4	&	95.0	\\
				& BSDe1& 52.0 & 58.0 & 85.6 & 48.6 & 60.6 & 86.4 \\
				& BSDe2& 94.2 & 94.6 & 95.0 & 96.2 & 95.0 & 95.2 \\ [1em]
				
				\multirow{11}{*}{$0.95$}
				&	Proposed	&	95.0	&	93.6	&	95.0	&	96.0	&	93.8	&	97.2	\\
				&	OldTwostep	&	95.0	&	93.6	&	95.4	&	96.0	&	93.8	&	97.2	\\
				&	Asym	&	95.0	&	93.6	&	91.6	&	96.0	&	93.8	&	92.2	\\
				&	MLE	&	90.0	&	91.6	&	91.2	&	87.8	&	87.8	&	86.8	\\
				&	Perturb	&	93.2	&	93.0	&	97.0	&	95.4	&	94.2	&	96.4	\\
				&EstEq &90.6 & 87.4 & 92.8 & 89.8 & 89.4 & 89.4 \\ 
				&	SdBS	&	93.8	&	93.8	&	95.6	&	93.4	&	93.4	&	95.6	\\
				&	SmBS	&	87.2	&	87.8	&	90.2	&	68.6	&	69.6	&	74.8	\\
				&	DeLasso	&	87.6	&	87.6	&	90.4	&	90.4	&	84.2	&	89.6	\\		
				& BSDe1& 23.0 & 26.0 & 34.8 & 17.8 & 15.4 & 26.4 \\ 
				& BSDe2& 94.8 & 95.6 & 97.4 & 94.4 & 95.0 & 95.6 \\ 
				\bottomrule
			\end{tabular*}
			\begin{tablenotes}	
				\item Note:  Proposed: the proposed two-step inference method; OldTwostep: the  two-step inference method based on  \cite{shi2017weak}, which does not construct confidence intervals for identified noise variables; Asym: the method based on the asymptotic theory using the one-step adaptive lasso estimator; MLE: the maximum likelihood estimation method; Perturb: the perturbation method; EstEq: the estimating equation-based method; SdBS: the standard bootstrap method; SmBS: the smoothed bootstrap method; DeLasso:  the de-biased lasso method; BSDe1: the type-I bootstrap de-biased lasso method; BSDe2: the type-II  bootstrap de-biased lasso method.
			\end{tablenotes}	
		\end{threeparttable}
	\end{footnotesize}
\end{table*}

\begin{table*}[ht]
	\caption{The coverage probabilities (\%) of  the $95\%$ confidence intervals when the sample size is $n=550$.}
	\label{t:tabletwo}
	\begin{footnotesize}
		\begin{threeparttable}
			\begin{tabular*}{\textwidth}{@{}c@{\extracolsep{\fill}}c@{\extracolsep{\fill}}c@{\extracolsep{\fill}}c@{\extracolsep{\fill}}c@{\extracolsep{\fill}}c@{\extracolsep{\fill}}c@{\extracolsep{\fill}}c@{}}	
				\toprule
				& &\multicolumn{3}{c}{$p=25$}&\multicolumn{3}{c}{$p=35$}\\
				\cline{3-5} \cline{6-8}\\
				$\theta$ &Method&$\rho=0$&$\rho=0.2$&$\rho=0.5$&$\rho=0$&$\rho=0.2$&$\rho=0.5$ \\
				\midrule
				
				\multirow{11}{*}{$0$}
				&	Proposed	&	95.4	&	94.8	&	95.4	&	94.6	&	94.2	&	95.8	\\
				&	OldTwostep	&	81.7	&	77.6	&	80.0	&	75.9	&	76.4	&	78.9	\\
				&	Asym	&	4.2	&	7.6	&	7.2	&	1.4	&	4.2	&	7.1	\\
				&	MLE	&	95.4	&	94.8	&	95.4	&	94.6	&	94.2	&	95.8	\\
				&	Perturb	&	99.8	&	100.0	&	100.0	&	100.0	&	100.0	&	100.0	\\
				&	EstEq &  95.6 & 93.8 & 95.6 & 95.2 & 95.0 & 96.8 \\ 
				&	SdBS	&	99.8	&	99.6	&	99.6	&	100.0	&	100.0	&	100.0	\\
				&	SmBS	&	99.8	&	100.0	&	100.0	&	100.0	&	100.0	&	100.0	\\
				&	DeLasso	&	96.6	&	95.4	&	97.0	&	96.4	&	95.8	&	97.4	\\
				&  BSDe1 & 99.8 & 100.0 & 99.8 & 100.0 & 100.0 & 100.0 \\ 
				& BSDe2 & 95.4 & 94.6 & 95.6 & 95.8 & 94.2 & 95.6 \\ [1em]
				
				\multirow{11}{*}{$0.25$}
				&	Proposed	&	94.4	&	95.6	&	95.0	&	95.4	&	93.8	&	95.6	\\
				&	OldTwostep	&	95.8	&	96.6	&	94.8	&	97.0	&	95.1	&	94.7	\\
				&	Asym	&	69.4	&	63.8	&	68.2	&	72.3	&	69.9	&	68.5	\\
				&	MLE	&	94.4	&	95.6	&	94.4	&	93.8	&	92.0	&	95.2	\\
				&	Perturb	&	57.4	&	52.8	&	56.2	&	54.8	&	55.2	&	54.6	\\
				& EstEq  & 93.6 & 95.0 & 93.8 & 93.4 & 91.4 & 94.8 \\ 
				&	SdBS	&	68.8	&	65.2	&	62.8	&	65.0	&	66.8	&	62.0	\\
				&	SmBS	&	67.8	&	66.0	&	63.6	&	61.6	&	62.8	&	64.4	\\
				&	DeLasso	&	93.0	&	94.8	&	94.4	&	94.0	&	93.0	&	95.8	\\
				& BSDe1 & 52.8 & 57.2 & 79.2 & 49.2 & 57.4 & 79.6 \\ 
				& BSDe2  & 94.2 & 96.4 & 94.8 & 95.2 & 96.0 & 96.0 \\ [1em]
				
				\multirow{11}{*}{$0.8$}
				&	Proposed	&	94.2	&	94.4	&	93.8	&	95.0	&	95.0	&	92.2	\\
				&	OldTwostep	&	94.2	&	94.4	&	93.8	&	95.0	&	95.0	&	92.2	\\
				&	Asym	&	94.2	&	94.4	&	90.6	&	95.0	&	95.0	&	89.0	\\
				&	MLE	&	93.6	&	94.4	&	92.6	&	90.4	&	89.4	&	91.2	\\
				&	Perturb	&	90.2	&	93.0	&	97.0	&	93.8	&	94.2	&	95.8	\\
				& EstEq  & 92.4 & 93.0 & 90.6 & 90.4 & 92.4 & 91.6 \\ 
				&	SdBS	&	91.2	&	93.8	&	96.2	&	91.8	&	91.6	&	94.2	\\
				&	SmBS	&	88.4	&	93.8	&	94.4	&	87.0	&	86.0	&	91.2	\\
				&	DeLasso	&	87.0	&	90.2	&	89.4	&	89.0	&	87.2	&	90.4	\\	
				& BSDe1 & 23.0 & 26.0 & 41.2 & 15.8 & 18.8 & 33.8 \\ 	
				& BSDe2 & 96.4 & 97.2 & 94.4 & 93.8 & 95.8 & 95.2 \\ 			
				\bottomrule
			\end{tabular*}
			\begin{tablenotes}	
				\item Note:   Proposed: the proposed two-step inference method; OldTwostep: the  two-step inference method based on  \cite{shi2017weak}, which does not construct confidence intervals for identified noise variables; Asym: the method based on the asymptotic theory using the one-step adaptive lasso estimator; MLE: the maximum likelihood estimation method; Perturb: the perturbation method; EstEq: the estimating equation-based method; SdBS: the standard bootstrap method; SmBS: the smoothed bootstrap method; DeLasso:  the de-biased lasso method; BSDe1: the type-I bootstrap de-biased lasso method; BSDe2: the type-II  bootstrap de-biased lasso method.
			\end{tablenotes}	
		\end{threeparttable}
	\end{footnotesize}
\end{table*}

\begin{table*}[ht]
	\caption{The widths ($\times 100$) of  the $95\%$ confidence intervals when the sample size is $n=350$}
	\label{t:tablethree}
	\begin{footnotesize}
		\begin{threeparttable}
			\begin{tabular*}{\textwidth}{@{}c@{\extracolsep{\fill}}c@{\extracolsep{\fill}}c@{\extracolsep{\fill}}c@{\extracolsep{\fill}}c@{\extracolsep{\fill}}c@{\extracolsep{\fill}}c@{\extracolsep{\fill}}c@{}}	
				\toprule
				& &\multicolumn{3}{c}{$p=25$}&\multicolumn{3}{c}{$p=35$}\\
				\cline{3-5} \cline{6-8}\\
				$\theta$ &Method&$\rho=0$&$\rho=0.2$&$\rho=0.5$&$\rho=0$&$\rho=0.2$&$\rho=0.5$ \\
				\midrule
				\multirow{11}{*}{$0$}
				&Proposed & 55.7 & 60.0 & 78.4 & 58.2 & 62.7 & 82.1 \\ 
				&OldTwostep & 55.9 & 60.8 & 79.7 & 58.6 & 63.2 & 82.8 \\ 
				&Asym & 19.6 & 21.6 & 22.3 & 19.7 & 18.9 & 23.3 \\ 
				&MLE & 55.7 & 60.0 & 78.4 & 58.2 & 62.8 & 82.1 \\ 
				&Perturb & 14.5 & 14.7 & 17.6 & 10.3 & 11.1 & 13.9 \\ 
				&EstEq & 50.4 & 53.9 & 70.1 & 51.1 & 54.7 & 71.0 \\ 
				&SdBS & 22.9 & 23.6 & 27.9 & 17.2 & 17.9 & 21.7 \\ 
				&SmBS & 16.6 & 16.8 & 19.4 & 11.4 & 11.9 & 14.3 \\ 
				&DeLasso & 48.7 & 51.9 & 66.8 & 49.4 & 52.6 & 67.5 \\ 
				&BSDe1 & 49.6 & 52.8 & 67.7 & 50.6 & 54.0 & 68.9 \\ 
				&BSDe2 & 58.7 & 63.2 & 82.8 & 63.6 & 69.0 & 90.6 \\ [1em]
				
				\multirow{11}{*}{$0.3$}
				& Proposed & 56.2 & 60.5 & 79.5 & 58.6 & 63.1 & 83.8 \\ 
				&OldTwostep & 56.2 & 60.6 & 79.1 & 58.6 & 63.0 & 83.9 \\ 
				&Asym & 33.5 & 34.0 & 35.0 & 30.2 & 32.8 & 35.8 \\ 
				&MLE & 57.0 & 61.6 & 80.7 & 59.5 & 64.5 & 84.9 \\ 
				&Perturb & 49.6 & 51.7 & 55.9 & 40.5 & 47.1 & 50.3 \\ 
				&EstEq &51.0 & 54.8 & 71.6 & 51.6 & 55.4 & 72.5 \\ 
				&SdBS & 51.6 & 53.4 & 58.4 & 41.1 & 45.8 & 49.5 \\ 
				&SmBS& 46.0 & 47.4 & 50.1 & 34.6 & 39.1 & 40.8 \\ 
				&DeLasso & 49.4 & 52.9 & 68.3 & 49.7 & 53.2 & 68.6 \\ 
				&BSDe1 & 51.2 & 54.9 & 70.3 & 52.7 & 56.4 & 72.5 \\ 
				&BSDe2 & 62.8 & 67.6 & 88.0 & 68.8 & 74.8 & 98.5 \\ [1em]
				
				\multirow{11}{*}{$0.95$}
				&Proposed & 60.9 & 63.9 & 73.4 & 62.0 & 64.9 & 75.1 \\ 
				&OldTwostep & 60.9 & 63.9 & 73.3 & 62.0 & 64.9 & 75.1 \\ 
				&Asym & 60.9 & 63.8 & 71.0 & 62.0 & 64.8 & 71.8 \\ 
				&MLE & 68.6 & 73.7 & 93.7 & 72.9 & 78.1 & 100.5 \\ 
				&Perturb & 67.4 & 70.4 & 91.6 & 71.1 & 76.1 & 103.2 \\ 
				&EstEq &  57.4 & 61.6 & 78.9 & 57.9 & 61.6 & 79.3 \\ 
				&SdBS & 67.6 & 70.4 & 87.4 & 67.2 & 70.4 & 86.4 \\ 
				&SmBS& 60.8 & 63.6 & 79.7 & 57.9 & 61.0 & 75.9 \\ 
				&DeLasso & 53.5 & 56.8 & 72.9 & 53.6 & 57.0 & 73.2 \\ 
				&BSDe1 & 56.0 & 60.2 & 77.7 & 58.0 & 61.8 & 80.6 \\ 
				&BSDe2 & 84.5 & 91.8 & 115.8 & 100.8 & 108.1 & 137.6 \\ 
				\bottomrule
			\end{tabular*}
			\begin{tablenotes}	
				\item  Note:  Proposed: the proposed two-step inference method; OldTwostep: the  two-step inference method based on  \cite{shi2017weak}, which does not construct confidence intervals for identified noise variables; Asym: the method based on the asymptotic theory using the one-step adaptive lasso estimator; MLE: the maximum likelihood estimation method; Perturb: the perturbation method; EstEq: the estimating equation-based method; SdBS: the standard bootstrap method; SmBS: the smoothed bootstrap method; DeLasso:  the de-biased lasso method; BSDe1: the type-I bootstrap de-biased lasso method; BSDe2: the type-II  bootstrap de-biased lasso method.
			\end{tablenotes}	
		\end{threeparttable}
	\end{footnotesize}
\end{table*}

\begin{table*}[ht]
	\caption{The widths ($\times 100$) of  the $95\%$ confidence intervals when the sample size is $n=550$}
	\label{t:tablefour}
	\begin{footnotesize}
		\begin{threeparttable}
			\begin{tabular*}{\textwidth}{@{}c@{\extracolsep{\fill}}c@{\extracolsep{\fill}}c@{\extracolsep{\fill}}c@{\extracolsep{\fill}}c@{\extracolsep{\fill}}c@{\extracolsep{\fill}}c@{\extracolsep{\fill}}c@{}}	
				\toprule
				& &\multicolumn{3}{c}{$p=25$}&\multicolumn{3}{c}{$p=35$}\\
				\cline{3-5} \cline{6-8}\\
				$\theta$ &Method&$\rho=0$&$\rho=0.2$&$\rho=0.5$&$\rho=0$&$\rho=0.2$&$\rho=0.5$ \\
				\midrule
				\multirow{11}{*}{$0$}
				&Proposed & 42.7 & 45.9 & 59.9 & 43.7 & 47.0 & 61.5 \\ 
				&OldTwostep & 42.8 & 46.2 & 60.3 & 44.0 & 47.2 & 61.6 \\ 
				&Asym & 14.8 & 15.3 & 17.0 & 13.7 & 14.7 & 17.0 \\ 
				&MLE & 42.7 & 45.9 & 59.9 & 43.7 & 47.0 & 61.5 \\ 
				&Perturb& 12.6 & 13.3 & 17.1 & 9.7 & 10.8 & 12.7 \\ 
				&EstEq & 39.8 & 42.7 & 55.5 & 40.1 & 42.8 & 55.8 \\ 
				&SdBS & 19.4 & 19.9 & 25.2 & 16.0 & 17.3 & 20.3 \\ 
				&SmBS & 15.1 & 15.3 & 19.4 & 11.8 & 12.9 & 14.7 \\ 
				&DeLasso & 38.6 & 41.2 & 53.2 & 38.8 & 41.4 & 53.5 \\ 
				&BSDe1 & 38.8 & 41.4 & 53.1 & 39.1 & 41.6 & 53.6 \\ 
				&BSDe2 & 43.0 & 46.1 & 60.3 & 44.6 & 48.0 & 62.9 \\ [1em]
				
				\multirow{11}{*}{$0.25$}
				& Proposed & 42.7 & 46.2 & 60.6 & 43.7 & 47.2 & 62.3 \\ 
				&OldTwostep & 42.7 & 46.2 & 60.6 & 43.7 & 47.2 & 62.0 \\ 
				&Asym & 25.7 & 25.2 & 28.9 & 25.8 & 26.4 & 27.5 \\ 
				&MLE & 43.4 & 46.7 & 61.2 & 44.5 & 48.0 & 62.9 \\ 
				&Perturb& 40.7 & 41.7 & 47.7 & 39.1 & 41.2 & 46.2 \\ 
				&EstEq & 40.2 & 43.1 & 56.3 & 40.4 & 43.3 & 56.7 \\ 
				&SdBS & 42.4 & 43.8 & 49.8 & 40.0 & 41.7 & 47.8 \\ 
				&SmBS & 40.2 & 41.4 & 46.0 & 37.3 & 39.0 & 43.6 \\ 
				&DeLasso & 39.0 & 41.7 & 54.0 & 39.2 & 41.7 & 54.2 \\ 
				&BSDe1 & 39.9 & 42.7 & 54.8 & 40.4 & 43.5 & 55.7 \\ 
				&BSDe2 & 45.1 & 48.3 & 62.8 & 47.3 & 51.0 & 66.5 \\ [1em]
				
				\multirow{11}{*}{$0.8$}
				&Proposed & 45.5 & 47.8 & 54.9 & 46.1 & 48.1 & 54.8 \\ 
				&OldTwostep & 45.5 & 47.8 & 54.9 & 46.1 & 48.1 & 54.8 \\ 
				&Asym & 45.5 & 47.8 & 53.6 & 46.1 & 48.1 & 53.6 \\ 
				&MLE & 49.4 & 53.1 & 68.0 & 51.1 & 54.7 & 70.2 \\ 
				&Perturb & 50.5 & 53.3 & 69.3 & 51.5 & 53.5 & 70.2 \\ 
				&EstEq &43.9 & 47.1 & 60.8 & 44.2 & 47.2 & 60.9 \\ 
				&SdBS& 49.3 & 52.0 & 66.2 & 48.9 & 50.9 & 64.4 \\ 
				&SmBS & 48.9 & 51.6 & 65.8 & 47.3 & 49.6 & 63.2 \\ 
				&DeLasso & 41.4 & 44.2 & 56.8 & 41.6 & 43.9 & 57.2 \\ 
				&BSDe1 & 42.9 & 45.8 & 59.2 & 43.3 & 46.7 & 60.4 \\ 
				&BSDe2 & 54.6 & 58.6 & 74.9 & 59.0 & 63.2 & 81.5 \\ 
				\bottomrule
			\end{tabular*}
			\begin{tablenotes}	
				\item  Note:  Proposed: the proposed two-step inference method; OldTwostep: the  two-step inference method based on  \cite{shi2017weak}, which does not construct confidence intervals for identified noise variables; Asym: the method based on the asymptotic theory using the one-step adaptive lasso estimator; MLE: the maximum likelihood estimation method; Perturb: the perturbation method; EstEq: the estimating equation-based method; SdBS: the standard bootstrap method; SmBS: the smoothed bootstrap method; DeLasso:  the de-biased lasso method; BSDe1: the type-I bootstrap de-biased lasso method; BSDe2: the type-II  bootstrap de-biased lasso method.
			\end{tablenotes}	
		\end{threeparttable}
	\end{footnotesize}
\end{table*}

\section{Additional Information in  Real-data Application \label{ss:addreal}}

Table \ref{t:tablefive} shows the candidate predictors used in the real-data application.

\begin{sidewaystable*}
	\caption{The candidate predictors used in the real-data analysis}
	\label{t:tablefive}
	\begin{tabular}{rl}
		\toprule
		Category & Predictor\\
		\hline
		\multirow{3}{*}{Basic information}	& year of birth\\
		& gender\\
		&3 predictors indicating whether a patient is from California, Texas, New York or  other states\\[3ex]
		\multirow{11}{*}{Transcript records}	&range of BMI\\
		& the median of weights\\
		& the median of heights\\
		& the median of systolic blood pressures\\
		& the medians of Diastolic blood pressures\\
		& the median of respiratory rates\\
		& the median of temperatures\\
		& 4 predictors corresponding to the numbers of transcripts for different physician specialties\\
		& number of physicians\\
		& number of transcripts with blank visit year\\
		& number of visits per weighted year\\[3ex]
		\multirow{5}{*}{Diagnosis information}	&69 predictors corresponding to the numbers of times being diagnosed with different diagnoses\\
		& number of diagnoses per weighted year\\
		& number of different 3 digits diagnostics groups in the icd9 table\\
		& number of different 3 digits diagnostics groups with medication\\[3ex]
		\multirow{4}{*}{Medication information}	 &23 predictors indicating the dose of active principle\\
		&number of prescriptions or the use of different medications\\
		&  number of medications without prescript\\
		& number of active principles\\[3ex]
		Lab result&1 binary variable indicating whether a patient has any lab test or not\\[3ex]
		Smoking status& 1 binary variable indicating whether a patient smoked in the past\\
		\bottomrule 		
	\end{tabular}
\end{sidewaystable*}

\end{document}